\documentclass [invmat,draft,numbook]{svjour} 
\usepackage[adobe-utopia]{mathdesign}

\usepackage[OMLmathbf,sfdefault=cmbr]{isomath}

\def\bos#1{{\mathbf{#1}}}
\def\mb#1{\hbox{{\mathversion{bold}$#1$}}}
\def\mbs#1{\hbox{{\footnotesize{\mathversion{bold}$#1$}}}}

\journalname{Communications in Mathematical Physics}

\begin{document}


\catcode`\@=11 

\global\newcount\nsecno \global\nsecno=0
\global\newcount\meqno \global\meqno=1
\def\newsec#1{\global\advance\nsecno by1
\eqnres@t
\section{#1}}
\def\eqnres@t{\xdef\nsecsym{\the\nsecno.}\global\meqno=1}
\def\sequentialequations{\def\eqnres@t{\bigbreak}}\xdef\nsecsym{}

\def\draftmode{\message{ DRAFTMODE }
\writelabels

{\count255=\time\divide\count255 by 60 \xdef\hourmin{\number\count255}
\multiply\count255 by-60\advance\count255 by\time
\xdef\hourmin{\hourmin:\ifnum\count255<10 0\fi\the\count255}}}
\def\nolabels{\def\wrlabeL##1{}\def\eqlabeL##1{}\def\reflabeL##1{}}
\def\writelabels{\def\wrlabeL##1{\leavevmode\vadjust{\rlap{\smash%
{\line{{\escapechar=` \hfill\rlap{\tt\hskip.03in\string##1}}}}}}}%
\def\eqlabeL##1{{\escapechar-1\rlap{\tt\hskip.05in\string##1}}}%
\def\reflabeL##1{\noexpand\llap{\noexpand\sevenrm\string\string\string##1}
}}

\nolabels

\def\eqn#1#2{
\xdef #1{(\nsecsym\the\meqno)}
\global\advance\meqno by1
$$#2\eqno#1\eqlabeL#1
$$}

\def\eqalign#1{\null\,\vcenter{\openup\jot\m@th
\ialign{\strut\hfil$\displaystyle{##}$&$\displaystyle{{}##}$\hfil
\crcr#1\crcr}}\,}

\def\foot#1{\footnote{#1}}

\catcode`\@=12 
%

\def\a{\alpha}
\def\b{\beta}
\def\c{\chi}
\def\d{\delta}  \def\D{\Delta}
\def\e{\varepsilon} \def\ep{\epsilon}
\def\f{\phi}  \def\F{\Phi}
\def\g{\gamma}  \def\G{\Gamma}
\def\k{\kappa}
\def\l{\lambda}  \def\La{\Lambda}
\def\m{\mu}
\def\n{\nu}
\def\r{\rho}
\def\vr{\varrho}
\def\o{\omega}  \def\O{\Omega}
\def\p{\psi}  \def\P{\Psi}
\def\s{\sigma}  \def\S{\Sigma}
\def\th{\theta}  \def\vt{\vartheta}
\def\t{\tau}
\def\w{\varphi}
\def\x{\xi}
\def\z{\zeta}
\def\U{\Upsilon}
\def\CA{{\cal A}}
\def\CB{{\cal B}}
\def\CC{{\cal C}}
\def\CD{{\cal D}}
\def\CE{{\cal E}}
\def\CF{{\cal F}}
\def\CG{{\cal G}}
\def\CH{{\cal H}}
\def\CI{{\cal I}}
\def\CJ{{\cal J}}
\def\CK{{\cal K}}
\def\CL{{\cal L}}
\def\CM{{\cal M}}
\def\CN{{\cal N}}
\def\CO{{\cal O}}
\def\CP{{\cal P}}
\def\CQ{{\cal Q}}
\def\CR{{\cal R}}
\def\CS{{\cal S}}
\def\CT{{\cal T}}
\def\CU{{\cal U}}
\def\CV{{\cal V}}
\def\CW{{\cal W}}
\def\CX{{\cal X}}
\def\CY{{\cal Y}}
\def\CZ{{\cal Z}}
%

\def\V{\mathbb{V}}
\def\E{\mathbb{E}}
\def\R{\mathbb{R}}
\def\C{\mathbb{C}}
\def\Z{\mathbb{Z}}
\def\A{\mathbb{A}}
\def\T{\mathbb{T}}
\def\L{\mathbb{L}}
\def\D{\mathbb{D}}
\def\Q{\mathbb{Q}}


\def\mJ{\mathfrak{J}}
\def\mq{\mathfrak{q}}
\def\mQ{\mathfrak{Q}}
\def\mP{\mathfrak{P}}
\def\mp{\mathfrak{p}}
\def\mH{\mathfrak{H}}
\def\mh{\mathfrak{h}}
\def\ma{\mathfrak{a}}
\def\mA{\mathfrak{A}}
\def\mC{\mathfrak{C}}
\def\mc{\mathfrak{c}}
\def\ms{\mathfrak{s}}
\def\mS{\mathfrak{S}}
\def\mm{\mathfrak{m}}
\def\mM{\mathfrak{M}}
\def\mn{\mathfrak{n}}
\def\mN{\mathfrak{N}}
\def\mt{\mathfrak{t}}
\def\ml{\mathfrak{l}}
\def\mT{\mathfrak{T}}
\def\mL{\mathfrak{L}}
\def\mo{\mathfrak{o}}
\def\mg{\mathfrak{g}}
\def\mG{\mathfrak{G}}
\def\mf{\mathfrak{f}}
\def\md{\mathfrak{d}}
\def\mD{\mathfrak{D}}
\def\mO{\mathfrak{O}}
\def\mk{\mathfrak{k}}
\def\mK{\mathfrak{K}}
\def\mR{\mathfrak{R}}
\def\sA{\mathscr{A}}
\def\sB{\mathscr{B}}
\def\sC{\mathscr{C}}
\def\sD{\mathscr{D}}
\def\sE{\mathscr{E}}
\def\sF{\mathscr{F}}
\def\sG{\mathscr{G}}
\def\sL{\mathscr{L}}
\def\sM{\mathscr{M}}
\def\sN{\mathscr{N}}
\def\sO{\mathscr{O}}
\def\sP{\mathscr{P}}
\def\sQ{\mathscr{Q}}
\def\sR{\mathscr{R}}
\def\sS{\mathscr{S}}
\def\sT{\mathscr{T}}
\def\sU{\mathscr{U}}
\def\sV{\mathscr{V}}
\def\sW{\mathscr{W}}
\def\sX{\mathscr{X}}
\def\sY{\mathscr{Y}}
\def\sZ{\mathscr{Z}}


\hyphenation{anom-aly anom-alies coun-ter-term coun-ter-terms
}

\def\tr{{\rm tr}} \def\Tr{{\rm Tr}}

\def\tilde{\widetilde} 
\def\hat{\widehat}
%

\def\grad#1{\,\nabla\!_{{#1}}\,}
\def\gradgrad#1#2{\,\nabla\!_{{#1}}\nabla\!_{{#2}}\,}
\def\ph{\varphi}
\def\psibar{\overline\psi}
\def\om#1#2{\omega^{#1}{}_{#2}}
\def\vev#1{\langle #1 \rangle}
\def\ha{{1\over2}}
\def\half{{\textstyle{1\over2}}} 
\def\roughly#1{\raise.3ex\hbox{$#1$\kern-.75em\lower1ex\hbox{$\sim$}}}

\def\rd{\partial}
\def\ha{{\textstyle{1\over2}}}
\def\fr#1#2{{\textstyle{#1\over#2}}}
\def\Fr#1#2{{#1\over#2}}
\def\fs#1{#1\!\!\!/\,}   
\def\Fs#1{#1\!\!\!\!/\,} 
\def\ato#1{{\buildrel #1\over\longrightarrow}}
\def\up#1#2{{\buildrel #1\over #2}}

\def\pr{\prime}
\def\ppr{{\prime\prime}}

\def\bari{\bar\imath}
\def\barj{\bar\jmath}
\def\mapr#1{\!\smash{\mathop{\longrightarrow}\limits^{#1}}\!}
\def\mapl#1{\!\smash{\mathop{\longleftarrow}\limits^{#1}}\!}
\def\mapbr{\!\smash{\mathop{\longrightarrow}\limits^{\bbs_+}}\!}
\def\mapbl{\!\smash{\mathop{\longleftarrow}\limits^{\bbs_-}}\!}
\def\mapd#1{\Big\downarrow\rlap{$\vcenter{#1}$}}
\def\mapu#1{\Big\uparrow\rlap{$\vcenter{#1}$}}
\def\maprd{\rlap{\lower.3ex\hbox{$\scriptstyle\bs_+$}}\searrow}
\def\mapld{\swarrow\!\!\!\rlap{\lower.3ex\hbox{$\scriptstyle\bs_-$}}}
\def\ne{\nearrow}
\def\se{\searrow}
\def\nw{\nwarrow}
\def\sw{\swarrow}
\def\xil{et al.}
\def\Ker{\hbox{Ker}\;}

\def\ket#1{\left|\bos{ #1}\right>}\vspace{.2in}
   \def\bra#1{\left<\bos{ #1}\right|}
\def\oket#1{\left.\bos{ #1}\right>}
\def\obra#1{\left<\bos{ #1}\right.}
\def\epv#1#2#3{\left<\bos{#1}\left|\bos{#2}\right|\bos{#3}\right>}
\def\qbvk#1#2{\bos{\left(\bos{#1},\bos{#2}\right)}}
\def\Hoch{{\tt Hoch}}
\def\rrd{\up{\rightarrow}{\rd}}
\def\lrd{\up{\leftarrow}{\rd}}


\def\mod{\hbox{ }mod\hbox{ }}

\def\mF{\matrixsym{F}}


\title{Algebraic Principles 
of Quantum Field Theory II}
\subtitle{Quantum Coordinates and WDVV Equation}
\author{Jae-Suk Park\thanks{
This work was supported by Mid-career Researcher Program through NRF grant funded by the MEST (No. 2010-0000497).}
}

\institute{
Department of Mathematics,
Yonsei University,
Seoul 120-749, Korea.\\
\email{jaesuk@yonsei.ac.kr}
}

\date{}

\maketitle

\begin{abstract}

This paper is about algebro-geometrical structures on a moduli space
$\CM$ of anomaly-free BV QFTs with finite number of inequivalent observables or in a finite superselection sector. 
We show that $\CM$ has
the structure of F-manifold -- a linear pencil of torsion-free flat connection with unity on the tangent space, in quantum
coordinates.  We study the notion of quantum coordinates for the family of QFTs, which determines the
connection $1$-form as well as  every quantum correlation function of the family 
in terms of the $1$-point functions of the initial theory. We then define free energy for
an unital BV QFT and show that it is another avatar of morphism of QFT algebra. 
These results are consequences of the solvability of refined quantum master equation of the theory.
We also introduce the notion of a QFT integral and study some
properties of BV QFT equipped with a QFT integral.
We show that  BV QFT with a non-degenerate QFT integral leads to  the WDVV
equation---the formal Frobenius manifold structure on $\CM$---if it admits
a semi-classical solution of quantum master equation.

\end{abstract}

\newsec{Introduction}

This the $2$nd installment of the series of papers in a quest to find algebraic principles of general quantum field theory.
A large part of this paper 
is a natural continuation  of the last chapter  of the previous one \cite{PI},  dealing with 
the same class of quantum field theory called anomaly-free BV QFT
with a finite number of equivalence classes 
of observables or  in a finite superselection sector.  
We have shown that such a theory comes with its family parametrized by 
a  formal smooth super (moduli) space  $\CM$ with a base point, which corresponds to the initial QFT,
 in quantum coordinates. It was  a natural consequence of  the existence of solution 
to  quantum master equation, which automatically gives a distinguished solution 
to the Batalin-Vilkovisky (BV) quantum master equation.  
Our quantum master equation was shown to govern the  
quantization of classical correlators such that its solution can be used to determine
every quantum correlation functions from a generating set of $1$-point functions.

In this paper we shall refine the previous notion of quantum master equation 
in its ``up to homotopy'' part and find and its solution. 
We shall also introduce a QFT integral, which is another piece of data
possibly carried by a BV QFT. 
An immediate consequence shall be that the tangent space $T\!\CM$ of the moduli space $\CM$ 
of a BV QFT with an QFT integral has certain algebraic structure, 
which, in special non-degenerate, semi-classical cases, reduces to that of
K.\ Saito \cite{Saito} and, independently,
 Witten-Dijkgraaf-Verlinde-Verlinde (WDVV) \cite{W,DVV}, 
a.k.a. the Frobenius manifold structure formalized and studied in details 
by Dubrovin \cite{Dubrovin}.

For us, the refined quantum master equation shall pave a way for understanding
both morphisms of QFT algebras up to homotopy and homotopy path
integrals, which 
are the subjects of the $3$rd paper in this series \cite{PIII}.
We also isolate the notion of QFT complex
from the definition of BV QFT algebra in \cite{PI}, which is the core structure 
of general QFT algebras to appear
in sequels.

The solvability of the refined quantum master equation implies that there is  
$3$-tensor  $A_{\a\b}{}^\g$, which is a formal power series in an affine coordinates 
$t_H=\{t^\a\}$ of the graded space vector space $H$ of equivalence class of observables, 
on $\CM$ satisfying

\begin{enumerate}
\item symmetry (graded commutativity and potentiality):
$$
A_{\a\b}{}^\g=(-1)^{|t^\a||t^\b|}A_{\b\a}{}^\g,\qquad
\Fr{\rd\! A_{\b\g}{}^\s}{\rd\! t^\a} -(-1)^{|t^\a||t^\b|}\Fr{\rd\! A_{\b\g}{}^\s}{\rd\! t^\a}=0,
 $$
 \item relation (associativity):
 $$
  \sum_\r A_{\a\b}{}^\r A_{\r\g}{}^\s = \sum_\r A_{\b\g}{}^\r A_{\a\r}{}^\s,
$$
\item unity:  there is distinguished   distinguish element $t^0$ such that $A_{0\b}{}^\g = \d_\b{}^\g$,

\item homogeneity (Euler vector): 
$$
\sum_\r |t^\r| t^\r \Fr{\rd}{\rd\! t^\r} A_{\a\b}{}^\g=(|t^\g|-|t^\b|-|t^\a|)A_{\a\b}{}^\g.
$$
\end{enumerate}

We remark that a formal super-manifold with such a $3$-tensor was
called an F-manifold by Herling-Manin and studied in \cite{HM,M}.
The symmetry of the $3$-tensor $A_{\a\b}{}^\g$ implies that
$
A_{\a\b}{}^\g = \Fr{\rd^2 \Phi^\g }{\rd\! t^\a\rd\! t^\b},
 $
 for some $\{\Phi^\g\}$, so that the associativity relation and unity imply that
 \eqn\wwdvv{
\eqalign{
\sum_\r \left( \Fr{\rd^2 \Phi^\r }{\rd\! t^\a\rd\! t^\b}\right)\left( \Fr{\rd^2 \Phi^\s }{\rd\! t^\r\rd\! t^\g}\right)
 &=\sum_\r \left( \Fr{\rd^2 \Phi^\r }{\rd\! t^\b\rd\! t^\g}\right)\left( \Fr{\rd^2 \Phi^\s }{\rd\! t^\a\rd\! t^\r}\right)
 ,\cr
 \Fr{\rd^2 \Phi^\g }{\rd\! t^0\rd\! t^\b}&=\d_\b{}^\g.
 } 
 }
 The above equation may be viewed as a weaker version of WDDV equation, though its solution
 $\{\Phi^\g\}$ is not particularly relevant for us. 
 Rather, the role of $A_{\a\b}{}^\g$ as quantum correlation functions 
 for the full family of BV QFT in 
 quantum coordinates is our greater interest.

The coefficients  of an expansion of $A_{\a\b}{}^\g$ at $t_H=0$:
 $$
 A_{\a\b}{}^\g 
 =m_{\a\b}{}^\s 
 +\sum_\r t^\r m_{\r\a\b}{}^\s 
 +\Fr{1}{2!}\sum_{\r_1,\r_2} t^{\r_2}t^{\r_1} m_{\r_1\r_2\a\b}{}^\g
 +\cdots
$$
correspond to the
structure constants of the sequence $m_2,m_3, m_4,\cdots$ of graded symmetric
products of ghost number zero on $H$, $m_n:S^n H\rightarrow H$, which
were discussed in the previous
paper and, together with the $1$-point functions
$\{\left<\bos{O}_\a\right>\}$ and the QFT cycle, were used 
to determine all of the $n$-point quantum correlation function
of the initial BV QFT.
In this paper we shall show that, in fact, $A_{\a\b}{}^\g$ can be used to
determine
all $n$-point correlation functions for every BV QFT in the family parametrized by
$\CM$.
The key concept is that of quantum coordinates $\{\mb{T}^\g\}$ for
the family, which are
formal power series in $t_H$ and $\hbar^{-1}$ involving the coefficients of $A_{\a\b}{}^\g$ 
in the following form
$$
\eqalign{
\mb{T}^\g =& t^\g 
-\Fr{1}{2\hbar}\sum_{\a_1,\a_2} t^{\a_2} t^{\a_1} m_{\a_1\a_2}{}^\g 
\cr
&
+\Fr{1}{6\hbar^2}\sum_{\a_1,\a_2,\a_3}t^{\a_3}t^{\a_2}t^{\a_1}
\left(-\hbar m_{\a_1\a_2\a_3}{}^\g + \sum_\r m_{\a_1\a_2}{}^\r m_{\r\a_3}{}^\g\right) + \cdots,
}
$$
such that 
$$
 \sum_\g(-\hbar)^n\Fr{\rd^n}{\rd\! t^{\a_1}\cdots \rd\! t^{\a_n}}\mb{T}^\g \left<\bos{O}_\g\right>
$$
are  quantum correlation functions for the family. Conversely
$\mb{T}^\g$ determine $A_{\a\b}{}^\g$ as follows
$$
A_{\a\b}{}^\g =-\hbar \sum_\r \Fr{\rd\mb{\sG}_\b{}^\r}{\rd\! t^\a} \mb{\sG}_\r^{-1\g},
$$
where $\mb{\sG}$ denotes the matrix with $\b\g$ entry 
$\mbs{\sG}_{\b}{}^\g:=\Fr{\rd}{\rd\! t^\b} \mb{T}^\g= \d_\b^\g 
+\cdots \in \Bbbk[[t_H,\hbar^{-1}]]$,
which is invertible.
Then (i) the graded commutativity of $A_{\a\b}{}^\g$ is obvious 
(ii) both the potentiality and associativity reduces to $d\! \mb{\sG}^{-1}\wedge d\! \mb{\sG}=0$,
where $d =\sum_\a d\!t^\a \Fr{\rd}{\rd\! t^\a}$ is the formal exterior derivative 
(iii) the unity becomes 
$\Fr{\rd}{\rd\! t^0}\mb{\sG} =-\Fr{1}{\hbar}\mb{\sG}$.

For an unital BV QFT, where the partition function $\left<1\right>$ is normalizable to $1$,
free energy $\bos{F}$ can be defined as follows
$$
e^{-\bos{F}/\hbar} = 1 -\Fr{1}{\hbar}\sum_\g \mb{T}^\g \tilde{\left<\bos{O}_\g\right>},
$$
where $\tilde{\left<\bos{O}_\a\right>}$ is normalized expectation value.
Then, the free energy is a formal power series in $t_H$ and $\hbar$ 
and satisfies the following system of differential equations:
$$
\hbar \Fr{\rd^2 \bos{F}}{\rd\! t^\a \rd\! t^\b} = 
\left(\Fr{\rd \!\bos{F}}{\rd\! t^\a}\right)\left(\Fr{\rd\! \bos{F}}{\rd\! t^\b}\right) 
- \sum_\g A_{\a\b}{}^\g\left(\Fr{\rd\! \bos{F}}{\rd\! t^\g}\right),
$$
where $-\hbar\Fr{\rd \!\bos{F}}{\rd\! t^\a}\big|_{t_H=0} = \tilde{\left<\bos{O}_\a\right>}$.
We shall see that the free energy is another avatar of a morphism of QFT algebra.
There is also
an exciting possibility to study phases of the moduli space $\CM$ via thermodynamical interpretation
of the free energy. For this, however, we should able to work with convergent power series instead of
formal one.

In Sect.\ $4.4$ of the previous paper \cite{PI}
we have argued that
the notion of quantum coordinates is a natural generalization 
of that of flat or special coordinates on moduli spaces of topological strings or conformal field theory
in $2$-dimensions \cite{W,DVV,COGP,BCOV}. 
For the mathematics side, it was demonstrated that both the  flat structure on
the moduli space of universal unfolding of simple singularities due to K.\ Saito \cite{Saito}
and  the flat coordinates on moduli space associated with  differential BV algebra
 with $\rd\!\bar\rd$-lemma due to Barannikov-Kontsevich \cite{BK}   are also examples of
quantum coordinates.   
We should, however, keep in mind that the associated moduli spaces of the above mentioned examples
 also come with a compatible flat metric
$g_{\a\b}$ on its tangent space:
$$
g_{\a\b}=(-1)^{|t^\a||t^\b|} g_{\b\a}, \qquad
\sum_\r A_{\a\b}{}^\r g_{\r\g} =\sum_\r A_{\b\g}{}^\r g_{\a\r}, \qquad \Fr{\rd g_{\b\g}}{\rd t^\a} =0,
$$ 
which is invertible. 
The additional data $\{g_{\a\b}\}$, then, implies that $A_{\a\b\g}:= \sum_\r A_{\a\b}{}^\r g_{\r\g} $
is totally graded symmetric for all $3$-indices and there is some  potential $\Phi$ such that
$A_{\a\b\g} =\Fr{\rd^3 \Phi}{ \rd\!t^\a \rd\!t^\b \rd\!t^\g}$. Then the associativity and unity conditions 
become the WDVV equation:
$$
\eqalign{
\sum_{\m,\n} 
\left( \Fr{\rd^3\Phi}{\rd\! t^\a\rd\! t^\b \rd\! t^\m}\right)g^{\m\n}
\left( \Fr{\rd^3\Phi}{\rd\! t^\n \rd\! t^\g\rd\! t^\r}\right)
 &=\sum_{\m,\n} 
\left( \Fr{\rd^3\Phi}{\rd\! t^\b\rd\! t^\g \rd\! t^\m}\right)g^{\m\n}
\left( \Fr{\rd^3\Phi}{\rd\! t^\a \rd\! t^\n\rd\! t^\r}\right)
,\cr
g_{\a\b} &= \Fr{\rd^2 }{\rd\!t^\a \rd\!t^\b}\left(\Fr{\rd \Phi}{\rd t^0}\right),
}
$$
where $g^{\a\b}$ denote the inverse  metric.

Our BV QFT package (BV QFT algebra plus QFT cycle) does not automatically lead to a compatible flat metric,
while the above historical examples come with additional data
such as suitable  version of the Poincar\'e metric or the
Barannikov-Kontsevich (BK) integral\cite{BK}.
In this paper we introduce the notion of QFT integral, which supplies new quantum homotopy invariants
in addition to those coming from a QFT cycle. A QFT integral shall be a QFT cycle with special property
which can also be viewed as a generalization of BK integral. We, then,
show that BV QFT with a QFT integral induces certain rich algebraic structure on $T\CM$,
which, in the semi-classical case, is exactly Frobenius manifold. 

These results  are further evidence of our assertion that the notion of quantum coordinates 
is a natural generalization of flat, or special, coordinates. The term
quantum coordinates is motivated by the following phenomena:
A solution to the quantum master equation  automatically gives a solution to 
quantum descendant equation (BV quantum master equation) whose
classical limit corresponds to a specific choice of a universal solution 
to the Maurer-Cartan equation (the classical BV master equation) governing the moduli space $\CM$. 
Via such the solution
the affine coordinates $t_H=\{t^\a\}$ on $H$ give a distinguished coordinates called quantum coordinates on 
$\CM$. In the semi-classical case such quantum coordinates has been
called flat since $g_{\a\b}$ happens  to be independent of $t_H$. 

We also note that the essential information of quantum correlations
is summarized by  the quantum coordinates $\{\mb{T}(t_H)^\g\}$ for the family, which
can be inverted after being regarded as a sequence of maps, parametrized by $\hbar^{-1}$, on $H$ to 
a formal neighborhood around the base point in $\CM$.
Such inversion is reminiscent of  the Abel-Jacobi inversion of period map on elliptic curve\foot{It is interesting
to note that an attempt to generalize this picture has led  K.\ Saito to his original discovery of Frobenius manifold
structure on a universal unfolding of  singularity (see his own account in\cite{SaitoP}).}
as well as the mirror map  and could be a universal problem associated with quantum field theory
without anomaly. For us it adds strength to the  mantra of our series that quantum field theory
is the study of morphisms of QFT algebras, with quasi-isomorphisms being physical equivalences.

This paper is organized as follows. In Section $2$ we setup notations and conventions
as well as a streamlined summary of the the basic mathematical setting of the previous paper
\cite{PI}. The notable additions are the definition of QFT complex and
two propositions about it.  The QFT complex 
is a core structure of general QFT algebras.
In Section $3$ we state the main theorem of this paper on solutions to refined quantum master
equation and study  various consequences, as summarized above. In Section $4$, we study
the notion of QFT integral and derive Frobenius manifold structure in the semi-classical case.
The last section is devoted to a proof of our main theorem.  Proofs of some technical propositions used
in the main text are in Appendix A and B.

I would like to thank John Terilla for a proofreading of this manuscript. 

\newsec{Preliminary}

This section is a brief  summary of the basic mathematical setting of the previous paper
\cite{PI}. We shall also fix notations and conventions.
The presentation here is in somewhat different order and, perhaps,
more streamlined since it is without the attempts to justify
whether our scheme is indeed about quantum field theory which complicated the previous document.

A notable addition in this paper is that we have isolated the
definition of a QFT complex from a BV QFT algebra as a core
structure. We prove two important propositions on QFT complex. 
General (both commutative and non-commutative) QFT algebra is  a QFT complex 
with certain compatible additional algebraic structure,
where the compatibility shall be always stated as  suitable $\hbar$-divisibility conditions.

In the remaining part of this paper we shall use the Einstein summation convention that
the repeated upper and lower indices are summed over without the summation notation.
It is also understood that an equality  labeled  by unrepeated indices  shall mean that it is valid 
for all the ranges of the indices, unless otherwise specified.\foot{For example, 
$A_{\a\b}{}^\r A_{\r\g}{}^s = A_{\b\r}{}^\r A_{\a\r}{}^\s$
means that $\sum_\r A_{\a\b}{}^\r A_{\r\g}{}^\s = \sum_\r A_{\b\r}{}^\r A_{\a\r}{}^\s$ for all $\a,\b,\g,\s$.}

Fix a ground field $\Bbbk$ of characteristic zero, usually $\R$ or $\C$.
Let $\sC$ be a $\Z$-graded $\Bbbk$-vector space and
set  
$$
\sC[[\hbar]]=\left\{\sum_{n\geq 0} \hbar^n a^{(n)} \big| a^{(n)}\in \sC\right\},
$$
where $\hbar$ is a formal parameter (the formal Planck constant).
The $\Z$-grading is specified by ghost number and $|\bos{a}|$ denotes the ghost number of $\bos{a}$.
We set  $|\hbar|=0$. 
We shall denote an element of $\sC[[\hbar]]$ by an upright {\bf bold} letter, 
i.e., $\bos{a}\in \sC[[\hbar]]$, and an element of $\sC$ by an {\itshape italic} letter, i.e.,
$a \in \sC$, such that  formal power series expansion of an element $\bos{a}\in\sC[[\hbar]]$ 
shall be denoted as
$\bos{a} = a^{(0)} + \hbar a^{(1)} +\hbar^2 a^{(2)}+\cdots$,
where $a^{(n)}\in \sC$ for all $n=0,1,2,\cdots$.
We shall often denote $a^{(0)}$ by $a$. 
Projection of any structure parametrized by $\hbar$ from $\sC[[\hbar]]$ 
to $\sC$ will be called taking classical limit. In general, anything in formal power series of $\hbar$
is denoted by an upright $\bos{bold}$ letter, such as $\bos{F}$, while a slanted $\mb{bold}$ letter, 
such as $\mb{T}$, denote  something in formal Laurent series of $\hbar$.

 \subsection{QFT Complex}

Consider a sequence $\bos{K} = K^{(0)} + \hbar {K}^{(1)} +\hbar^2 K^{(2)}+\cdots$
of $\Bbbk$-linear maps on $\sC$ into $\sC$ parametrized by $\hbar$ with ghost number $1$.
The action of $\bos{K}$ on $\sC[[\hbar]]$ is defined by the $\Bbbk[[\hbar]]$-linearity
and $\hbar$-adic continuity:
$$
\bos{K}\bos{a}:=\sum_{n=0}^\infty\sum_{j=0}^n K^{(j)} a^{(n-j)}. 
$$
We shall usually denote $K^{(0)}$ by $Q$.

\begin{definition}
The pair $\big(\sC[[\hbar]], \bos{K}\big)$ defines the structure of a 
QFT complex  on $\sC$ if
$\bos{K}^2=0$. 
\end{definition}

The structure of QFT complex on $\sC$ is considered modulo the natural automorphism
on $\sC[[\hbar]]$, which is a sequence $\bos{g}= 1 +\hbar g^{(1)} +\hbar^2 g^{(2)}+\cdots
$ of $\Bbbk$-linear maps on $\sC$ into itself parametrized by $\hbar$ with ghost number $0$
such that $\bos{g}|_{\hbar =0}=1$.
The action of  $\bos{g}$ on $\sC[[\hbar]]$ is defined the $\Bbbk[[\hbar]]$-linearity
and $\hbar$-adic continuity. Such an automorphism send
$\bos{K}$ to $\bos{K}^\pr = \bos{g}\bos{K} \bos{g}^{-1}$:
$$
\bos{K}=Q+\hbar K^{(1)}+\cdots \quad\rightarrow\quad 
\bos{K}^\pr = Q +\hbar \left(K^{(1)} + g^{(1)}Q -Q g^{(1)}\right) +\cdots.
$$

The leading relations for the condition $\bos{K}^2=0$ are
$$
\eqalign{
Q^2&=0,\cr
Q K^{(1)} +K^{(1)}Q&=0,\cr
K^{(1)}K^{(1)} + Q K^{(2)} + K^{(2)}Q&=0,\cr
}
$$
etc. Thus $(\sC, Q)$ is a cochain complex over $\Bbbk$ called 
the underlying classical cochain complex (in the QFT complex).
Note that the structure $(\sC,Q)$ is invariant under the natural automorphisms on $\sC[[\hbar]]$.

\begin{definition}\label{cohom}
The cohomology of a QFT complex $\big(\sC[[\hbar]],\bos{K}\big)$
is the cohomology $H$ of the underlying classical cochain complex $(\sC, Q)$, 
which is fixed by the automorphism on $\sC[[\hbar]]$. 
\end{definition}

 A morphism of QFT complex from $(V[[\hbar]], \bos{K}_V)$
to $(W[[\hbar]], \bos{K}_W)$ is a sequence 
$
\bos{f}_{VW}=f_{VW}^{(0)} + \hbar f_{VW}^{(1)}+\cdots
$ 
of $\Bbbk$-linear maps on $V$ to $W$ parametrized by $\hbar$ of ghost number $0$, such
that 
$
\bos{f}_{VW}\bos{K}_V = \bos{K}_W \bos{f}_{VW}.
$
We shall often denote $f_{VW}^{(0)}$ by $f_{VW}$.
The leading relations are
$$
\eqalign{
{f}_{VW}{Q}_V &= {Q}_W {f}_{VW}
,\cr
{f}_{VW}{K}_V^{(1)}
+{f}^{(1)}_{VW}{Q}_V &= {K}_W^{(1)} {f}_{VW}
+{Q}_W {f}^{(1)}_{VW}.\cr
}
$$
Thus, in particular, $f_{VW}$ is a cochain map of the underlying classical cochain complex.

\begin{definition}\label{qftqiso}
A quasi-isomorphism of QFT complexes is a morphism of QFT complexes
whose underlying classical cochain map 
induces an isomorphism on the cohomology.
\end{definition}

A quantum homotopy is a sequence 
$$
\bos{s}_{VW}=s_{VW} +\hbar s^{(1)}_{VW} +\hbar^2 s^{(2)}_{VW}+\cdots
$$
 of $\Bbbk$-linear maps on $V$ into $W$ with ghost number $-1$
parametrized by $\hbar$. We shall often denote $s_{VW}^{(0)}$ by $s_{VW}$.
Morphisms of QFT complex $\bos{f}_{VW}$ and
$\bos{f}_{VW}^\pr$ are quantum homotopic if 
$$
\bos{f}_{VW}^\pr - \bos{f}_{VW}=\bos{s}_{VW}\bos{K}_V+ \bos{K}_W \bos{s}_{VW}
$$
for some quantum homotopy $\bos{s}_{VW}$. The leading relations are
$$
\eqalign{
{f}_{VW}^\pr - {f}_{VW}&={s}_{VW}{K}_V+ {Q}_W {s}_{VW}
,\cr
{f}_{VW}^{\pr(1)} - {f}_{VW}^{(1)}&={s}_{VW}{K}_V^{(1)}+ {K}_W^{(1)} {s}_{VW}+{s}_{VW}^{(1)}{Q}_V
+ {Q}_W {s}_{VW}^{(1)}.
}
$$
Thus, in particular, ${f}_{VW}^\pr$ and ${f}_{VW}$ are cochain homotopic.
Both quantum morphism and quantum homotopy are defined modulo automorphisms
$\bos{g}_{VV}$ and $\bos{g}_{WW}$ such as
$
\bos{f}_{VW} \rightarrow \bos{g}_{WW}\bos{f}_{VW} \bos{g}_{VV}^{-1}. 
$

\begin{definition}
A QFT complex $\big(\sC[[\hbar]], \bos{K}\big)$ is called {\it on-shell} if $\bos{K}|_{\hbar=0}=0$.
\end{definition}

\begin{remark}
The prefix ``on-shell'' is motivated by an analogy with the classical equation of motion,
which corresponds to the vanishing loci of $Q$ regarded as an odd vector field
over the space of classical fields.
\end{remark}

The following is the theorem in section $3$ of the previous paper \cite{PI}:

\begin{theorem}
\label{anomaly}
On cohomology $H$ of a QFT complex $\Big(\sC[[\hbar]],\bos{K}=Q +\hbar K^{(1)}+\cdots\Big)$
there is the structure $\left(H[[\hbar]], \bos{\k}=\hbar\k^{(1)} +\hbar^2 \k^{(2)} +\cdots \right)$ 
of an on-shell QFT complex.
Also, a quasi-isomorphism
$f:(H,0)\rightarrow (\sC, Q)$ of the classical complex which induces
the identity map on $H$ 
has an extension
to a morphism $\bos{f} = f+\hbar f^{(1)} +\hbar f^{(2)}+\cdots$
of QFT complexes, 
$\bos{K} \bos{f}=\bos{f}\bos{\k}$,
which is defined uniquely up to quantum homotopy and automorphisms.
\end{theorem}

The leading conditions for $\bos{K} \bos{f}=\bos{f}\bos{\k}$ are
$$
\eqalign{
Qf &=0
,\cr
K^{(1)}f + Q f^{(1)} &=f\k^{(1)}
.\cr
}
$$
Let $\bos{\k}=\hbar^n \k^{(n)}+\hbar^{n+1}\k^{(n+1)}+\cdots$ such that $\k^{(n)}\neq 0$.
Then $\k^{(n)}$ remains invariant under automorphism on $H[[\hbar]]$.
The quasi-isomorphism $f:(H,0)\rightarrow (\sC, Q)$ in the above proposition is
nothing but a $\Bbbk$-linear way of choosing representative of each and every cohomology
class. Then $Qf(x)=0$ for any $x\in H$ and the $Q$-cohomology class of $f(x)$ is $x$, 
i.e., $\left[f(x)\right]=x$,
so that $f$ induces the identity map on $H$.
Theorem \ref{anomaly} implies that  the structures 
$\big(H[[\hbar]],\bos{\k}\big)$ and $\big(\sC[[\hbar]], \bos{K}\big)$ on $H$ and $\sC$, respectively,
are quasi-isomorphic as QFT complex via $\bos{f}$. 

\begin{remark}
Note that the both$\big(H[[\hbar]],\bos{\k}\big)$ and $\big(\sC[[\hbar]], \bos{K}\big)$
are also cochain complexes over the formal power series ring $\Bbbk[[\hbar]]$ 
and $\bos{f}$ as a $\Bbbk[[\hbar]]$-linear map
on $H[[\hbar]]$ into $\sC[[\hbar]]$ is a cochain map since $\bos{K}\bos{f}= \bos{f}\bos{\k}$. Recall that
a cochain map induces a morphism on the cohomology.  
The cohomology of a cochain 
complex $\big(V[[\hbar]], \bos{K}_V\big)$ over $\Bbbk[[\hbar]]$ is, of course, the quotient 
$\Bbbk[[\hbar]]$-module $\Ker \bos{K}_V/\hbox{Im }\bos{K}_V$.
A cochain map is a quasi-isomorphism  (of cochain complex) if it induces an isomorphism on the cohomology. 
Hence, our definitions of the cohomology of QFT complex and quasi-isomorphism of QFT complex could be
vexing. The following two propositions shall be clarifying,  which proofs are located in Appendix A for the
sake of streamlined presentation.

\end{remark}

\begin{proposition}\label{flemma}
Any $\bos{\eta}\in \sC[[\hbar]]^{|\bos{\eta}|}$ satisfying 
$$
\bos{K}\bos{\eta}=0,
$$
can be expressed as
$$
\bos{\eta} =\bos{f}(\bos{x}) + \bos{K} \bos{\l},
$$
for certain pair $\left(\bos{x}, \bos{\l}\right) \in H[[\hbar]]^{|\bos{\eta}|}\oplus \sC[[\hbar]]^{|\bos{\eta}|-1}$
such that $$\bos{\k}\bos{x}=0.$$
Let $\left(\bos{x}^\pr, \bos{\l}^\pr\right) \in H[[\hbar]]^{|\bos{\eta}|}\oplus \sC[[\hbar]]^{|\bos{\eta}|-1}$
be any other pair satisfying $\bos{\eta} =\bos{f}(\bos{x}^\pr) + \bos{K} \bos{\l}^\pr$. Then there is
certain pair $\left(\bos{y}, \bos{\zeta}\right) \in H[[\hbar]]^{|\bos{\eta}|-1}\oplus \sC[[\hbar]]^{|\bos{\eta}|-2}$
such that
$$
\eqalign{
\bos{x}^\pr - \bos{x} &=\bos{\k}\bos{y}
,\cr
\bos{\l}^\pr - \bos{\l}&=\bos{f}(\bos{y}) + \bos{K}\bos{\zeta}
,
}
$$
\end{proposition}

\begin{proposition}\label{flemmb}

A pair $\left\{\bos{x}, \bos{\l}\right\} \in H[[\hbar]]^{|\bos{x}|}\oplus \sC[[\hbar]]^{|\bos{x}|-1}$ satisfies
$$
\bos{f}(\bos{x}) =\bos{K}\bos{\l}
$$
if and only if there is a pair $\left\{\bos{y},\bos{\zeta}\right\} 
\in H[[\hbar]]^{|\bos{x}|-1}\oplus \sC[[\hbar]]^{|\bos{x}|-2}$
such that
$$
\eqalign{
\bos{x}&=\bos{\k}\bos{y}
,\cr
\bos{\l}&=\bos{f}\left(\bos{y}\right)+\bos{K}\bos{\zeta}.
}
$$
\end{proposition}

Combining the above two propositions, we have

\begin{lemma}\label{elemma}
The structures $\big(H[[\hbar]],\bos{\k}\big)$ and $\big(\sC[[\hbar]], \bos{K}\big)$
of QFT complex on $H$ and $\sC$, respectively, are quasi-isomorphic (via $\bos{f}$) also as 
cochain complex over $\Bbbk[[\hbar]]$.
\end{lemma}

\begin{remark}
Why didn't we define cohomology of a QFT complex $\big(\sC[[\hbar]],\bos{K}\big)$,
which is automatically a cochain complex over $\Bbbk[[\hbar]]$, by the standard one?
There are at least three reasons
not to do that. One reason is that the natural automorphism on $\sC[[\hbar]]$
does not fix $\bos{K}$ so that $\bos{K}$-cohomology may not have intrinsic meaning for our purpose.
Another reason is that we can hardly expect the $\bos{K}$-cohomology group is
free $\Bbbk[[\hbar]]$-module in general, while both $H[[\hbar]]$ and $\sC[[\hbar]]$ are free. The final and perhaps
the most important reason for us is that we are loosing some crucial information, encoded by $\bos{\k}$, 
by taking $\bos{K}$-cohomology. It may be plausible that the totality of free resolutions of 
the $\bos{K}$-cohomology module may recover the lost information, while 
the on-shell QFT complex $\big(\sC[[\hbar]],\bos{\k}\big)$ could be "the best" model.

\end{remark}

The purposes of propositions \ref{flemma} and \ref{flemmb} are  for more than lemma \ref{elemma}.
Consider, for example, proposition \ref{flemma}. Note that its classical limit
is completely standard.
Let $\eta \in \sC^{|\eta|}$ satisfying $Q\eta=0$. 
Then there is certain pair $(x,\l) \in H^{|\eta|}\oplus \sC^{|\eta|-1}$ 
such that $$\eta = f(x) + Q\l.$$ Let $(x^\pr,\l^\pr) \in H^{|\eta|}\oplus \sC^{|\eta|-1}$ 
is any other
pair satisfying $\eta = f(x^\pr) + Q\l^\pr$ then there is a pair 
$(y,\zeta) \in H^{|\eta|-1}\oplus \sC^{|\eta|-2}$  such that
$$
\eqalign{
x^\pr - x &=0,\cr
\l^\pr -\l&= f(y) + Q \zeta,
}
$$
since $\bos{\k}|_{\hbar=0}=0$.
The above statements mean, simply, that  $x=[\eta] = \left[f(x)\right]$ (uniqueness), while the space of
ambiguities of $\l$ is $\Ker Q$, which has a decomposition $\hbox{Im }f\oplus \hbox{Im }Q$. 
And there is no "correlation" between the (zero) ambiguity of $x$ and
the ambiguity of $\l$. The situation is rather different in general. 
The ambiguities of $\bos{x}$ and $\bos{\l}$
are correlated and the space of ambiguities of $\bos{\l}$ is not necessarily $\Ker \bos{K}$, since 
$\bos{K}\left(\bos{\l}^\pr -\bos{\l}\right)= \bos{f}\left(\bos{\k}\bos{y}\right)\neq 0$ in general.
Such correlation vanishes if $\bos{\k}=0$. 

\begin{corollary}\label{misca}
Let $\bos{\k}=0$. Then any $\bos{\eta}\in \sC[[\hbar]]^{|\bos{\eta}|}$ satisfying 
$\bos{K}\bos{\eta}=0$ can be written
as $\bos{\eta}=\bos{f}(\bos{x}) +\bos{K}\bos{\zeta}$ for unique $\bos{x} \in H[[\hbar]]$
and some $\bos{\zeta}\in \sC[[\hbar]]$, which is defined modulo $\Ker \bos{K}$.
\end{corollary}

\begin{proof}
Proposition \ref{flemma} with $\bos{\k}=0$. \qed
\end{proof}

\begin{corollary}\label{miscb}
Let $\bos{\k}=0$ and let $\bos{f}(\bos{x})=\bos{K}\bos{\l}$. Then $\bos{x}=0$ and $\bos{K}\bos{\l}=0$.
\end{corollary}

\begin{proof}
Proposition \ref{flemmb} with $\bos{\k}=0$. \qed
\end{proof}

Finally we state a simple proposition, which deserve some attention.

\begin{proposition}\label{aisiden}
Let $\bos{\k}=0$.
Assume that we have the following type of equality
$$
\hbar\bos{\eta} = \bos{K}\bos{\l},
$$
where $\bos{\eta}\in \sC[[\hbar]]^{|\bos{\eta}|}$ and $\bos{\l}\in \sC[[\hbar]]^{|\bos{\eta}|-1}$. 
Then there exist some $\bos{\xi} \in \sC[[\hbar]]$ such that
$\bos{\eta} =\bos{K}\bos{\xi}$.
\end{proposition}

\begin{proof}

From the classical limit we have $Q\l=0$. Hence, $\l = f\left([\l]\right) + Q \vr$ for
some $\vr \in \sC$. It follows that the expression
$\bos{\l}-\bos{f}\left([\l]\right) - \bos{K} \vr$ is divisible by $\hbar$.
Define 
$$
\hbar \bos{\xi} := \bos{\l}-\bos{f}\left([\l]\right) - \bos{K} \vr
$$
so that $\bos{\xi} \in  \sC[[\hbar]]^{|\bos{M}|-1}$. Applying $\bos{K}$ to the above we obtain that
$\hbar\bos{K} \bos{\xi} =\bos{K} \bos{\l}$.
Hence we have
$\hbar \bos{M} = \hbar\bos{K} \bos{\xi}$, which is equivalent to $\bos{M} =\bos{K}\bos{\xi}$.
\qed

\end{proof}

\begin{remark}
We should emphasize that the condition $\hbar\bos{\eta} = \bos{K}\bos{\l}$ does not necessarily  implies
that $\bos{\eta}=\bos{K}\bos{\xi}$ for some $\bos{\xi} \in \sC[[\hbar]]$, since the obvious candidate
$\bos{\xi}=\Fr{1}{\hbar}\bos{\l}$ may not belong to $\sC[[\hbar]]$ -
$\bos{\l}$ may not be divisible by $\hbar$
in $\sC[[\hbar]]$. The notion of QFT complex and  the condition divisibility by $\hbar$ shall plays prominent
roles in developing our theory.

\end{remark}

\subsection{Observable,  QFT Cycle and Quantum Expectation Value}

\begin{definition}
An observable $x$ is an element of $H$ such that $\bos{\k}x=0$, i.e., $\k^{(\ell)}x=0$
for $\forall \ell=1,2,3,\cdots$.
\end{definition}
Let $x\in H$ be an observable. Then $Q f(x) =0$, where $f(x)\in \sC^{|x|}$ and the $Q$-cohomology
class $[f(x)]$ of $f(x)$ is $x$. We say $f(x)$ a {\it classical representative} of the observable $x$.
Then theorem \ref{anomaly} means that $\bos{f}(x)=f(x) +\hbar f^{(1)}(x)+\cdots \in \sC[[\hbar]]^{|x|}$
satisfies $\bos{K} \bos{f}(x)=0$. We say $\bos{f}(x)$ a {\it quantum representative} of the observable $x$.
Now consider an element $y \in H$ such that $\bos{\k}y\neq 0$.
 We still have
$Q f(y) =0$, while theorem \ref{anomaly} means that $\hbar$-correction to $f(y)$ such that
it is annihilated by $\bos{K}$ does not exist. We say  QFT complex  is {\it anomaly-free} if $\bos{\k}=0$
on its cohomology.  We call an element $y \in H$ with $\bos{\k}y\neq 0$ an invisible.
Note that observable and invisible are indistinguishable in the classical limit. The existence of
invisibles shall be identified with that of fundamental quantum symmetry in a sequel.

\begin{definition}
A QFT cycle of dimension $N$ for a QFT complex $\big(\sC[[\hbar]],\bos{K}\big)$
 is a sequence $\mb{\mc}=\mc^{(0)} +\hbar \mc^{(1)} +\hbar \mc^{(2)}
+\cdots$ of $\Bbbk$-linear maps $c^{(\ell)}$, $\ell=0,1,2,\cdots$, on $\sC$ into $\Bbbk$ of ghost number $-N$
parametrized by $\hbar$ satisfying
$$
\mb{\mc}\bos{K} =0.
$$
Two QFT cycles $\mb{\mc},\mb{\mc}^\pr$ of dimension $N$ are quantum homotopic if
there is a sequence $\bos{r}=r^{(0)} +\hbar r^{(1)} +\hbar^2 r^{(2)}+\cdots$
of $\;\Bbbk$-linear maps $r^{(\ell)}$, $\ell=0,1,2,\cdots$, on $\sC$ into $\Bbbk$ of ghost number $-N-1$
parametrized by $\hbar$ such that
$\mb{\mc}^\pr -\mb{\mc} = \bos{r}\bos{K}.$

\end{definition}

\begin{remark}
Note that the ghost number of $\Bbbk$ and $\Bbbk[[\hbar]]$ is concentrated zero. Hence
the sequence maps ${\mc}^{(0)},\mc^{(1)},\cdots$ should be zero maps on $\sC^j$ for $j\neq N$.
\end{remark}

We recall that theorem \ref{anomaly} gives
a sequence $\bos{f}:=f +\hbar f^{(1)}+\hbar^{2} f^{(2)}+\cdots$ of $\Bbbk$-linear maps
parametrized by $\hbar$ on $H$ into $\sC$ defined up to 
quantum  homotopy satisfying $\bos{K}\,\bos{f}= \bos{f}\,\bos{\k}$.
We can compose the map $\bos{f}$, regarded as a $\Bbbk[[\hbar]]$-linear map
on $H[[\hbar]]$ into $\sC[[\hbar]]$, with
the map $\mb{\mc}:= \mc^{(0)}+\hbar \mc^{(1)}+\hbar^{2}\mc^{(2)}+
\cdots$, regarded as a $\Bbbk[[\hbar]]$-linear map
on $\sC[[\hbar]]$ into $\Bbbk[[\hbar]]$, 
to obtain a sequence\foot{The composition symbol
is omitted throughout this paper.}
$$
\bos{\iota}:=\mb{\c}\,\bos{f}=\iota^{(0)}+\hbar \iota^{(1)}+\hbar^{2}\iota^{(2)}+\cdots
$$ 
of $\Bbbk$-linear maps parametrized by $\hbar$ on $H$ into $\Bbbk$  such that
$\iota^{(n)}=\sum_{\ell=0}^{n}\mc^{(n-\ell)} f^{(\ell)}$.
The ambiguity of $\bos{\iota}$ due to the ambiguities of $\bos{f}$ and $\mb{\mc}$
up to quantum homotopy, 
$\bos{f}\,\bos{\sim}\,\bos{f}^{\pr}=\bos{f} + \bos{K}\bos{s} +\bos{s}\bos{k}$ and 
$\mb{\mc}\,\bos{\sim}\,\mb{\mc}^{\pr}=\mb{\mc} + \bos{r}\bos{K}$,
is $
\bos{\iota}^{\pr}-\bos{\iota}\equiv \mb{\mc}^{\pr}\bos{f}^{\pr} -\mb{\mc}\,\bos{f}
= \left(\mb{\mc}\,\bos{s}+\bos{r}\,\bos{f} +\bos{r}\,\bos{K}\,\bos{s}\right)\bos{\k}
$.

\begin{definition}
Let $x \in H$ be an observable. The quantum expectation value of
the observable $x$ is $\bos{\iota}(x):=\mb{\mc}\bos{f}(x)$, which is a quantum homotopy invariant.
\end{definition}

\subsection{BV QFT Algebra and BV QFT}

Now we turn to BV QFT algebra and BV QFT, which notions were motivated by the celebrated 
Batalin-Vilkovisky quantization scheme \cite{BV}.
Let $\big(\sC,\hbox{ }\cdot\hbox{ }\big)$ be a $\Z$- graded $\Bbbk$-vector
space with a bilinear product $\cdot$ of ghost number zero. Then there is a canonical
$\Bbbk[[\hbar]]$-bilinear product on $\sC[[\hbar]]$, denoted by the same notation $\cdot$,
induced from $\sC$ by $\hbar$-adic continuity, i.e., 
$$
\bos{a}\cdot \bos{b}
= a^{(0)}\cdot b^{(0)} +\sum_{n=1}^\infty\hbar^n\sum_{i+j=n}  a^{(i)}\cdot b^{(j)}.
$$

\begin{definition}
A BV QFT algebra (with unit $1$) is a triple 
$
\big(\bos{\sC}[[\hbar]],\bos{K}, \hbox{ }\cdot\hbox{ }\big)$, 
where the pair $\big(\bos{\sC}[[\hbar]],\bos{K}\big)$ is QFT complex
and the bilinear product $\cdot$ is graded-commutative and associative
such that
\begin{itemize}
\item quantum unit: $\bos{K}1=0$ and $1\cdot \bos{a}=\bos{a}$ for all 
$\bos{a}\in\sC[[\hbar]]$.

\item $\hbar$-condition: the failure of $\bos{K}$ being  a derivation of the product
$\cdot$ is {\it divisible} by $\hbar$ and the binary operation measuring
the failure is a derivation of the product.

\end{itemize}
\end{definition}

From the $\hbar$-condition,
it is convenient to introduce a $\Bbbk[[\hbar]]$-bilinear map
$$
\left(\hbox{ },\hbox{ }\right)_{\hbar}:\sC[[\hbar]]^{k_{1}}\otimes \sC[[\hbar]]^{k_{2}}
\longrightarrow \sC[[\hbar]]^{k_{1}+k_{2}+1}
$$
called BV bracket
by the formula
\eqn\bvbracket{
-\hbar (-1)^{|\bos{a}|}\big(\bos{a},\bos{b}\big)_{\hbar}:= 
\bos{K}\left(\bos{a}\cdot \bos{b} \right)
- \bos{K a}\cdot \bos{b} -(-1)^{|\bos{a}|}\bos{a}\cdot \bos{Kb}.
}
Then, by definition, the bracket is a (graded) derivation of the product (Poisson-law)
\eqn\poisson{
(\bos{a}, \bos{b}\cdot \bos{c})_{\hbar}= (\bos{a},\bos{b})_{\hbar}\cdot \bos{c} 
+ (-1)^{(|\bos{a}|+1)|\bos{b}|}\bos{b}\cdot (\bos{a},\bos{c})_{\hbar},
}
and the unit $1$ is in its center.
It follows that the bracket satisfies the graded commutativity and
the graded  Jacobi-identity such that $\bos{K}$ is a  derivation of it:
\eqn\dglie{
\eqalign{
\big(\bos{a}, \bos{b}\big)_{\hbar} 
&=-(-1)^{(|\bos{a}|+1)(|\bos{b}|+1)}\big(\bos{b}, \bos{a}\big)_{\hbar}
,\cr
\big(\bos{a},\big(\bos{b},\bos{c}\big)_{\hbar}\big)_{\hbar} 
&= \big(\big(\bos{a},\bos{b}\big),\bos{c}\big)_{\hbar} 
+(-1)^{(|\bos{a}|+1)(|\bos{b}|+1)}(\bos{b}, \big(\bos{a},\bos{c}\big)_{\hbar}\big)_{\hbar}
,\cr
\bos{K}\big(\bos{a}, \bos{b}\big)_{\hbar}&=\big(\bos{K}\bos{a}, \bos{b}\big)_{\hbar}
+(-1)^{|\bos{a}|+1}\big(\bos{a}, \bos{K}\bos{b}\big)_{\hbar}.
}
}
We call the triple
$$
\big(\sC[[\hbar]], \bos{K},(\hbox{ },\hbox{ })_{\hbar}\big)
$$
the quantum descendant algebra (of the the BV QFT algebra).  

\begin{remark}
Note that $(\hbox{ },\hbox{ })_{\hbar}=(\hbox{ },\hbox{ })
+\hbar (\hbox{ },\hbox{ })^{(1)} +\hbar (\hbox{ },\hbox{ })^{(2)}+\cdots$ is a sequence 
of $\Bbbk$-bilinear maps,
parametrized by $\hbar$, on $\sC\otimes \sC$ into $\sC$. 
Strictly speaking the traditional BV bracket does not depends on $\hbar$. 
In the previous paper we have abused notation by not distinguishing 
$(\hbox{ },\hbox{ })_{\hbar}$ with its classical limit $(\hbox{ },\hbox{ })$.
\end{remark}

\begin{remark}
The $\hbar$-condition in the definition of BV QFT algebra can  be relaxed 
such that the failure of the binary operation being a derivation of the product is
divisible by $\hbar^2$, the failure of the resulting ternary operation being a derivation of the
product is divisible by $\hbar^3$, to be repeated ad infinitum. Such QFT algebra may be called
(graded)-commutative binary QFT algebra and a similar story as in this paper can be developed.
It is just another (not-general) example of QFT algebra, which may be studied elsewhere.
For us two of the essential properties of general QFT algebra are that the underlying QFT complex and
its interplay with additional algebraic structure with the compatibility stated in term of sequences of 
divisibility condition by $\hbar$.

\end{remark}

By definition the classical limit $Q$ of $\bos{K}$ 
satisfies $Q^2=0$ and is a derivation of the
product, i.e., $Q(a\cdot b)=Q a\cdot b +(-1)^{|a|}a\cdot Q b$. Also $Q1=0$.
Thus the classical limit 
$
\big(\sC, Q,\,\cdot\,\big)
$
of a BV QFT algebra is a differential graded commutative and associative 
algebra (CDGA) over $\Bbbk$ with unit $1$.
It also follows that  the classical limit
$
\big(\sC, Q,\,(\hbox{ },\hbox{ })\big)
$
of the descendant algebra is a DG0LA
 over $\Bbbk$.\foot{In our convention, a DG0LA
$\big(\sC, Q,\,(\hbox{ },\hbox{ })\big)$ is a differential graded algebra after
shifting the ghost number by $-1$.} 
Hence
the quadruple $\big(\sC, Q,\,\cdot\,,(\hbox{ },\hbox{ })\big)$ is a differential
$0$-algebra since the bracket is a derivation of the product. 
The cohomology $H$ of $(\sC,Q)$ is also the cohomology of 
the CDGA and the DG0LA as well as  the BV QFT algebra.
BV QFT algebra is also defined up to natural automorphism on $\sC[[\hbar]]$,
which fix all the classical limits.

Observables of a BV QFT algebra are the observables in its QFT
complex, 
elements in
$H$ annihilated by $\bos{\k}=\hbar \k^{(1)}+\hbar^2 \k^{(2)} +\cdots$.
From the quantum unity $\bos{K}1={Q}1=0$, in the definition of BV QFT algebra, 
there is a distinguished element $e\in H^0$ corresponding to the cohomology class $[1]$ of the unit 
$1$ such that $\bos{\k}e=0$. It is natural to fix
$f$ and its extension $\bos{f}$ to a morphism of QFT complex 
such that $f(e)=\bos{f}(e)=1$, the unit in $\sC$.

\begin{definition}

A BV QFT with ghost number anomaly $N \in \Z$ is a BV QFT  algebra
(with unit $1$) together with a QFT
cycle $\mb{\mc}=\mc^{(0)}+\hbar \mc^{(1)}+\cdots$ of dimension $N$. 

\end{definition}

We call the quantum expectation value $\mb{\mc}\bos{f}(e)=\mb{\mc}(1) \in \Bbbk[[\hbar]]$
of $e\in H^0$ the partition function $\bos{Z}$ of BV QFT. It follows that
the partition function vanishes unless $N=0$. 
We may consider a class of BV QFT with vanishing ghost number
anomaly such that $\mb{\mc}(1)\big|_{\hbar=0}\equiv \mc^{(0)}(1)\neq 0$. 
Then the quantum expectation values can be normalized by multiplying 
$\bos{Z}^{-1}$. Such a theory is called unital BV QFT, with the normalization being understood.

\begin{definition}

An unital BV QFT is a BV QFT algebra (with unit $1$) together with a QFT
cycle $\mb{\mc}=\mc^{(0)}+\hbar \mc^{(1)}+\cdots$ with dimension $N=0$ 
such that $\mb{\mc}(1)=1 \in \Bbbk$.

\end{definition}

From now on we are going to adopt the time honored notation for expectation value
such that, for $\forall x\in H$,
$$
\big<\bos{f}(x)\big>=\mb{\mc}\bos{f}(x),
$$
where $\bos{f}(x)\in \sC[[\hbar]]^{|x|}$.
We emphasize that the expectation value $\big<\bos{f}(x)\big>$ depends only on
the quantum homotopy class of the QFT cycle $\mb{\mc}$ if and only if $\bos{\k}(x)=0$.
We also emphasis that the condition $\bos{\k}(x)=0$
implies that $\bos{K}\bos{f}(x)=0$ and the expectation value 
$\big<\bos{f}(a)\big>$ depends only on the quantum homotopy class of $\bos{f}$.

\begin{remark}

It is perhaps useful to tweak the notation as follows. Let $O \in \sC$
be such that
$QO=0$, then $O= f([O])$ up to homotopy $Q \La$ for some $\La \in \sC$. Assume that $\bos{\k}[O]=0$.
Set $\bos{O}= \bos{f}([O])\in \sC[[\hbar]]$
up to quantum homotopy $\bos{K}\bos{\La}$ for some $\bos{\La}\in \sC[[\hbar]]$. 
Then $\bos{K}\bos{O}=0$ and 
$$
\big< \bos{O} \big>= \big< \bos{O} + \bos{K} \bos{\La}\big>,
$$
since $\big<\bos{K} \bos{\La}\big>=\mb{\mc}\bos{K}\bos{\La}=0$.
The variation of the QFT cycle $\mb{\mc}$ preserving its quantum homotopy class
correspond to homologous deformation of Lagrangian subspace on where the BV-Feynman path integral
is defined in the BV quantization scheme. 

\end{remark}

\newsec{Statement of the Main Theorem and Its Consequences}

Fix a BV QFT algebra (with unit $1$) 
$\big(\bos{\sC}[[\hbar]],\bos{K}, \hbox{ }\cdot\hbox{ }\big)$
with quantum descendant algebra $\big(\sC[[\hbar]], \bos{K},(\hbox{ },\hbox{ })_{\hbar}\big)$
and let $\big(\sC, Q,\hbox{ }\cdot\hbox{ }\big)$ and
$\big(\sC, Q,(\hbox{ },\hbox{ })\big)$, respectively, be their classical limits.
In this paper we consider the case that
(i) the cohomology $H$ 
is finite-dimensional for each ghost number 
and (ii) $\bos{\k}=0$ on $H$.

From the assumption (i),
it is convenient to introduce a homogeneous basis $\{e_\a\}$ of $H$.
Define a $\Bbbk$-linear
map $f: H\longrightarrow \sC$ of ghost number $0$ by choosing 
a representative $O_\a$ of each $e_\a \in H$  such that 
$$
f(e_\a)= O_\a \in \sC^{|e_\a|},
$$
i.e., $QO_\a=0$ and the $Q$-cohomology class $[O_{\a}]$ 
of $O_\a$ is $e_\a$. Then $Q f=0$.
It is convenient to  fix a basis $\{e_\a\}$ of $H$ such that one of its component, say $e_0$, 
is the distinguished element $e \in H^0$, i.e., 
$$
f(e_0) =O_0=1.
$$
It follows that  $f$ is  a quasi-isomorphism of cochain complexes $f:(H, 0)\longrightarrow (\sC, Q)$,
which induces the identity map on $H$.
The  graded-commutative and  associative product $m_2:H\otimes H \longrightarrow H$ 
of ghost number $0$ on $H$ can be specified structure constants 
$m_{\a\b}{}^\g\in \Bbbk$:
$$
m_2(e_\a, e_\b)=m_{\a\b}{}^\g e_\g.
$$
satisfying 
\begin{itemize}
\item  graded symmetry: $m_{\a\b}{}^{\g}=(-1)^{|\a||\b|}m_{\b\a}{}^{\g}$,
\item  associativity: $m_{\a\b}{}^{\r}m_{\r\g}{}^{\s}=m_{\b\g}{}^{\r}m_{\a\r}{}^{\s}$,
\item  identity: $m_{0\b}{}^\g =\d_{\b}{}^\g$.
\end{itemize} 
At the cochain level, we have
\eqn\caa{
f(e_{\a})\cdot f(e_{\b})=f(m_{2}(e_{\a},e_{\b}))+ K \l_{2}(e_{\a},e_{\b}),
}
where $\l_{2}:H\otimes H\longrightarrow \sC$ is a $\Bbbk$-bilinear map of ghost
number $1$. That is, $f$ is a DGA map  
$f:(H,0,m_{2})\longrightarrow (\sC, Q, \hbox{ }\cdot\hbox{ })$
up to homotopy.

Let $t_H=\{t^\a\}$ be the dual basis of $H^*$ such that $|t^\a| + |e_\a|=0$, which is 
an affine coordinates system on $H$ with a distinguished coordinate $t^0$.
We denote $\{\rd_\a= \rd/\rd\! t^\a\}$ be the
corresponding formal partial derivatives acting on $\Bbbk[[t_H]]$ as derivations.
To save notation we replace $(-1)^{|e_\a|}=(-1)^{-|t_\a|}=(-1)^{|O_\a|}=(-1)^{|\bos{O}|_{\a}}$ 
by $(-1)^{|\a|}$. We shall also use notation $t^{\bar\a}$ for $(-1)^{|\a|}t^\a$ and
$\rd_{\bar\a}= \rd/\rd\! t^{\bar\a}$.

From the assumption (ii) and theorem \ref{anomaly}, we have a sequence 
$\bos{f}=f +\hbar f^{(1)}+\cdots$ of $\Bbbk$-linear maps on $H$ into $\sC$,
parametrized by $\hbar$,
of ghost number zero such that $\bos{K} \bos{f}=0$.
The image of $e_{\a}\in H$ of the map $\bos{f}$ will be denoted by
$\bos{O}_{\a}$:
 $$
 \bos{f}(e_\a)=\bos{O}_{\a}=O_\a +\hbar O^{(1)}_\a +\hbar^2 O^{(2)}_\a +\cdots 
 \in \sC[[\hbar]]^{|e_{\a}|}
 $$
such that $\bos{K}\bos{O}_{\a}=0$ and 
$$
\bos{f}(e_0)= \bos{O}_0=1.
$$
We use the following terminology:  
$O_\a$ is a classical representative of the observable $e_\a$,
$\bos{O}_\a$ is a quantum representative of the observable $e_\a$,
$O_\a$ is the classical limit $\bos{O}_\a \big|_{\hbar=0}$ of $\bos{O}_\a$,
and $\bos{O}_\a$ is the quantization of $O_\a$ with respect 
to the quantization map $\bos{f}$.

\begin{theorem}
There is the structure of $\Bbbk[[t_H]]$-algebra on $H\otimes \Bbbk[[t_H]]$
defined by a formal power series $3$-tensor $A_{\b\g}{}^\s \in \Bbbk[[t_{H}]]$,
which satisfies

\begin{itemize}
\item {\it graded symmetricity}:
$$
\eqalign{
A_{\b\g}{}^\s &= (-1)^{|\b||\g|} A_{\g\b}{}^\s
,\cr
\rd_\a A_{\b\g}{}^\s &= (-1)^{|\a||\b|} \rd_\b A_{\a\g}{}^\s
,
}
$$ 
\item {\it associativity}:
$$
A_{\a\b}{}^\r A_{\r\g}{}^\s = A_{\b\g}{}^\r A_{\a\r}{}^\s.
$$
\item {\it identity}:
$$
A_{0\b}{}^\g =\d_{\b}{}^\g,
$$

\item {\it The homogeneity}: 
$$ 
|e_\r| t^\r\rd_\r A_{\a\b}{}^\g =\left(|e_\g|-|e_\b| -|e_\a|\right)A_{\a\b}{}^\g.
$$
\end{itemize}
And, there is a distinguished solution
$\bos{\Theta}$ of the Maurer-Cartan equation
$$
\bos{K}\bos{\Theta} +\Fr{1}{2}\big(\bos{\Theta},\bos{\Theta}\big)_{\hbar}=0,
$$
where
$$
\bos{\Theta} = t^\a \bos{O}_\a 
+ \sum_{n=2}^\infty \Fr{1}{n!}t^{\a_n}\cdots t^{\a_1} \bos{O}_{\a_1\cdots\a_n} \in  \sC[[t_H,\hbar]]^0,
$$
which satisfies
\begin{enumerate}
\item {\it quantum master equation}:
$$
\hbar\rd_{\b}\rd_{\g}\bos{\Theta}
= \rd_{\b}\bos{\Theta}\cdot \rd_{\g}\bos{\Theta} - A_{\b\g}{}^\s \rd_{\s}\bos{\Theta} 
- \bos{K}\bos{\La}_{\b\g} - 
\left(\bos{\Theta},\bos{\La}_{\b\g}\right)_{\hbar},
$$
for some $\bos{\La}_{\b\g}
\in \sC[[t_H,\hbar]]^{|\b|+|\g|-1}$ in quantum gauge.

\item {\it quantum unity}:
$\rd_0 \bos{\Theta} =1$.

\end{enumerate}

\end{theorem}

\begin{remark}
The quantum gauge condition for $\bos{\La}_{\a\b}$ shall be stated later.
\end{remark}

\begin{remark} The Maurer-Cartan equation 
$\bos{K}\bos{\Theta}+\Fr{1}{2}\big(\bos{\Theta},\bos{\Theta}\big)_{\hbar}=0$ 
is called quantum descendant equation, which is an automatic consequence of the quantum master
equation. We remark that the quantum descendant equation is the Batalin-Vilkovisky quantum master equation
for a family of QFT. 
\end{remark}

The quantum master equation together with  the unity should be regraded as a system of formal differential equations
for $\bos{\Theta} \in \sC[[t_H,\hbar]]^0$, $A_{\a\b}{}^\g \in \Bbbk[[t_H]]$ and
$\bos{\La}_{\a\b} \in \sC[[t_H,\hbar]]^{|\a|+|\b|-1}$ with the initial
condition that $\bos{\Theta}= t^\a \bos{f}(e_\a) \mod t^2_H$, where   $\bos{f}(e_\a)=\bos{O}_\a$
and $\bos{f}(e_0)=1$. The theorem claims that the only obstruction to solve the quantum master equation
with the unity is $\bos{\k}=\hbar \k^{(1)} +\hbar^2 \k^{(2)}+\cdots$ on $H$, which vanishes
in the present case.

This is illustrated for the $1$st order solution in the following
example.

\begin{example}
To begin with, consider the quantum master equation modulo $t_H$:
\eqn\qmerva{
\eqalign{
\hbar \bos{O}_{\a\b} &=\bos{O}_\a\cdot\bos{O}_\b - m_{\a\b}{}^\g \bos{O}_\g - \bos{K}\bos{\l}_{\a\b}
}
}
where $\bos{O}_{\a\b}= \rd_\a\rd_\b \bos{\Theta} \mod t_H$,  $m_{\a\b}{}^\g =A_{\a\b}{}^\g \mod t_H$
and $\bos{\l}_{\a\b} =\bos{\La}_{\a\b} \mod t_H$ are unknowns. 
Then we have the following consistency conditions for \qmerva:

(1) classical limit: 
$$
O_\a\cdot O_\b = m_{\a\b}{}^\g O_\g + Q \l_{\a\b},
$$

(2) graded commutativity of the product $\cdot$: 
$$
\left( m_{\a\b}{}^\g -(-1)^{|\a||\b|}m_{\b\a}{}^\g\right)O_\g =
- Q\left(\l_{\a\b}-(-1)^{|\a||\b|}\l_{\b\a}\right),
$$

(3) unity: 
$$
O_\b = m_{0\b}{}^\g O_\g + Q \l_{0\b}, \qquad O_\a = m_{\a 0}{}^\g O_\g + Q \l_{\a 0}.
$$

For $(1)$, we note that $O_\a\cdot O_\b \in \Ker Q$ since $Q O_\a =0$ and
$Q$ is a derivation of the product. Hence $O_\a\cdot O_\b$ can be expressed as 
$m_{\a\b}{}^\g O_\g + Q \l_{\a\b}$ for unique $m_{\a\b}{}^\g$ and for some $\l_{\a\b} \in \sC^{|\a|+|\b|-1}$
modulo $\Ker Q$. From (2) and $(3)$, we conclude that  
$$
m_{\a\b}{}^\g -(-1)^{|\a||\b|}m_{\b\a}{}^\g=0, \qquad m_{0\b}{}^\g =\d_{\b}{}^\g,
$$
as well as
$Q\left(\l_{\a\b}-(-1)^{|\a||\b|}\l_{\b\a}\right)=0$ and $Q\l_{0\b}=Q\l_{\a 0}=0$.
For $\l_{\a\b}$, we may choose them, without loss of generality,
 to be graded-symmetric, $\l_{\a\b} =(-1)^{|\a||\b|}\l_{\b\a}$,
and satisfy $\l_{0\a}=0$.  
Let $\bos{\l}_{\a\b}=\l_{\a\b}$ such that
\eqn\qguab{
\bos{\l}_{\a\b} =(-1)^{|\a||\b|}\bos{\l}_{\b\a},\qquad \bos{\l}_{0\b} =0.  
}
Then the expression $\bos{L}_{\a\b}=\bos{O}_\a\cdot\bos{O}_\b 
- m_{\a\b}{}^\g \bos{O}_\g - \bos{K}\bos{\l}_{\a\b}$
is divisible by $\hbar$ and graded symmetric. It also satisfies that $\bos{L}_{0\b}=\bos{L}_{\a 0}=0$.
Once we  define $\bos{O}_{\a\b}:=\Fr{1}{\hbar}\bos{L}_{\a\b}\in \sC[[\hbar]]^{|\a|+|\b|}$,
$\bos{O}_{\a\b}$ 
is graded symmetric $\bos{O}_{\a\b}=(-1)^{|\a|||\b|}\bos{O}_{\b\a}$
and satisfies $\bos{O}_{0\b}=0$. Hence we have just solved the modulo $t_H$ quantum master equation 
\qmerva\ by setting $\bos{\Theta} =t^\a\bos{O}_\a +\Fr{1}{2}t^{\b}t^\a\bos{O}_{\a\b}\mod t_H^3$,
$A_{\a\b}{}^\g = m_{\a\b}{}^\g \mod t_H$ and $\bos{\La}_{\a\b} =\bos{\l}_{\a\b} \mod t_H$
such that the quantum identity modulo $t_H^2$ is satisfied:
$$
\rd_0\bos{\Theta}=1 \mod t_H^2,
$$
and
$$
A_{\a\b}{}^\g -(-1)^{|\a||\b|} A_{\b\a}{}^\g =0\mod t_H.
$$
For the quantum descendent equation, apply $\bos{K}$ to the modulo $t_H$ quantum master equation 
\qmerva\ to have $\bos{K} \bos{O}_{\a\b} = -\hbar (-1)^{|\a|} \big(\bos{O}_\a,\bos{O}_\b\big)_{\hbar}$,
which implies, together with the initial condition $\bos{K}\bos{O}_\a =0$, that
$$
\bos{K} \bos{\Theta} +\Fr{1}{2} \big(\bos{\Theta},\bos{\Theta} \big)_{\hbar}=0 \mod t_H^3.
$$
Hence we have the quantum descendent equation modulo $t_H^3$. 

\end{example}

The above, perhaps, is not enough as the example for demonstration of our method of solving 
quantum master equation modulo $t_H^n$ for general $n\geq 2$.  
Here comes a step by step description of our solution modulo $t_H^2$ featuring general behavior
for higher $n$. Most of necessary propositions shall be stated as claims,
referring to the actual proof in Sect.\ $4$. The following example is useful for pedagogical purpose
but some readers may want to skip it in the first reading.

\begin{example}
Now we consider the quantum master equation modulo $t_H^2$, which is eq. \qmerva\ 
and
\eqn\qmervb{
\hbar \bos{O}_{\a\b\g}
=\bos{C}_{\a\b\g} -m_{\a\b\g}{}^\r \bos{O}_\r - \bos{K}\bos{\l}_{\a\b\g},
}
where $\bos{O}_{\a\b\g}=\rd_{\a}\rd_\b \rd_\g \bos{\Theta} \mod t_H$, $m_{\a\b\g}{}^\r$, 
and $\bos{\l}_{\a\b\g}$ are unknowns, while the expression $\bos{C}_{\a\b\g}$ is determined
by the previous data $\bos{O}_\a,\bos{O}_{\a\b}, m_{\a\b}{}^\g, \bos{\l}_{\a\b}$ as follows:
$$
\bos{C}_{\a\b\g}:= 
\bos{O}_{\a\b} \cdot\bos{O}_\g +(-1)^{|\a||\b|} \bos{O}_\b\cdot \bos{O}_{\a\g}
-{m}_{\b\g}{}^\r \bos{O}_{\a\r} -\big(\bos{O}_\a, \bos{\l}_{\b\g}\big)_{\hbar}.
$$
The consistency of \qmervb\ in the classical limit requires that the classical limit $C_{\a\b\g}$
of $\bos{C}_{\a\b\g}$ satisfies $Q C_{\a\b\g}=0$ (see claim (1) below) such that 
\eqn\exppb{
C_{\a\b\g} = m_{\a\b\g}{}^\r O_\r + Q \l_{\a\b\g}
}
for unique $m_{\a\b\g}{}^\r$ and some $\l_{\a\b\g}\in \sC^{|\a|+|\b|+|\g|-1}$ defined modulo $\Ker Q$.
Note also that 
$$
\bos{C}_{0\b\g}=\bos{C}_{\a 0\g}= \bos{C}_{\a\b 0}=0, 
\qquad \bos{C}_{\a\b\g}=(-1)^{|\b||\g|}\bos{C}_{\a\g\b}.
$$
Hence, we have
\eqn\exppa{
\eqalign{
m_{0\b\g}{}^\r O_\r + Q \l_{0\b\g}
=m_{\a 0\g}{}^\r O_\r + Q \l_{\a 0\g}=m_{\a\b 0}{}^\r O_\r + Q \l_{\a\b 0}=0
,\cr
\left( m_{\a\b\g}{}^\r- (-1)^{|\b||\g|} m_{\a\g\b}{}^\r\right)O_\r =-Q\left(
 \l_{\a\b\g}{}^\r- (-1)^{|\b||\g|} \l_{\a\g\b}{}\right)
.
}
}
Another consistency condition is that the RHS of \qmervb\ should be graded symmetric 
for the all $3$ indices 
$\a,\b,\g$ since $\bos{O}_{\a\b\g}$ has such property.

We claim that 

\begin{claim}[1]
$\bos{K} \bos{C}_{\a\b\g}= \hbar\bos{F}_{\a\b\g}$, where
$$
\bos{F}_{\a\b\g}:= 
- (-1)^{|\a|} \left\{
(\bos{O}_\a,\bos{O}_{\b\g})_{\hbar} 
+ (-1)^{|\b|}(\bos{O}_{\a\b},\bos{O}_\g)_{\hbar} 
-(-1)^{(|\a|+1)(|\b|+1)} (\bos{O}_\b, \bos{O}_{\a\g})_{\hbar}
\right\},
$$
so that $Q C_{\a\b\g}=0$.
\end{claim}

\begin{claim}[2] 
$
\bos{K}\bos{N}_{\a\b\g}
+\left({m}_{\b\g}{}^\r {m}_{\a\r}{}^\s-(-1)^{|\a||\b|}{m}_{\a\g}{}^\r {m}_{\b\r}{}^\s\right)\bos{O}_\s
=-\hbar \bos{C}_{[\a\b]\g}$,
where 
$$
\eqalign{
\bos{C}_{[\a\b]\g}:=&\bos{C}_{\a\b\g}-(-1)^{|\a||\b|}\bos{C}_{\b\a\g}
,\cr
\bos{N}_{\a\b\g}
:=&{m}_{\b\g}{}^\r \bos{\l}_{\a\r}
-(-1)^{\a\b}{m}_{\a\g}{}^\r \bos{\l}_{\b\r} 
+(-1)^{\a} \bos{O}_\a \cdot \bos{\l}_{\b\g}
-(-1)^{\a\b +\b}\bos{O}_\b\cdot \bos{\l}_{\a\g}.
}
$$
\end{claim}

From the classical limit 
$QN_{\a\b\g}+\left({m}_{\b\g}{}^\r {m}_{\a\r}{}^\s-(-1)^{|\a||\b|}{m}_{\a\g}{}^\r {m}_{\b\r}{}^\s\right){O}_\s=0$ 
of claim (2)  we conclude that 
$$
\eqalign{
{m}_{\b\g}{}^\r {m}_{\a\r}{}^\s-(-1)^{|\a||\b|}{m}_{\a\g}{}^\r {m}_{\b\r}{}^\s&=0.\cr
}
$$
Hence claim $(2)$ reduce to 
$$
\bos{K}\bos{N}_{\a\b\g}=-\hbar\bos{C}_{[\a\b]\g}.
$$
Actually we have a stronger result:
\begin{claim}[3]
There exist some $\bos{\xi}_{\a\b\g} \in \sC[[\hbar]]^{|\a|+|\b|+|\g|-1}$
satisfying $\bos{C}_{[\a\b]\g} =\bos{K}\bos{\xi}_{\a\b\g}$ and
$$
\eqalign{
\bos{\xi}_{0\b\g}=\bos{\xi}_{\a 0\g}=\bos{\xi}_{\a\b 0}=0,\cr
\bos{\xi}_{\a\b\g}+(-1)^{|\a||\b|}\bos{\xi}_{\b\a\g}=0,\cr
\bos{\xi}_{\a\b\g}-(-1)^{|\b||\g|}\bos{\xi}_{\a\g\b} +(-1)^{|\a|(|\b|+|\g|)}\bos{\xi}_{\b\g\a}=0,
}
$$
such that the classical limit $\xi_{\a\b\g}$ of $\bos{\xi}_{\a\b\g}$ is
$$
\xi_{\a\b\g}=\l_{\a\b\g} -(-1)^{|\a||\b|}\l_{\b\a\g}.
$$
\end{claim}
From the above claim we have $\left( m_{\a\b\g}{}^\r- (-1)^{|\a||\b|} m_{\b\a\g}{}^\r\right)O_\r =0$.
Hence, together with \exppa\ we deduce that
$$
m_{\a\b\g}{}^\r =(-1)^{|a||\b|}m_{\b\a\g}{}^\r = (-1)^{|\b||\g|}m_{\a\g\b}, \qquad
m_{0\b\g}{}^\r=0
$$
and
$$
\l_{\a\b\g} = (-1)^{|\b||\g|}\l_{\a\g\b}, \qquad
\l_{0\b\g}=\l_{\a 0\g}=\l_{\a\b 0}=0.
$$
From claim (3), the expression $\bos{\xi}_{\a\b\g} - \left(\l_{\a\b\g} -(-1)^{|\a||\b|}\l_{\b\a\g}\right)$
is divisible by $\hbar$ so that we can define $\bos{\eta}_{\a\b\g} \in \sC[[\hbar]]^{|\a|+|\b|+|\g|-1}$ by
the formula
$$
\hbar\bos{\eta}_{\a\b\g} :=\bos{\xi}_{\a\b\g} - \left(\l_{\a\b\g} -(-1)^{|\a||\b|}\l_{\b\a\g}\right).
$$
Define
$$
\bos{\l}_{\a\b\g}:= \l_{\a\b\g} -\Fr{\hbar}{3}\left(\bos{\eta}_{\a\b\g} +(-1)^{|\b||\g|}
\bos{\eta}_{\a\g\b}\right),
$$
such that $\bos{\l}_{\a\b\g}\big|_{\hbar=0} =\l_{\a\b\g}$.
Then, from claim (3), we obtain that
\eqn\qgauc{
\eqalign{
\bos{\l}_{0\b\g} =\bos{\l}_{\a 0\g} =\bos{\l}_{\a\b 0} &=0,\cr
\bos{\l}_{\a\b\g} -(-1)^{|\b||\g|}\bos{\l}_{\a\g\b}&=0,\cr
\bos{\l}_{\a\b\g} -(-1)^{|\a||\b|}\bos{\l}_{\b\a\g}&=\bos{\xi}_{\a\b\g}.\cr
}
}
We refer the above choice for $\bos{\l}_{\a\b\g}$ together with the previous choice \qguab\ for
$\l_{\a\b}$ "quantum gauge" (see the forthcoming remark).

Now consider the expression
$\bos{L}_{\a\b\g} := \bos{C}_{\a\b\g} -m_{\a\b\g}{}^\r \bos{O} -\bos{K}\bos{\l}_{\a\b\g}$.
Then (i) $\bos{L}_{\a\b\g}$ is divisible by $\hbar$, (ii)  $\bos{L}_{\a\b\g}$
is graded symmetric for the all $3$ indices  $\a,\b,\g$, (iii) $\bos{L}_{\a\b 0}=0$.
Hence we can define $\bos{O}_{\a\b\g} :=\Fr{1}{\hbar}\bos{L}_{\a\b\g}$, i.e.,
$$
\hbar\bos{O}_{\a\b\g}=\bos{C}_{\a\b\g} -m_{\a\b\g}{}^\r \bos{O} -\bos{K}\bos{\l}_{\a\b\g},
$$
such that $\bos{O}_{\a\b\g}=(-1)^{|\a||\b|}\bos{O}_{\b\a\g}=(-1)^{|\b||\g|}\bos{O}_{\a\g\b}$ and
$\bos{O}_{\a\b 0}=0$. From claim (1), it also follows that
$$
\bos{K}\bos{O}_{\a\b\g}= 
- (-1)^{|\a|} \left\{
\big(\bos{O}_\a,\bos{O}_{\b\g}\big)_{\hbar} + (-1)^{|\b|}\big(\bos{O}_{\a\b},\bos{O}_\g\big)_{\hbar} 
-(-1)^{(|\a|+1)(|\b|+1)} \big(\bos{O}_\b, \bos{O}_{\a\g}\big)_{\hbar}
\right\}.
$$
Let 
$$
\eqalign{
\bos{\Theta} &:= t^\a + \Fr{1}{2}t^\b t^\a\bos{O}_{\a\b} +\Fr{1}{3!}t^\g t^\b t^\a\bos{O}_{\a\b\g} \mod t_H^4
,\cr 
A_{\b\g}{}^\g &:= m_{\b\g}{}^\r + t^\a m_{\a\b\g}{}^\r \mod t_H^2
,\cr
\bos{\La}_{\b\g} &:= \bos{\l}_{\a\b} + t^{\bar\a} \bos{\l}_{\a\b\g} \mod t_H^2.
}
$$
Then we just have solved the quantum master equation modulo $t_H^2$:
$$
\hbar \rd_\b\rd_\g\bos{\Theta} =\rd_\b\bos{\Theta}\cdot \rd_\g\bos{\Theta}_\g - A_{\b\g}{}^\r\bos{O}_\r
+ \bos{K}\bos{\La}_{\a\b\g} + \left(\bos{\Theta},\bos{\La}_{\a\b\g}\right)_{\hbar} \mod t_H^3,
$$
such that 
$$
\eqalign{
\rd_0 \bos{\Theta}=1 \mod t_H^3,\cr
\bos{K}\bos{\Theta} +\Fr{1}{2}\big(\bos{\Theta},\bos{\Theta}\big)_{\hbar}=0 \mod t_H^4.
}
$$
and
$$
\eqalign{
A_{\b\g}{}^\r -(-1)^{|\b||\g|}A_{\g\b}{}^\r =0\mod t_H^2
,\cr
\rd_\a A_{\b\g}{}^\r -(-1)^{|\a||\b|}\rd_\b A_{\a\g}{}^\r =0\mod t_H
,\cr
A_{\b\g}{}^\s A_{a\s}{}^\r -(-1)^{|\a||\b|}A_{\a\g}{}^\s A_{b\s}{}^\r=0\mod t_H.
}
$$
\end{example}

\begin{remark}
The choice \qguab\  we have made for $\bos{\La}_{\a\b}=\bos{\l}_{\a\b} \mod t_H$ 
and  the choices \qguab\ and \qgauc\ for 
 $\bos{\La}_{\b\g}=\bos{\l}_{\b\g} +t^{\bar\a}\bos{\l}_{\a\b\g} \mod t_H$ 
are  what we call "quantum gauge".   
In this paper the quantum gauge is merely a convenient choice of $\bos{\La}_{\a\b}$
in solving quantum master equation and its full meaning is the subject of the next paper in this series.
To explain the quantum gauge in more details, let's sketch a proof of  claim (3) in example $3.2$.
Consider the expression $\bos{N}_{\a\b\g}$ in claim (2). We note that
\eqn\egmaxa{
\eqalign{
\bos{N}_{0\b\g}=\bos{N}_{\a 0\g}=\bos{N}_{\a\b 0}=0,\cr
\bos{N}_{\a\b\g}+(-1)^{|\a||\b|}\bos{N}_{\b\a\g}=0,\cr
\bos{N}_{\a\b\g}-(-1)^{|\b||\g|}\bos{N}_{\a\g\b} +(-1)^{|\a|(|\b|+|\g|)}\bos{N}_{\b\g\a}=0.
}
}
As a corollary of claim (2) we have obtained that $\bos{K}\bos{N}_{\a\b\g}=-\hbar \bos{M}_{[\a\b]\g}$,
which implies that $Q N_{\a\b\g}=0$, where $N_{\a\b\g}$ is the classical limit of $\bos{N}_{\a\b\g}$
given by 
$$
{N}_{\a\b\g}:={m}_{\b\g}{}^\r {\l}_{\a\r}
-(-1)^{\a\b}{m}_{\a\g}{}^\r {\l}_{\b\r} 
+(-1)^{\a} {O}_\a \cdot {\l}_{\b\g}
-(-1)^{\a\b +\b}{O}_\b\cdot {\l}_{\a\g}.
$$
Hence we have
\eqn\egmassy{
N_{\a\b\g} = n_{\a\b\g}{}^\r O_\g + Q x_{\a\b\g},
}
for unique $n_{\a\b\g}{}^\r \in \Bbbk$ and some $x_{\a\b\g} \in \sC^{|\a|+|\b|+|\g|-2}$
defined modulo $\Ker Q$. From \egmaxa\ we deduce that
$$
\eqalign{
{n}_{0\b\g}={n}_{\a 0\g}=n_{\a\b 0}=0,\cr
{n}_{\a\b\g}+(-1)^{|\a||\b|}{n}_{\b\a\g}=0,\cr
{n}_{\a\b\g}-(-1)^{|\b||\g|}{n}_{\a\g\b} +(-1)^{|\a|(|\b|+|\g|)}{n}_{\b\g\a}=0,
}
$$
and we can make a choice for $x_{\a\b\g}$ such that
$$
\eqalign{
{x}_{0\b\g}={x}_{\a 0\g}=x_{\a\b 0}=0,\cr
{x}_{\a\b\g}+(-1)^{|\a||\b|}{x}_{\b\a\g}=0,\cr
{x}_{\a\b\g}-(-1)^{|\b||\g|}{x}_{\a\g\b} +(-1)^{|\a|(|\b|+|\g|)}{x}_{\b\g\a}=0.
}
$$
From \egmassy, we deduce that the expression 
$\bos{N}_{\a\b\g} - n_{\a\b\g}{}^\r \bos{O}_\g - \bos{K} x_{\a\b\g}$
is divisible by $\hbar$, so that we define $\bos{\xi}_{\a\b\g} \in \sC[[\hbar]]^{|\a|+|\b|+|\g|-1}$
by the formula
$$
\hbar\bos{\xi}_{\a\b\g}
:=\bos{N}_{\a\b\g} - n_{\a\b\g}{}^\r \bos{O}_\g - \bos{K} x_{\a\b\g},
$$
and all the properties of $\bos{\xi}_{\a\b\g}$ listed in claim (3) follows.
Then, from the $3$rd relation in \qgauc, we have 
$\hbar\left(\bos{\l}_{\a\b\g} -(-1)^{|\a||\b|}\bos{\l}_{\b\a\g}\right)
=\bos{N}_{\a\b\g} - n_{\a\b\g}{}^\r \bos{O}_\g - \bos{K} x_{\a\b\g}$. More explicitly, 
we have
$$
\eqalign{
\hbar\left(\bos{\l}_{\a\b\g} -(-1)^{|\a||\b|}\bos{\l}_{\b\a\g}\right) 
=&{m}_{\b\g}{}^\r \bos{\l}_{\a\r}
-(-1)^{\a\b}{m}_{\a\g}{}^\r \bos{\l}_{\b\r} 
\cr
&
+(-1)^{\a} \bos{O}_\a \cdot \bos{\l}_{\b\g}
-(-1)^{\a\b +\b}\bos{O}_\b\cdot \bos{\l}_{\a\g}
\cr
&
- n_{\a\b\g}{}^\r \bos{O}_\g - \bos{K} x_{\a\b\g}.
}
$$

Now consider the definition
$\bos{\La}_{\b\g} = \bos{\l}_{\a\b} + t^{\bar\a} \bos{\l}_{\a\b\g} \mod t_H^2$.
Then quantum gauge conditions \qguab\ and \qgauc\ are summarized as
follows
$$
\eqalign{
\bos{\La}_{\b 0}=&0 \mod t_H^2,
\cr
\bos{\La}_{\a_2\a_1}=&(-1)^{|\a_2||\a_1|}\bos{\La}_{\a_1\a_2} \mod t_H^2
,\cr
\rd_{\bar\a}\bos{\La}_{\b\g}=& 0 \mod t_H
,\cr
\hbar\left(\rd_{\bar\a}\bos{\La}_{\b\g}-(-1)^{|\a||\b|}\rd_{\bar\b}\bos{\La}_{\a\g} \right)
=
&
-A_{\b\g}{}^\r \bos{\La}_{\a\r}
+(-1)^{|\a||\b|}A_{\a\g}{}^\r \bos{\La}_{\b\r} 
\cr
&-(-1)^{|\a|} \rd_{\a}\bos{\Theta} \cdot\bos{\La}_{\b\g}
+(-1)^{|\a||\b| +|\b|}\rd_{\b}\bos{\Theta}\cdot \bos{\La}_{\a\g}
\cr
&
-B_{\a\b\g}{}^\r \rd_\r \bos{\Theta}
-\bos{K}{X}_{\a\b\g}-\big(\bos{\Theta},{X}_{\a\b\g}\big)_{\hbar}
\cr
&
\mod t_H,
}
$$
where $B_{\a\b\g}{}^\r =n_{\a\b\g}{}^\r \mod t_H$ and $X_{\a\b\g}=x_{\a\b\g}\mod t_H$.
In general we are going to solve the quantum master equation with quantum gauge choice for 
$\bos{\La}_{\a\b}$,
which is the above condition without modulo $t_H^2$ or $t_H$.
\end{remark}

\begin{remark}\label{claimdh}
We note that there is some ambiguity in solutions for  $\bos{\Theta}$ and $\bos{\La}_{\a\b}$.
There are two sources of ambiguity. 
With the fixed  initial condition $\bos{\Theta}=t^\a \bos{f}(e_\a) \mod t_H$, 
namely the quantization map $\bos{f}$,
$\bos{\l}_{\a\b}=\bos{\La}_{\a\b}\mod t_H$ is not uniquely determined, which  effects 
$\bos{O}_{\a\b}$ in $\bos{\Theta}=t^\a \bos{O}_\a +\Fr{1}{2}t^\b t^\a\bos{O}_{\a\b} \mod t_H^2$
etc.
We may also vary the
initial condition within the same quantum homotopy class $\bos{f}\;\bos{\sim}\;\bos{f}^\pr=\bos{f}
+ \bos{K} \bos{s}$ such that $\bos{f}^\pr(e_0)=\bos{f}(e_0)=1$. 
Hence the above theorem
does not necessarily imply that the $3$-tensor $A_{\a\b}{}^\g$ is independent of those ambiguities.
We claim, however, that every  possible ambiguity does not effect the
the $3$-tensor $A_{\a\b}{}^\g$: this claim implies that quantum correlation functions are quantum homotopy invariants.
We shall need to introduce notion of quantum 
descendant homotopy, which deserves a separate consideration to appear elsewhere,
to establish the above claim.
\end{remark}

\begin{remark}
Consider  the coefficients  of expansion of $A_{\a\b}{}^\g$ at $t_H=0$:
 $$
 A_{\a\b}{}^\g =
 m_{\a\b}{}^\s 
 +\sum_\r t^\r m_{\r\a\b}{}^\s 
 +\Fr{1}{2!}\sum_{\r_1,\r_2} t^{\r_2}t^{\r_1} m_{\r_1\r_2\a\b}{}^\g
 +\cdots
$$
The two conditions for graded symmetry  of $A_{\a\b}{}^\g$, then, imply that
$m_{\a_1\cdots \a_n}{}^\g$ is totally graded symmetric for all the lower indices
for all $n=2,3,\cdots$. Hence  there is a sequence
$m_2,m_3, m_4,\cdots$ of graded symmetric
products of ghost number zero on $H$, $m_n:S^n H\rightarrow H$,
such that $m_n(e_{\a_1},\cdots, e_{\a_n})=m_{\a_1\cdots \a_n}{}^\g$.
The identity $A_{0\b}{}^\g =\d_\b{}^\g$ imply that $m_2(e_0, e_\a)= e_\a$ 
and $m_n(e_0,e_{\a_2},\cdots, e_{\a_{n-1}})=0$ for all $n=3,4,5,\cdots$. 
A similar structure on $H$ was discussed in section $4$ of the previous paper \cite{PI} with a different
presentation of quantum master equation. Once the claim in remark \ref{claimdh} is established it
is trivial to show that  those two structures
is identical. What was not clear in the previous paper is
the associativity (an easy part of  proof in this paper)
of the $A_{\a\b}{}^\g$ summarizing an infinite sequence of relations among the sequence 
$m_2, m_3, m_4, \cdots$ of multilinear products. The solution in the pervious paper
is suffice to determine quantum correlation functions. The new solution in this paper with
quantum gauge is a preparation to define and study homotopy  quantum correlation functions
in the forthcoming paper.

\end{remark}

\subsection{Quantum coordinates and linear pencil of torsion-free flat connection on $\CM$.} 

One of the immediate consequence of our main result is that
that the classical limit $\Theta$ of $\bos{\Theta}$, i.e., $\Theta=\bos{\Theta}|_{\hbar =0}$:
$$
{\Theta} = t^\a {O}_\a 
+ \sum_{n=2}^\infty \Fr{1}{n!}t^{\a_n}\cdots t^{\a_1} {O}_{\a_1\cdots\a_n}
\in\sC[[t_H]]^0,
$$
where $O_{\a_1\cdots \a_n} =\left. \bos{O}_{\a_1\cdots \a_n}\right|_{t=0}$,
is a distinguished versal solution
 to the Maurer-Cartan equation of the classical limit 
 $\big(\sC, Q, (\hbox{ },\hbox{ })\big)$
 of the quantum descendant DG0LA:
\eqn\cldeform{
Q\Theta + \Fr{1}{2}(\Theta,\Theta)=0.
}

\begin{itemize}
\item
It is a versal solution since the cohomology classes of $\{O_\a\}$ form a basis of $H$ 
- the cohomology of the complex $(\sC,Q)$. It follows that the natural extended
moduli space $\CM$ of solutions modulo the natural equivalence is smooth -
minimal $L_{\infty}$-structure on $H$ is trivial and is quasi-isomorphic
to the DG0LA $\big(\sC,Q,(\bullet,\bullet)\big)$ as an $L_{\infty}$-algebra.
(See \cite{Kontsevich} for a comprehensive and lucid introduction of $L_\infty$-algebra and morphism 
and references therein.)

\item
It is a distinguished solution since not every versal solution 
of \cldeform\ arises as the classical limit of solution of quantum master equation, i.e.,
the special solution to the quantum descendant equation. Equivalently it is
a distinguished quasi-isomorphism of $L_{\infty}$-algebras. 
\end{itemize}

Any versal solution to \cldeform\ identify $\CM$ with the affine space $H$ with
the affine coordinates $t_H=\{t^\a\}$ and give a coordinates system on $\CM$.
We call the coordinates system on $\CM$ induced by the distinguished solution
$\Theta$ quantum coordinates on $\CM$ (around the base point of $\CM$).
To the tangent space to $\CM$, our distinguished solution 
induces a linear pencil of  connection 
$\nabla_{\hbar}=-\hbar d + A$, where $d:= d\!t^\a \Fr{\rd}{\rd t^\a}$ and 
$\{A\}_{\b}{}^\g := d\!t^\a A_{\a\b}{}^\g$,
which is torsion-free, $A_{\b\g}{}^\s = (-1)^{|\b||\g|} A_{\g\b}{}^\s$,
and flat $\nabla_{\hbar}^2=0$:
$$
d A_{\b}{}^\g =A_{\b}{}^\r A_{\r}{}^\g=0,
$$
by combining the {\it graded symmetricity} and the {\it associativity relation}.

\begin{remark}

The MC equation \cldeform\
implies that $Q_{\!\Theta}:= Q + (\Theta,\hbox{ })$ 
satisfies $Q_\Theta^2=0$, 
the Jacobi-identity for the bracket $(\hbox{ },\hbox{ })$ implies 
that $Q_{\!\Theta}$ is a derivation of the bracket. 
Thus we have a distinguished 
family $\big(\sC[[t_H]], Q_{\!\Theta}, (\hbox{ }, \hbox{ })\big)$ of DGLAs . 
Furthermore  the super-Poisson law implies that $Q_{\!\Theta}$ is a derivation of the product. 
Thus the quadruple $$\big(\sC[[t_H]], Q_{\!\Theta}, \;\cdot\;, (\hbox{ }, \hbox{ })\big)$$
is a differential $0$-algebra.
Applying $\Fr{\rd}{\rd t^\a}$ to the MC equation \cldeform, we obtain that
$$
Q\Theta_\a + (\Theta, \Theta_\a)=0\quad\Longleftrightarrow \quad  Q_{\!\Theta}\Theta_\a=0.
$$
where $\Theta_\a := \rd_\a \Theta_\a$.
In Appendix B, we shall establish that  $\{\Theta_\a\}$ is a set of representative
of a basis of the cohomology $\mH$ of the  cochain complex 
$(\sC[[t_H]],Q_{\!\Theta})$ such that
\eqn\ewa{
\Theta_\a \cdot \Theta_\b = A_{\a\b}{}^\g\Theta_\g + Q_{\!\Theta}\La_{\a\b},
}
for unique $3$-tensor $A_{\a\b}{}^\g$ in $\Bbbk[[t_H]]$
and for some 
$\La_{\a\b} \in\left(\sC[[t_H]]\right)^{|\a|+|\b|-1}$, which is defined modulo $\Ker Q_{\!\Theta}$. 
Then,
the graded commutativity and the associativity of the product $\cdot$
imply  that $\mH$ is a super-commutative and associative
$\Bbbk[[t_H]]$-algebra with the structure "constants" $A_{\a\b}{}^\g$,
that is,
$A_{\a\b}{}^\g=(-1)^{|\a||\b|}A_{\b\a}{}^\g$ and
$A_{\a\b}{}^\r A_{\r \g}{}^\s = A_{\b\g}{}^\r A_{\a\r}{}^\s$.
Note that \ewa\ is the classical limit of the quantum master equation.
Forgotten its quantum origin the potentiality 
$\rd_\a A_{\b\g}{}^\s -(-1)^{|\a||\b|}\rd_\b A_{\a\g}{}^\s=0$
is obscured.

\end{remark}

\begin{remark}

Note that we can repeat the similar story as the above remark for an
arbitrary versal solution $\Theta^\pr=t^\a O^\pr_\a + \Fr{1}{2}t^{\b}t^\a O_{\a\b}^\pr
+\cdots$ to \cldeform\ such that $\{O^\pr_\a\}$ is an another set of representative of
the same basis $\{e_\a\}$ of the cohomology $H$.
Then the result proved in Appendix B does
not necessarily  imply that we have the same $3$-tensor $A_{\a\b}{}^\g$. It only implies that
there is unique $3$-tensor $A^\pr_{\a\b}{}^\g$ with respect to the solution $\Theta^\pr$ such that
\eqn\ewaz{
\Theta_\a^\pr \cdot \Theta_\b^\pr = A^\pr_{\a\b}{}^\g\Theta^\pr_\g + Q_{\!\Theta}\La_{\a\b}^\pr,
}
for some 
$\La^\pr_{\a\b} \in\left(\sC[[t_H]]\right)^{|\a|+|\b|-1}$, which is defined modulo $\Ker Q_{\!\Theta^\pr}$. 
The only relation between $A_{\a\b}{}^\g$ and $A^\pr_{\a\b}{}^\g$ is that
$A^\pr_{\a\b}{}^\g=A_{\a\b}{}^\g=m_{\a\b}{}^\g \mod t_H$. It can be, actually, shown that
the difference between $A_{\a\b}{}^\g$ and $A^\pr_{\a\b}{}^g$ in higher order is arbitrary
(see the next remark).
Note also that the associativity and the commutativity of the product $\cdot$ do imply that
$A^\pr_{\a\b}{}^\g=(-1)^{|\a||\b|}A^\pr_{\b\a}{}^\g$ and
$A^\pr_{\a\b}{}^\r A^\pr_{\r \g}{}^\s = A^\pr_{\b\g}{}^\r A^\pr_{\a\r}{}^\s$, while there is absolutely 
no reason to expect that $\rd_\a A^\pr_{\b\g}{}^\s -(-1)^{|\a||\b|}\rd_\b A^\pr_{\a\g}{}^\s=0$.
\end{remark}

\begin{remark} Here is a brief comparison:
Modulo $t_H$, \ewaz\ is
\eqn\ewaza{
O^\pr_\a\cdot O^\pr_\b = m^\pr_{\a\b}{}^\g O^\pr_\g + Q\l_{\a\b}^\pr
}
where $\l_{\a\b}^\pr$ is defined modulo $\xi_{\a\b}^\pr$ satisfying $Q\xi_{\a\b}^\pr=0$.
Modulo $t_H^2$, \ewaz\ is the above equation together with the following
\eqn\ewazb{
O^\pr_{\a\b}\cdot O^\pr_\g +(-1)^{|\a||\b|} O^\pr_{\a}\cdot O^\pr_{\b\g}
-m^\pr_{\b\g}{}^\r O^\pr_{\a\r} -(O^\pr_\a, \l_{\b\g}^\pr)
= m^\pr_{\a\b\g}{}^\s O^\pr_\s + Q\l^\pr_{\a\b\g}
}
where $\l_{\a\b\g}^\pr$ is defined modulo $\xi_{\a\b\g}^\pr$ satisfying $Q\xi_{\a\b\g}^\pr
+(-1)^{|\a|}\big(O^\pr_{\a}, \xi_{\b\g}^\pr\big)=0$. 
Then the only content of 
proposition $B.1$ to this order is that the cohomology class of the LHS of \ewazb, thus the structure
constant $m^\pr_{\a\b\g}{}^\s$, does not depend
on the ambiguity of $\l_{\a\b}^\pr$. 
Now we compare \ewa\ with \ewaz\ for the leading two terms. In general
$O^\pr_\a$ differs from $O_\a$
at most by $Q \l_\a$ for some $\l_\a \in \sC^{|\a|-1}$. It follows that
$[O^\pr_\a\cdot O^\pr_\b]= [O_\a\cdot O_\b]$ such that $m_{\a\b}{}^\g =m^\pr_{\a\b}{}^\g$.
For simplicity set $O^\pr_\a = O_\a$.
Let, for example, $O^\pr_{\a\b} = O_{\a\b} + b_{\a\b}{}^\g O_\g $, where $\{b_{\a\b}^\g\}$ is
arbitrary. Then $\Theta^\pr =t^\a O_\a + \Fr{1}{2}t^\b t^\a O^\pr_{\a\b} \mod t_H^3$ 
solves the MC equation \cldeform\ modulo $t_H^3$ since
$\Theta =t^\a O^\pr_\a + \Fr{1}{2}t^\b t^\a O_{\a\b} \mod t_H^3$ does.
On the  other hand, the LHS of \ewazb\ is
$$
\eqalign{
O^\pr_{\a\b}\cdot O^\pr_\g &+(-1) O^\pr_{\a}\cdot O^\pr_{\b\g}
-m^\pr_{\b\g}{}^\r O^\pr_{\a\r} -(O^\pr_\a, \l^\pr_{\b\g})
\cr
= &C_{\a\b\g} 
+ \left(b_{\a\b}{}^\r m_{\r\g}{}^\s
(-1)^{|\a||\b|}  b_{\b\g}{}^\r m_{\a\r}{}^\s - m_{\b\g}{}^\r b_{\a\r}{}^\s\right)O_\s
\mod \hbox{Im } Q,
}
$$
where
$C_{\a\b\g}=O_{\a\b}\cdot O_\g +(-1) O_{\a}\cdot O_{\b\g}
-m_{\b\g}{}^\r O_{\a\r} -(O_\a, \l_{\b\g})$ such that the word-length $2$ in $t_H$ term
of \ewa\ is $C_{\a\b\g} = m_{\a\b\g}{}^\s O_\s + Q \l_{\a\b\g}$. Hence  the difference 
between $m_{\a\b\g}{}^\s$ and $m^\pr_{\a\b\g}{}^\s$ is arbitrary.

\end{remark}

\subsection{Quantum coordinates and generating function of quantum correlations for the family}

Consider the solution $\bos{\Theta}=t^\a \bos{O}_\a +\Fr{1}{2}t^\b t^\a\bos{O}_{\a\b}+\cdots$
of quantum master equation, which implies that $\bos{\Theta}$ also solve
the quantum descendant equation 
$\bos{K}\bos{\Theta}+\Fr{1}{2}\big(\bos{\Theta},\bos{\Theta}\big)_{\hbar}=0$.
Note that quantum descendant equation 
is equivalent to
\eqn\useqd{
\bos{K} e^{-\bos{\Theta}/\hbar}=0,
}
 due to the identity
$
\hbar^2 e^{\bos{\Theta}/\hbar}\bos{K}e^{-\bos{\Theta}/\hbar}=
\bos{K}\bos{\Theta}+\Fr{1}{2}\big(\bos{\Theta},\bos{\Theta}\big)_{\hbar}$.
Define $\bos{\Pi}_{\a_{1}\cdots\a_{n}}\in\sC[[t_H,\hbar]]^{|\a_1|+\cdots+ |\a_n|}$ for $\forall n=1,2,\cdots$
by the formula
\eqn\mtra{
\bos{\Pi}_{\a_1\cdots \a_n}= (-\hbar)^n
\rd_{\a_1}\cdots \rd_{\a_n}e^{-\bos{\Theta}/\hbar}.
}
Then quantum descendant equation \useqd\ implies that
\eqn\mtrb{
\bos{K}_{\!\bos{\Theta}}\bos{\Pi}_{\a_1\cdots \a_n}=0,
}
due to the identity that $-\hbar e^{\bos{\Theta}/\hbar} \bos{K}\left( \bos{Y}
e^{-\bos{\Theta}/\hbar}\right) = \bos{K}_{\!\bos{\Theta}}\bos{Y}$ for any $\bos{Y}
\in \sC[[t_H,\hbar]]$:
$$
\eqalign{
\bos{K}\left(\bos{Y}\cdot e^{-\bos{\Theta}/\hbar}\right)
&=e^{-\bos{\Theta}/\hbar}\Big(\bos{K}\bos{Y}+(\bos{\Theta},\bos{Y}\big)_{\hbar}\Big)
+(-1)^{|\bos{Y}|}\bos{Y}\cdot \bos{K}e^{-\bos{\Theta}/\hbar}
\cr
&=e^{-\bos{\Theta}/\hbar}\Big(\bos{K}\bos{Y}+(\bos{\Theta},\bos{Y}\big)_{\hbar}\Big),
}
$$
where we used the definition of \bvbracket\ for the first equality
and the quantum descendant equation for the second equality.
We also have $\bos{\Pi}_{0}=1$ and $\bos{\Pi}_{0\a_1\cdots \a_n}=\bos{\Pi}_{\a_1\cdots \a_n}$ 
for all $\forall n\geq 1$, due to the quantum identity $\rd_0 \bos{\Theta}=1$.

 We call 
$\bos{\Pi}_{\a_1\cdots \a_n}$ quantum $n$-point correlators of the family (of BV QFT
parametrized by $\CM$ in quantum coordinates).
Note that 
$$
\eqalign{
\bos{\Pi}_\a =& \rd_\a \bos{\Theta}
,\cr
\bos{\Pi}_{\a\b} =&\rd_{\a}\bos{\Theta}\cdot \rd_{\b}\bos{\Theta}  -\hbar \rd_\a\rd_\b \bos{\Theta} 
,\cr
\bos{\Pi}_{\a\b\g} =&\rd_{\a}\bos{\Theta}\cdot\rd_{\b}\bos{\Theta}\cdot \rd_{\g}\bos{\Theta}  
-\hbar\rd_{\a}\rd_\b\bos{\Theta}\cdot \rd_{\g}\bos{\Theta}  
-\hbar\rd_{\a}\bos{\Theta}\cdot \rd_\b\rd_{\g}\bos{\Theta}  
-\hbar(-1)^{|\a||\b|}\rd_\b\bos{\Theta}\cdot\rd_\a \rd_{\g}\bos{\Theta}  
\cr
&+\hbar^2\rd_\a \rd_\b\rd_\g \bos{\Theta} 
,\cr
}
$$
etc.

We recall, from \cite{PI}, that
a solution $\bos{\Theta}$ of the quantum master equation 
was used to define generation function of
quantum correlators of the initial theory by the formula
\eqn\genad{
e^{-\bos{\Theta}/\hbar}
=1 + \sum_{n=1}^\infty \Fr{1}{n!}\Fr{(-1)^n}{\hbar^{n}}\bos{\Theta}^n
=1+\sum_{n=1}^\infty \Fr{(-1)^n}{\hbar^n} \bos{\Omega}_n \in \sC[[t_H]](\!(\hbar)\!)^0,
}
where the sequence $\bos{\Omega}_1,\bos{\Omega}_2, \cdots$  is defined 
by matching the word-lengths in $t_H$. Then
$\bos{\Omega}_{n}$ generates $n$-point quantum correlators of the initial theory:
$$
\bos{\Omega}_n=\Fr{1}{n!}t^{\a_n}\cdots t^{\a_1}\bos{\pi}_{\a_1\cdots \a_n}
\quad\hbox{where}\quad \bos{\pi}_{\a_1\cdots \a_n}\in \sC[[\hbar]]^{|\a_1|+\cdots+|\a_n|},
$$
such that $\bos{K} \bos{\pi}_{\a_1\cdots \a_n}=0$ and $\bos{\pi}_{\a_1\cdots \a_n}\big|_{\hbar=0}
=O_{\a_1}\cdots O_{\a_n}$.
We observe, from the definition in \mtra, that
\eqn\mtre{
\bos{\pi}_{\a_1\cdots \a_n}=\bos{\Pi}_{\a_1\cdots \a_n}\big|_{t_H=0}.
}

Fix a QFT cycle $\mb{\mc}$ of dimension $N$ and denote $\mb{\mc}(\bos{a})=\left<\bos{a}\right>$.
The generating
functional $\mb{\sZ}(t_H)$ of  quantum correlation functions of the initial theory
was defined by the formula
\eqn\mtrd{
\eqalign{
\mb{\sZ}=\left<e^{-\bos{\Theta}/\hbar}\right>
&:= <1> + \sum_{n=1}^\infty \Fr{(-1)^n}{\hbar^{n}}\left<\bos{\O}_n\right>
\cr
&=<1> + \sum_{n=1}^\infty \Fr{1}{n!}\Fr{(-1)^n}{\hbar^{n}}
t^{\a_{n}}\cdots t^{\a_{1}}\left<\bos{\pi}_{\a_{1}\cdots\a_{n}}\right>
}
}
such that an arbitrary $n$-point correlation function 
$\left<\bos{\pi}_{\a_1\cdots\a_n}\right>$
is obtained as follows:
$$
\left<\bos{\pi}_{\a_1\cdots\a_n}\right>
\equiv (-\hbar)^n \rd_{\a_1}\cdots \rd_{\a_n}\mb{\sZ}(t_H)\bigl|_{t=0} \in \Bbbk[[\hbar]].
$$
Remark that $\mb{\sZ}$ is a formal power series in $t_H$ and formal Laurent series in $\hbar$.

Now we introduce the notion of quantum correlation function for the family.
We use the notation $\big< \bos{Y} \big>_{t_H}$ for any $\bos{Y}\in \sC[[t_H,\hbar]]$ such that
$$
\big< \bos{Y} \big>_{t_H}= \left< \bos{Y}\cdot e^{-\bos{\Theta}/\hbar}\right>
:= \left<\bos{Y}\right> + \sum_{n=1}^\infty \Fr{(-1)^n}{\hbar^{n}}\left<\bos{Y}\cdot\bos{\O}_n\right>.
$$
Remark that $\big< \bos{Y} \big>_{t_H}$ is a formal Laurent series $\hbar$,
while  $\big< \bos{Y} \big>_{t_H=0}=\big< \bos{Y} \big>$ is a formal power series $\hbar$.
Then $\big< \bos{K}_{\!\bos{\Theta}}\bos{\Xi}\big>_{t_H}=0$ for all $\bos{\Xi}\in \sC[[t_H,\hbar]]$
since
$$
\eqalign{
\left<\bos{K}_{\!\bos{\Theta}}\bos{\Xi}\cdot e^{-\bos{\Theta}/\hbar}\right>
\equiv\left<\bos{K}\left(\bos{\Xi}\cdot e^{-\bos{\Theta}/\hbar}\right)\right>
:=\big<\bos{K}\bos{\Xi}\big> 
+\sum_{n=1}^\infty \Fr{(-1)^n}{\hbar^{n}}\big<\bos{K}\left(\bos{\Xi}\cdot\bos{\O}_n\right)\big>
=0.
}
$$
We define $n$-point quantum correlation functions in the family by
$\big<\bos{\Pi}_{\a_1\cdots\a_n}\big>_{t_H}$ for all $n=1,2,3,\cdots$. 
Then we deduce that
$$
\left<\bos{\Pi}_{\a_1\cdots\a_n}\right>_{t_H}
\equiv (-\hbar)^n \rd_{\a_1}\cdots \rd_{\a_n}\mb{\sZ} \in \Bbbk[[t_H]](\!(\hbar)\!).
$$
and 
\eqn\qcfa{
\left<\bos{\Pi}_{\a_1\cdots\a_n}\right>_{t_H=0}=\left<\bos{\pi}_{\a_1\cdots\a_n}\right>.
}

Note that the quantum master equation
$$
\hbar\rd_{\b}\rd_{\g}\bos{\Theta}
= \rd_{\b}\bos{\Theta}\cdot \rd_{\g}\bos{\Theta} - A_{\b\g}{}^\s \rd_{\s}\bos{\Theta} 
- \bos{K}\bos{\La}_{\b\g} - 
\left(\bos{\Theta},\bos{\La}_{\b\g}\right)_{\hbar},
$$
is equivalent to $\bos{\Pi}_{\a\b} = A_{\a\b}{}^\g \bos{\Pi}_\g + \bos{K}_{\!\bos{\Theta}}\bos{\La}_{\a\b}$,
so that
\eqn\qcftp{
\big<\bos{\Pi}_{\a\b}\big>_{t_H} = A_{\a\b}{}^\g \big<\bos{\Pi}_\g\big>_{t_H},
}
which relates quantum $2$-point functions to quantum $1$-point correlation functions 
both in the family. For higher correlations,
it is useful to consider the following form of the quantum master equation:
\eqn\hea{
\Big(\hbar^2 \rd_\b \rd_\g +\hbar A_{\b\g}{}^\r \rd_\r\Big)e^{-\bos{\Theta}/\hbar}
=\bos{K}\left(\bos{\La}_{\a\b} e^{-\bos{\Theta}/\hbar} \right),
}
since 
$$
\eqalign{
LHS:
\Big(\hbar^2 \rd_\b \rd_\g +\hbar A_{\b\g}{}^\r \rd_\r\Big)e^{-\bos{\Theta}/\hbar}
&=e^{-\bos{\Theta}/\hbar}\Big(\rd_\b \bos{\Theta}\cdot \rd_\g \bos{\Theta}
-\hbar \rd_\b \rd_\g\bos{\Theta}
- A_{\b\g}{}^\r \rd_\g \bos{\Theta}\Big)
,\cr
RHS: \bos{K}\left(\bos{\La}_{\a\b} e^{-\bos{\Theta}/\hbar} \right)
&=e^{-\bos{\Theta}/\hbar}\Big(\bos{K}\bos{\La}_{\a\b} 
+\big(\bos{\Theta},\bos{\La}_{\b\g}\big)_{\hbar}\Big).
}
$$
Similarly, the quantum identity $\rd_0 \bos{\Theta}=1$ is equivalent to
\eqn\heb{
-\hbar \rd_0 e^{-\bos{\Theta}/\hbar} = e^{-\bos{\Theta}/\hbar}.
}
Note  the relations \hea\ and \heb\ after using 
$\left<\bos{K}\left(\bos{\La}_{\a\b} e^{-\bos{\Theta}/\hbar} \right)\right>=0$ imply that 

\begin{lemma} \label{tivaxu}
The generating function $\mb{\sZ} \in \Bbbk[[t_H]](\!(\hbar)\!)$
satisfies the following system of differential equations:
$$
\eqalign{
\left(\hbar\Fr{\rd^2}{\rd\! t^\a \rd\! t^\b} + A_{\a\b}{}^\g \Fr{\rd}{\rd\! t^\g}\right)\mb{\sZ}&=0,
\cr
\Big(\hbar\Fr{\rd}{\rd\! t^0} + 1\Big)\mb{\sZ}&=0,
\cr
\Big(|e_\a| t^\a \Fr{\rd}{\rd\! t^\a} - N\Big)\mb{\sZ}&=0, 
}
$$
where $N$ is the dimension of QFT cycle.
\end{lemma}

\begin{proof}
The first two relations are consequences of \hea\ and \heb, respectively.
The last relation is a direct consequence of $\left< \bos{Y}\right>=0$
 for all $\bos{Y}\in \sC[[\hbar]]$ with ghost number
not equals to $N$. Consider the definition $\mb{\sZ}$ in \mtrd:
$$
\mb{\sZ}=\left<e^{-\bos{\Theta}/\hbar}\right>
=<1> + \sum_{n=1}^\infty \Fr{1}{n!}\Fr{(-1)^n}{\hbar^{n}}
\sum_{\a_1,\ldots,\a_n}t^{\a_{n}}\cdots t^{\a_{1}}\left<\bos{\pi}_{\a_{1}\cdots\a_{n}}\right>,
$$
where we temporarily abandon Einstein summation convention.
Then the relation follows from
the identity
$$
\sum_\a |e_\a| t^\a \Fr{\rd}{\rd t^\a}\left(t^{\a_{n}}\cdots t^{\a_{1}}
\left<\bos{\pi}_{\a_{1}\cdots\a_{n}}\right>\right)
= \left(\sum_{i=1}^n |e_{\a_i}|\right)t^{\a_{n}}\cdots t^{\a_{1}}\left<\bos{\pi}_{\a_{1}\cdots\a_{n}}\right>
$$
and
the condition that
$$
t^{\a_{n}}\cdots t^{\a_{1}}\left<\bos{\pi}_{\a_1\cdots\a_n}\right> =0
$$
for
$
|\bos{\pi}_{\a_1\cdots\a_n}|\equiv -\sum_{i=1}^n |t^{\a_i}|\equiv \sum_{i=1}^n |e_{\a_i}|\neq N$.

\qed

\end{proof}

Let $\{\bos{P}_{\a_{1}\a_{2}}{}^{\g},\bos{P}_{\a_{1}\a_{2}\a_{3}}{}^{\g},\cdots\}$
be the set of infinite sequences in $\Bbbk[[t_{H},\hbar]]$ defined recursively 
with the initial condition
$\bos{P}_{\a_1\a_2}{}^\g= A_{\a_{1}\a_{2}}{}^{\g}$ and, for $n=3,4,\ldots$,
$$
\bos{P}_{\a_1\a_2\cdots\a_n}{}^\g
=-\hbar\rd_{\a_{1}} \bos{P}_{\a_2\cdots\a_n}{}^\g
+\bos{P}_{\a_2\cdots\a_n}{}^\r A_{\a_1\r}{}^\g.
$$
We note that $\bos{P}_{\a_1\a_2\cdots\a_n}{}^\g$
is at most of degree
$(n-2)$ polynomial in $\hbar$ with coefficient in $\Bbbk[[t_H]]$.
Note also that $\bos{P}_{0\b}{}^\g=\d_{\b}{}^\g$ and
$\bos{P}_{0\a_1\a_2\cdots\a_n}{}^\g= \bos{P}_{\a_1\a_2\cdots\a_n}{}^\g$
for $\forall n\geq 2$. Then

\begin{lemma}
Any $n$-point quantum correlation function 
$\left<\bos{\Pi}_{\a_{1}\cdots \a_{n}}\right>_{t_H}$, for $n=2,3,\ldots$, for the family
is given by the following formula
$$
\left<\bos{\Pi}_{\a_{1}\cdots \a_{n}}\right>_{t_H}=
\bos{P}_{\a_{1}\cdots\a_{n}}{}^{\g}\left<\bos{\Pi}_\g\right>_{t_H}.
$$

\end{lemma}

\begin{proof}
From the quantum master equation in the form of \hea:
\eqn\qcfb{
\eqalign{
\hbar^2 \rd_\b \rd_\g e^{-\bos{\Theta}/\hbar}
&=-\hbar A_{\b\g}{}^\r \rd_\r e^{-\bos{\Theta}/\hbar}
+\bos{K}\left(\bos{\La}_{\b\g} \cdot e^{-\bos{\Theta}/\hbar}\right)
..\cr
}
}
as a direct consequence of the quantum master equation. 
Applying $-\hbar \rd_{\a}$ to the both hand sides of \qcfb,
and using \qcfb\ one more time, we obtain that 
\eqn\qcfc{
\eqalign{
-\hbar^3 \rd_\a \rd_\b\rd_\g e^{-\bos{\Theta}/\hbar}
=&
\left( -\hbar \rd_\a A_{\b\g}{}^\s + A_{\b\g}{}^\r A_{\a\r}{}^\s\right)\bos{\Theta}_\s \cdot
e^{-\bos{\Theta}/\hbar}
\cr
&
+ \bos{K}\left(\bos{X}_{\a\b\g}\cdot
e^{-\bos{\Theta}/\hbar}\right),
}
}
where 
$$
\bos{X}_{\a\b\g}=-\hbar  (-1)^{|a|} \rd_\a\bos{\La}_{\b\g}
+ (-1)^{|a|}\bos{\Theta}_\a\cdot \bos{\La}_{\b\g}
+A_{\b\g}{}^\r \bos{X}_{\a\r}.
$$
In general we deduce that, for $n=3,4,\ldots$,
\eqn\qcfd{
\eqalign{
(-\hbar)^n \rd_{\a_1}\rd_{\a_2}\cdots \rd_{\a_n} e^{-\bos{\Theta}/\hbar}
=\bos{P}_{\a_1\a_2\cdots\a_n}{}^\s {\bos{\Theta}}_\s\cdot e^{-\bos{\Theta}/\hbar}
+ \bos{K}
\left({\bos{X}}_{\a_1\a_2\cdots\a_n}\cdot
e^{-\bos{\Theta}/\hbar}\right),
}
}
where  
$$
\eqalign{
\bos{P}_{\a_1\a_2\cdots\a_n}{}^\s 
&=-\hbar\rd_\a \bos{P}_{\a_2\cdots\a_n}{}^\s
+\bos{P}_{\a_2\cdots\a_n}{}^\r A_{\a_1\r}{}^\s,
\cr
\bos{X}_{\a_1\a_2\cdots \a_n}
&=\hbar  (-1)^{|\a_1|} \rd_{\a_1}\bos{X}_{\a_2\cdots\a_n}
+ (-1)^{|\a_1|}\bos{\Theta}_{\a_1} \cdot\bos{X}_{\a_2\cdots\a_n}
+\bos{P}_{\a_2\cdots\a_n}{}^\r \bos{X}_{\a_1\r}
}
$$
with the initial conditions 
$\bos{P}_{\a_1\a_2}{}^\g= A_{\a_1\a_2}{}^\g$
and $\mathbb{X}_{\a_1\a_2}=\bos{\La}_{\a_1\a_2}$.
From \mtra, the relation \qcfd\ is equivalent to
\eqn\qcfdz{
\bos{\Pi}_{\a_1\cdots\a_n}{}^\r =\bos{P}_{\a_1\cdots\a_n}{}^\g\bos{\Pi}_\g + 
\bos{K}_{\!\bos{\Theta}}\bos{X}_{\a_1\cdots \a_n},
}
which implies our lemma.
\qed
\end{proof}

Now set $t_H=0$ in the identity \qcfdz\ and use the conditions that
$\bos{\Theta}\bigl|_{t_H=0}=0$ and $\bos{\Theta}_{\a}\bigl|_{t_H=0}=\bos{O}_{\a}$ 
as well as the relation \qcfa\ to obtain that
\eqn\qcfd{
\bos{\pi}_{\a_1\cdots \a_n}= \bos{p}_{\a_1\cdots \a_n}{}^\g \bos{O}_\g
+ \bos{K} \,\bos{x}_{\a_1\cdots \a_n}
}
where $\bos{p}_{\a_1\cdots \a_n}{}^\g :=\bos{P}_{\a_1\cdots \a_n}{}^\g\big|_{t_H=0}$
and $\bos{x}_{\a_1\cdots \a_n} :=\bos{X}_{\a_1\cdots \a_n}\big|_{{t_H=0}}$.
Note that $\bos{p}_{\a_1\cdots \a_n}{}^\g\in \Bbbk[[\hbar]]$ is at most of degree $(n-2)$ polynomial
 in $\hbar$ with coefficients in $\Bbbk$.
Hence we have 
$$
\left<\bos{\pi}_{\a_1\cdots \a_n}\right>
= \bos{p}_{\a_1\cdots \a_n}{}^\g \left<\bos{O}_\g\right>.
$$
which implies that
the set $\{A_{\a\b}{}^{\g}\}$ can be used to determine the generating function $\mb{\sZ}$ completely 
once we fix every $1$-point correlation function 
$\left<\bos{O}_{\a}\right>$.

\begin{example}
The first few quantum correlators are 
$$
\eqalign{
\bos{\pi}_\a:=&\bos{O}_\a
,\cr
\bos{\pi}_{\a_{1}\a_{2}}:=&\bos{O}_{\a_{1}} \bos{O}_{\a_{2}} 
-\hbar \bos{O}_{\a_{1}\a_{2}}
,\cr
\bos{\pi}_{\a_{1}\a_{2}\a_{3}}=&
 \bos{O}_{\a_{1}} \bos{O}_{\a_{2}} \bos{O}_{\a_{3}}
-\hbar  \bos{O}_{\a_{1}\a_{2}}  \bos{O}_{\a_{3}}
-\hbar  \bos{O}_{\a_{1}} \bos{O}_{\a_{2}\a_{3}}
-\hbar (-1)^{|\a_{1}||\a_{2}|} \bos{O}_{\a_{2}}  \bos{O}_{\a_{1}\a_{3}}
+\hbar^2\bos{O}_{\a_{1}\a_{2}\a_{3}},
}
$$
and, just for fun,
$$
\eqalign{
\bos{\pi}&_{\a_{1}\a_{2}\a_{3}\a_{4}}
=
 \bos{O}_{\a_{1}} \bos{O}_{\a_{2}}  \bos{O}_{\a_{3}}  \bos{O}_{\a_{4}}
 -\hbar \bos{O}_{\a_{1}\a_{2}}  \bos{O}_{\a_{3}}  \bos{O}_{\a_{4}}
 -\hbar  \bos{O}_{\a_{1}} \bos{O}_{\a_{2}\a_{3}}  \bos{O}_{\a_{4}}
 -\hbar  \bos{O}_{\a_{1}} \bos{O}_{\a_{2}}  \bos{O}_{\a_{3}\a_{4}}
 \cr
 &
 -\hbar(-1)^{|\a_{1}||\a_{2}|}  \bos{O}_{\a_{2}}  \bos{O}_{\a_{1}\a_{3}}  \bos{O}_{\a_{4}}
-\hbar(-1)^{|\a_{1}|(|\a_{2}|+|\a_{3}|)} 
\bos{O}_{\a_{2}}  \bos{O}_{\a_{3}}  \bos{O}_{\a_{1}\a_{4}} 
-\hbar (-1)^{|\a_{2}||\a_{3}|}  \bos{O}_{\a_{1}}\bos{O}_{\a_{3}}  \bos{O}_{\a_{2}\a_{4}}
 \cr
 &
+\hbar^{2}  \bos{O}_{\a_{1}\a_{2}}  \bos{O}_{\a_{3}\a_{4}}
+\hbar^{2} (-1)^{|\a_{2}||\a_{3}|} \bos{O}_{\a_{1}\a_{3}}  \bos{O}_{\a_{2}\a_{4}}
+\hbar^{2} (-1)^{|\a_{1}|(|\a_{2}|+|\a_{3}|)}  \bos{O}_{\a_{2}\a_{3}}  \bos{O}_{\a_{1}\a_{4}}
\cr
&
+\hbar^2 \bos{O}_{\a_{1}}\bos{O}_{\a_{2}\a_{3}\a_{4}}
+\hbar^{2}(-1)^{|\a_{1}||\a_{2}|}  \bos{O}_{\a_{2}}  \bos{O}_{\a_{1}\a_{3}\a_{4}}
+\hbar^{2}(-1)^{(|\a_{1}|+|\a_{2}|)|\a_{3}|} \bos{O}_{\a_{3}}  \bos{O}_{\a_{1}\a_{2}\a_{4}}
\cr
&
+\hbar^{2}  \bos{O}_{\a_{1}\a_{2}\a_{3}}  \bos{O}_{\a_{4}}
-\hbar^3\bos{O}_{\a_{1}\a_{2}\a_{3}\a_{4}}
}
$$
etc. 
Expand $A_{\a\b}{}^\g= m_{\a\b}{}^{\g} + t^\r m_{\r\a\b}{}^\g + \ldots$
we have
$$
\eqalign{
\bos{p}_{\a_{1}\a_{2}}{}^{\g}
=& m_{\a_{1}\a_{2}}{}^\g,\cr
\bos{p}_{\a_{1}\a_{2}\a_{3}}{}^{\g}
= &  m_{\a_{2}\a_{3}}{}^\r m_{\a_{1}\r}{}^\g-\hbar m_{\a_{1}\a_{2}\a_{3}}{}^\g 
,\cr
\bos{p}_{\a_{1}\a_{2}\a_{3}\a_{4}}{}^{\g}
= &  
m_{\a_{3}\a_{4}}{}^\r m_{\a_{2}\r}{}^\s m_{\a_{1}\s}{}^{\g}
\cr
&
-\hbar\left( m_{\a_{1}\a_{3}\a_{4}}{}^\r m_{\a_{2}\r}{}^\g
	+ m_{\a_{3}\a_{4}}{}^\r m_{\a_{1}\a_{2}\r}{}^\g
 	+m_{\a_{2}\a_{3}\a_{4}}{}^\r m_{\a_{1}\r}{}^{\g}
	\right)
\cr
& 
+ \hbar^{2} m_{\a_{1}\a_{2}\a_{3}\a_{4}}{}^\g,
}
$$
etc. 
\end{example}

Combining \genad\ and \qcfd, the expression $e^{-\bos{\Theta}/\hbar}$ has the following expansion at
$t_H$
\eqn\mainzx{
e^{-\bos{\Theta}/\hbar}
= 1 -\Fr{1}{\hbar} \mb{T}^{\g}\bos{O}_{\g} +\sum_{n=2}^\infty \Fr{(-1)^n}{\hbar^{n}}\bos{K}\bos{x}^{[n]}
\in \sC[[t_H]](\!(\hbar)\!),
}
where
\eqn\mainy{
\mb{T}^{\g}:=t^{\g} 
+\sum_{n=2}^\infty \Fr{1}{n!}\Fr{(-1)^{n-1}}{\hbar^{n-1}}
 t^{\a_{n}}\cdots  t^{\a_{1}}\bos{p}_{\a_{1}\cdots\a_{n}}{}^{\g},
}
and $\bos{x}^{[n]} =\Fr{1}{n!}\bar t^{\a_{n}}\cdots \bar t^{\a_{1}}\bos{x}_{\a_{1}\cdots\a_{n}}{}^{\g}$.
Note that $\mb{T}^\g =t^\g \mod t_H^2$ and a formal power series in $t_H$ and $\hbar^{-1}$,
since $\bos{p}_{\a_{1}\cdots\a_{n}}{}^{\g}$ is at most degree $n-2$ polynomial in $\hbar$.

We call $\{\mb{T}^\g\}$ the quantum coordinates for the family (in formal neighborhood of the base point $o$ in $\CM$). Note that this notion is independent to QFT cycle.

From \mainzx,  the generating function $\mb{\sZ}$ can be  expressed as, with respect to a QFT cycle,
\eqn\mainz{
\mb{\sZ} =<1> -\Fr{1}{\hbar} \mb{T}^{\g}\left<\bos{O}_{\g}\right>.
}
Applying $-\hbar\rd_\b$ to \mainz, we have
$-\hbar\rd_\b \mb{\sZ} = \rd_\b\mb{T}^{\g}\left<\bos{O}_{\g}\right>$.
Combing with the relation that $-\hbar\rd_\b \mb{\sZ}=\left<\bos{\Pi}_{\b}\right>_{t_H}$ we obtain
a crucial formula: 
\eqn\qto{
\left<\bos{\Pi}_{\b}\right>_{t_H}= \Fr{\rd \mb{T}^\g}{\rd t^\b}\left<\bos{O}_{\g}\right>,
}
which relates the set of expectation value of observables
of the theory at the basepoint $\CM$ with set
of expectation value of observables of  theory 
at a general point in $\CM$. From the definition of
$\left<\bos{\Pi}_{\a_1\ldots\a_n}\right>_{t_H}$ the following relation is obvious
\eqn\qtoa{
\left<\bos{\Pi}_{\a_1\ldots\a_n}\right>_{t_H} = 
(-\hbar)^{n-1} \Fr{\rd^{n}\mb{T}^\g}{\rd t^{\a_1}\ldots\rd t^{\a_n}}
\left<\bos{O}_\g\right>.
}
Thus the equation \qto\ determine every correlation function of the
family of the theory.  In particular, the relation
$\left<\bos{\Pi}_{\a\b}\right>_{t_H} =\hbar \rd_\a\rd_\b  \mT^\g \left<\bos{O}_\r\right>$
after combined with \qcftp\ and \qto\ implies that
$
\left(\hbar\rd_\a \rd_\b+ A_{\a\b}{}^\r \rd_\r \right)\mb{T}^\g\left<\bos{O}_\g\right>=0$.
Actually we have a better result than that:

\begin{lemma}\label{gtopa}
Each $\mb{T}^\g$, which is in $\Bbbk[[t_H,\hbar^{-1}]]$, satisfies the following 
system of formal differential equations

(1)$\quad
\left(\hbar\Fr{\rd^2}{\rd\! t^\a\rd\! t^\b} +A_{\a\b}{}^\r\Fr{\rd}{\rd\! t^\r}\right)\mb{T}^\g=0$,

(2)$\quad
\left(\hbar\Fr{\rd}{\rd\!t^0} +1\right) \mb{T}^\g = \hbar\d_0{}^\g$,

(3)$\quad
\left(|e_\a| t^\a\Fr{\rd}{\rd\! t^\a} -|e_\g|\right)\mb{T}^\g=0$.

\end{lemma}

\begin{proof} $\hbox{ }$

$(1)$ From \mainzx, the quantum master equation \hea\ implies that
$$
\left(\hbar\rd_\a\rd_\b + A_{\a\b}{}^\r\rd_\r\right) \mb{T}^\g \bos{O}_\g =\bos{K} \mathfrak{X}_{\a\b} 
$$
for some $\mathfrak{X}_{\a\b} \in \sC[[t_H]](\!(\hbar)\!)^{|\a|+|\b|-1}$. 
Consider the above identity modulo $t_H^{n+1}$ and multiply $\hbar^{n-1}$:
\eqn\hoska{
\hbar^{n-1}\left(\hbar\rd_\a\rd_\b + A_{\a\b}{}^\r\rd_\r\right) \mb{T}^\g \bos{O}_\g 
= \hbar^{n-1} \bos{K}\mathfrak{X}_{\a\b} \mod t_H^{n+1}.
}
Then both $\hbar^{n-1}\left(\hbar\rd_\a\rd_\b + A_{\a\b}{}^\r\rd_\r\right) \mb{T}^\g \mod t_H^{n+1}$
and $\hbar^{n-1}\mathfrak{X}_{\a\b}$ have no negative power in $\hbar$. Now consider corollary
\ref{miscb}, which state that "if $\bos{f}(\bos{x})=\bos{K}\bos{\l}$ for some $\bos{x} \in H[[\hbar]]$
and $\bos{\l} \in \sC[[\hbar]]$, then
$\bos{x}=0$ and $\bos{K}\bos{\l}=0$". With our basis $\{e_\a\}$ of $H$, any $\bos{x}\in H[[\hbar]]$ 
can be expressed
as $\bos{x}=\bos{b}^\g e_\g$ for some set $\{\bos{b}^\g\}$, where $\bos{b}^\g \in \Bbbk[[\hbar]]$.
Hence $\bos{f}(\bos{x})= \bos{b}^\g \bos{f}\left(e_\g\right)= \bos{b}^\g \bos{O}_\g$ by our convention.
Consequently the equality $\bos{b}^\g \bos{O}_\g=\bos{K}\bos{\l}$ implies that 
$\bos{b}^\g e_\g=0$, which is equivalent to $b^{(\ell)\g} e_\g=0$ for all $\ell=0,1,2,\ldots$.
It follows that $\bos{b}^\g =0$ for $\forall \g$. The above quoted corollary is also valid
for $\bos{x}\in H[[t_H,\hbar]]$ and $\bos{\l}\in \sC[[t_H,\hbar]]$. Hence, from \hoska, we conclude
that $\hbar^{n-1}\left(\hbar\rd_\a\rd_\b + A_{\a\b}{}^\r\rd_\r\right) \mb{T}^\g =0 \mod t_H^{n+1}$,
which implies that $\left(\hbar\rd_\a\rd_\b + A_{\a\b}{}^\r\rd_\r\right) \mb{T}^\g =0 \mod t_H^{n+1}$.
Then take $n\rightarrow \infty$ limit to conclude that 
$\left(\hbar\rd_\a\rd_\b + A_{\a\b}{}^\r\rd_\r\right) \mb{T}^\g =0$. 

$(2)$ The result follows from the quantum unity $\rd_0 \bos{\Theta}=1$, 
hence $-\hbar \rd_0 e^{-\bos{\Theta}/\hbar} = e^{-\bos{\Theta}/\hbar}$,  and \mainzx\ after adopting
the similar argument as in $(1)$.

$(3)$ The result follows from \mainzx\ by matching the ghost number: $|\mb{T}^\r|= -|e_\g|$ since
$e^{-\bos{\Theta}/\hbar}$
has ghost number $0$ and $|\bos{O}_\g|=|e_\g|$.

\qed
\end{proof}
\label{tivaxu}

Define
$$
\mb{\sG}_\b{}^\g :=\Fr{\rd \mb{T}^\g}{\rd\! t^\b} =\d_\b{}^\g -\Fr{1}{\hbar} t^\a m_{\a\b}{}^\g
+\cdots
$$
and let $\mb{\sG}$ denote the matrix over formal power series ring $\Bbbk[[t_H,\hbar^{-1}]]$
with  $\mb{\sG}_\b{}^\g$ its $\b\g$ entry. 
\begin{lemma} The matrix $\mb{\sG}$ is invertible and satisfies

(1)$\quad
d\!\mb{\sG}^{-1}\wedge d\!\mb{\sG}=0$,

(2)$\quad
\rd_0 \mb{\sG} =-\Fr{1}{\hbar}\mb{\sG}$,

(3)$\quad
 |e_\r| t^\r \rd_\r \mb{\sG}_{\b}{}^\g=\left(|e_\g| -|e_\b|\right)\mb{\sG}_{\b}{}^\g$.

\end{lemma}

\begin{proof} The matrix $\mb{\sG}$ is invertible since its constant part
is the identity matrix. 

$(1)$
From property $(1)$ in lemma \ref{gtopa}, we have 
$\hbar\rd_\a \mb{\sG}_\b{}^\g +A_{\a\b}{}^\r \mb{\sG}_\r{}^\g=0$.
 It follows that 
$$
A_{\a\b}{}^\g =-\hbar\rd_\a \!\mb{\sG}_\b{}^\r \mb{\sG}^{-1}_\r{}^\g
=\hbar \mb{\sG}_\b^{-1\r}  \rd_\a \mb{\sG}_\r{}^\g \in \Bbbk[[t_H]].
$$
Let $A$ denote matrix valued $1$-form which $\b\g$ entry is $A_\b{}^\g =d\!t^\a A_{\a\b}{}^\g$.
 Then
$$
A = \hbar \mb{\sG}^{-1} d\!\mb{\sG},
$$
and the both conditions $dA=A^2=0$ reduce to
$d\!\mb{\sG}^{-1}\wedge d\!\mb{\sG}=0.$

$(2)$
From property $(2)$ in lemma \ref{gtopa}, we have $\hbar\rd_0\mb{\sG}_\b{}^\g+\mb{\sG}_\b{}^\g=0$,
which is equivalent to property $(2)$.

$(3)$
From property $(3)$ in lemma \ref{gtopa}, we have 
$|e_\r|t^\r \mb{\sG}_\r{}^\g -|e_\g|\mb{T}^\g=0$. Applying $\rd_\b$ to the relation we obtain
that $ \left(|e_\s| t^\s \rd_\s -(|e_\g| -|e_\b|)\right)\mb{\sG}_{\b}{}^\g=0$.
\qed
\end{proof}

\begin{corollary}
Let $A_{\a\b}{}^\g :=-\hbar\rd_\a \!\mb{\sG}_\b{}^\r \mb{\sG}^{-1}_\r{}^\g$. Then the $3$-tensor
$A_{\a\b}{}^\g$ is in formal power series of $t_H$  independent to $\hbar$ and satisfies
$$
\eqalign{
A_{\a\b}{}^\g &=(-1)^{|\a||\b|}A_{\b\a}{}^\g 
,\cr
\rd_\a A_{\b\g}{}^\r &=(-1)^{|\a||\b|}\rd_\b A_{\a\g}{}^\r
,\cr
A_{\a\b}{}^\s A_{\s\g}{}^\g &= A_{\b\g}{}^\s A_{\a\s}{}^\r
,\cr
A_{0\b}{}^\g &=\d_\b{}^\g
,\cr
|e_\r| t^\r \rd_\r A_{\a\b}{}^\g&=(|e_\g|-|e_\b|-|e_\a|)A_{\a\b}{}^\g.
}
$$
\end{corollary}

\subsubsection{Free energy as generating function of  morphism of QFT algebra.}

This subsection is about some preliminary understanding of underlying algebraic
structures of QFT as is uncovered by this paper. 
We consider an unital BV QFT, whose partition function is normalizable
to $1$. We shall see that the free energy is another avatar of the
notion of QFT algebra morphism.

Consider a BV QFT with QFT cycle $\bos{c}: \sC\rightarrow
\Bbbk[[\hbar]]$ and suppose that $|\bos{c}|=0$, and the partition function $\bos{c}(\bos{f}(e_0))=\bos{c}(1)=\left<1\right>$ 
is normalizable to $1$. We define free energy $\bos{F}$  as follows:
\eqn\frena{
e^{-\bos{F}/\hbar} = 1 -\Fr{1}{\hbar} \mb{T}^\g \tilde{\left<\bos{O}_\g\right>},
}
where $\tilde{\left<\bos{O}_\a\right>}\in \Bbbk[[\hbar]]$ is normalized expectation value.
Note that $\bos{F}\big|_{t_H=0}=0$. Applying $-\hbar\rd_\g$ to \frena, we have
$$
-\hbar \rd_\g e^{-\bos{F}/\hbar}=\rd_\g\!\bos{F} \cdot e^{-\bos{F}/\hbar}
= \rd_\g \mb{T}^\s  \tilde{\left<\bos{O}_\g\right>}
=\tilde{\left<\bos{\Pi}_\g\right>}_{t_H}.
$$
It follows that 
$$
\bos{\w}_{\a} := \rd_\a \bos{F}\big|_{t_H=0}=\tilde{\left<\bos{O}_\a\right>} \in \Bbbk[[\hbar]].
$$
Applying $\hbar^2\rd_\b\rd_\b$ to \frena\ and using the above, we  also have 
\eqn\frenb{
\eqalign{
(-\hbar)^2 \rd_\b \rd_\g e^{-\bos{F}/\hbar}
&=\left(-\hbar\rd_\b\rd_\g\bos{F} +\rd_\b\! \bos{F} \cdot\rd_\g\! \bos{F}\right) e^{-\bos{F}/\hbar}
\cr
&=A_{\b\g}{}^\s\tilde{\left<\bos{\Pi}_\s\right>}_{t_H}
\cr
&= A_{\b\g}{}^\s \rd_\s \!\bos{F} \cdot e^{-\bos{F}/\hbar}.
}
}
Hence we obtain 
the following system of differential equations:
\eqn\effb{
\eqalign{
\hbar \rd_{\a}\rd_{\b}\bos{F}
&=\rd_{\a}\bos{F}\cdot \rd_{\b}\bos{F}
- A_{\a\b}{}^{\g}\rd_{\g}\bos{F}
,\cr
\rd_0 \bos{F}&=1,
}
}
which should be compared with the quantum master equation with unity:
\eqn\qmede{
\eqalign{
\hbar\rd_\a\rd_\b \bos{\Theta}
&= \rd_\a\bos{\Theta}\cdot \rd_\b\bos{\Theta}- A_{\a\b}{}^\g \rd_\g\bos{\Theta} 
- \bos{K}\bos{\La}_{\a\b} - \left(\bos{\Theta},\bos{\La}_{\a\b}\right)_{\hbar},
\cr
\rd_0 \bos{\Theta}&=1.
}
}

The $\Bbbk[[t_H]]$-algebra on $H\otimes \Bbbk[[t_H]]$ defined by
the set $\{A_{\a\b}{}^\g\}$ in our main theorem can be, using the {\it graded symmetry}, 
identified with a sequences
$\underline{m}=m_2,m_3,\cdots$ of multi-linear  maps 
$m_n: S^n H\longrightarrow H$, of ghost number $0$, defined by
$$
m_n(e_{a_1},\ldots,e_{\a_n}) = m_{\a_1\cdots \a_n}{}^\g e_\g.
$$
for $n=2,3,\ldots$, where $S^n H$ denote the graded symmetric products of $H$
and
$$
A_{\a\b}{}^\g = 
m_{\a\b}{}^\g + \sum_{\ell=1}^\infty \Fr{1}{\ell!}t^{\r_\ell}\cdots t^{\r_1}m_{\r_1\cdots \r_\ell \a\b}{}^\g.
$$
The sequence of $\underline{m}=m_2,m_3,\ldots$ of multi-linear  maps satisfy
the infinite set of relations summarized by  
the {\it relation} $A_{\a\b}{}^\r A_{\r\g}{}^\s = A_{\b\g}{}^\r A_{\a\r}{}^\s$. 
From these data we build a triple
$$
\bigl(H[[\hbar]], 0, \underline{m}\bigr),
$$ 
where  $m_n$ is understood to be extended by $\hbar$-adic continuity to a $\Bbbk[[\hbar]]$-multi-linear map
 $m_n: S^n \left(H[[\hbar]]\right)\longrightarrow H[[\hbar]]$.
We call the above triple a structure of 
{\it on-shell  QFT algebra}\foot{This is just a special class of on-shell QFT algebra 
and general definition of on shell QFT algebra
will be give in the $4$-th paper in this series.} with zero differential on $H$
with trivial quantum descendant algebra:
$$
\bigl(H[[\hbar]], \underline{0}\bigr).
$$

Recall that $\bos{O}_\a$ is the image of $e_\a$ under the quantization map
$$
\bos{f}: H\longrightarrow \sC[[\hbar]],
$$
which is a morphism of QFT complexes $(H[[\hbar]],0)\longrightarrow (\sC[[\hbar]],\bos{K})$, 
i.e., $\bos{K} {\bf{f}}=0$ and ${\bf{f}}(e_\a)=\bos{O}_\a$ and $\bos{K}\bos{O}_\a=0$.
Now, we like to interpret ${\bf{f}}$ as a quasi-isomorphism from the on-shell QFT algebra  
$\left(H[[\hbar]],0, \underline{m}\right)$ to the BV QFT algebra 
$\left(\sC[[\hbar]],\bos{K}, \;\cdot\;\right)$ as QFT algebra up to homotopy.\foot{
We didn't define general QFT algebra and its morphism, so the above
statement 
is not
valid as mathematical notion yet. 
But the whole purpose of this series of papers is to reach to 
the "correct" definition of QFT algebra and its morphism, which contains all relevant physical information.}

For the above purpose, consider the
the {\it quantum master equation} \qmede,
a solution to which $\bos{\Theta}=t^\a\bos{O}_\a + \Fr{1}{2}t^\b t^\a\bos{O}_{\a\b}+\cdots$ shall be
relabeled as follows
\eqn\vkav{
\bos{\Theta} = t^\a {\bf{\phi}}_1(e_\a) + \Fr{1}{2!} t^\b t^\a {\bf{\phi}}_2(e_\a, e_\b) 
+ \sum_{n=3}^{\infty} \Fr{1}{n!} t^{\a_n}\cdots t^{\a_1}{\bf{\phi}}_n(e_{\a_1},\ldots, e_{\a_n}),
}
where $\bos{\phi}_1=\bos{f}$ and ${\bf{\phi}}_n(e_{\a_1},\ldots,e_{\a_n})=\bos{O}_{\a_1\cdots\a_n}$
such that $\bos{\phi}_n =\phi_n + \hbar \phi^{(1)} +\hbar \phi^{(2)}_n+\cdots$ is interpreted as a sequence
of $\Bbbk$-multilinear maps, parametrized by $\hbar$, on $S^n H$ into $\sC$ with ghost number $0$.
We also denote
$\bos{\La}_{\a\b} =\sum_{n=2}^\infty  \Fr{1}{n!} t^{\bar\a_n}\cdots t^{\bar\a_1}\bos{\l}_n(e_{\a_1},\ldots, e_{\a_n})$, 
where 
$t^{\bar\a}:=(-1)^{|a|}t^\a$, such that $\bos{\l}_n =\l_n + \hbar \l^{(1)} +\hbar \l^{(2)}_n+\cdots$ is 
interpreted as a sequence
of $\Bbbk$-multilinear maps, parametrized by $\hbar$, on $S^n H$ into $\sC$ with ghost number $-1$.
Then the equation \qmede\ implies that $\bos{f}$ induces
the sequence $\underline{\bf{\phi}}= {\bf{\phi}}_1,{\bf{\phi}}_2,{\bf{\phi}}_3, \cdots$  of $\Bbbk[[\hbar]]$-multilinear 
as follows:
\eqn\vka{
\eqalign{
\hbar {\bf{\phi}}_2(e_\a, e_\b) = &{\bf{\phi}}_1(e_\a) \cdot{\bf{\phi}}_1(e_\b) -{\bf{\phi}}_1(m_2(e_\a,e_\b))
- \bos{K}\bos{\l}_2(e_\a, e_\b)
,\cr
\hbar {\bf{\phi}}_3(e_\a, e_\b, e_\g) = &{\bf{\phi}}_2(e_\a,e_\b)\cdot {\bf{\phi}}_1(e_\g) 
+(-1)^{|e_\a||e_\b|}{\bf{\phi}}_1(e_\b)\cdot {\bf{\phi}}_2(e_\a, e_\g) 
\cr
&
-{\bf{\phi}}_2(e_\a, m_2(e_\b,e_\g))
-{\bf{\phi}}_1(m_3(e_\a,e_\b,e_\g))
\cr
&
- \bos{K}\bos{\l}_3(e_\a,e_\b,e_\g)
-\left({\bf{\phi}}_1(e_\a), \bos{\l}_2(e_\b, e_\g)\right)_{\hbar}
,\cr
\vdots
} 
}
We note that the sequence  $\underline{\bf{\phi}}= {\bf{\phi}}_1,{\bf{\phi}}_2,{\bf{\phi}}_3, \ldots$ 
is determined by $\bos{f}$, the product $\cdot$ in $\sC$
and $\underline{m}=m_2,m_3,\ldots$ in $H$ up to homotopy. 
In particular $\hbar {\bf{\phi}}_2$ measures the failure
of $\bos{f}=\bos{\phi}_1$ be an algebra map 
$\big(H[[\hbar]],m_2\big)\rightarrow \big(\sC[[\hbar]],\hbox{ }\cdot\hbox{ }\big)$ up to homotopy.
In other words, the first condition for  the morphism ${\bf{f}}$ of QFT complex being a morphism of 
QFT algebra is that the failure of ${\bf{f}}$ from being an algebra map 
$(H[[\hbar]],m_2)\longrightarrow (\sC[[\hbar]],\;\cdot\;)$ up to homotopy must be divisible by $\hbar$. 
Similarly, the quantum master equation governs an elaborate sequence of divisibility conditions 
by $\hbar^n$ of  ${\bf{f}}$ with respect to the product $\cdot$ in $\sC[[\hbar]]$ 
and $\underline{m}=m_2,m_3,\ldots$ in $H[[\hbar]]$ up to homotopy. 

We now recall that the \vkav\ also solves quantum descendant equation:
$$
\bos{K}\bos{\Theta} +\Fr{1}{2}\left(\bos{\Theta},\bos{\Theta}\right)_{\hbar}=0,
$$
as an {\it automatic} consequence of quantum master equation. This means 
that the sequence
$\underline{\bf{\phi}}=
{\bf{\phi}}_1,{\bf{\phi}}_2,{\bf{\phi}}_3, \ldots$ is a distinguished $L_\infty$-morphism 
from the trivial quantum descendant
algebra  $\big(H[[\hbar]],\underline{0}\big)$ 
to  the quantum descendant DGLA $\big(\sC[[\hbar]], \bos{K}, (\hbox{ }, \hbox{ })_{\hbar}\big)$ 
at the chain level, both regarded
as $L_\infty$-algebras.
So we call ${\bf{\phi}}=({\bf{\phi}}_1,{\bf{\phi}}_2,{\bf{\phi}}_3, \ldots)$ quantum descendant morphism,
or descendant morphism of quantum descendant algebra. It should be clear that not every morphism 
from $(H[[\hbar]], \underline{0})$ to  $\big(\sC[[\hbar]], \bos{K}, (\hbox{ },\hbox{ })_{\hbar}\big)$ 
as $L_\infty$-algebra over $\Bbbk[[\hbar]]$ is
 quantum descendant morphism.
 
It also follows that the classical limit 
$\underline{\phi}=\phi_1, \phi_2, \phi_3,\cdots$ of 
$\underline{\bf{\phi}}= {\bf{\phi}}_1,{\bf{\phi}}_2,{\bf{\phi}}_3, \cdots$ 
is a distinguished $L_\infty$-quasi-isomorphism from $H$  to $\sC$.
Thus the moduli space $\CM$ defined by the MC equation of the DGLA $\big(\sC, Q, (\hbox{ },\hbox{ })\big)$ 
is smooth-formal.
It should be clear that not every $L_\infty$ quasi-isomorphism from $H$ to 
$\sC$ is the classical limit of a quantum descendant morphism.
Equivalently, not every versal solution to the MC equation of 
the DGLA $(\sC, Q, (\hbox{ },\hbox{ }))$ is originated from solution to the quantum master equation.
We have called the versal solution obtained from the classical limit of quantum descendant of morphism of QFT
algebra the quantum coordinates on $\CM$.

Now we turn to the similar interpretation of the free energy $\bos{F}$. From \effb,
we deduce that
\eqn\exta{
\hbar \bos{\w}_{\a\b} = \bos{\w}_\a \cdot \bos{\w}_\b -m_{\a\b}{}^\s \bos{\w}_\s,
}
 where
$\bos{\w}_{\a\b} := \rd_\a\rd_\b \mF\big|_{t_H=0}$ and $\cdot$ now denotes the multiplication of the
ground field $\Bbbk$. The above equation is to be compared with the first relation in \vka.
We can interpret $\bos{\w}_\a$ as the image of $e_\a\in H$ be a map $\bos{\w}_1: H\rightarrow \Bbbk[[\hbar]]$,
i.e., $\bos{\w}_1(e_\a)=\bos{\w}_\a$, defined as $\bos{\w}_1=\tilde\bos{\iota} = \tilde\bos{c} \circ\bos{f}$,
where $\tilde\bos{c}$ denote the normalized QFT cycle such that $\tilde\bos{c}(1)=1$.
Then \exta\ can be viewed as the definition of $\bos{\w}_2(e_\a,e_\b)= \bos{\w}_{\a\b}$:
\eqn\exta{
\hbar \bos{\w}_2(e_\a,e_\b) := \bos{\w}_1(e_\a) \cdot \bos{\w}_1(e_\b) -\bos{\w}_1(m_2(e_\a,e_\b)),
}
which measures the failure of $\bos{\w}_1=\tilde\bos{\iota}$ being an algebra map
$(H[[\hbar]] , m_2) \rightarrow (\Bbbk[[\hbar]],\hbox{ }\cdot\hbox{ })$.
In general, we have the following infinite sequence of relations (compare with \vka)
\eqn\freeg{
\eqalign{
\hbar \bos{\w}_2(e_a, e_b) = &\bos{\w}_1(e_\a) \cdot\bos{\w}_1(e_\b) -\bos{\w}_1(m_2(e_\a,e_\b))
,\cr
\hbar \bos{\w}_3(e_\a, e_\b, e_\g) = &\bos{\w}_2(e_\a,e_\b)\cdot \bos{\w}_1(e_\g) 
+(-1)^{|e_\a||e_\b|}\bos{\w}_1(e_\b)\cdot \bos{\w}_2(e_\a, e_\g) 
\cr
&
-\bos{\w}_2(e_\a, m_2(e_\b,e_\g))
-\bos{\w}_1(m_3(e_\a,e_\b,e_\g))
,\cr
\vdots
} 
}
which is formal power series expansion of the differential relations \effb\ at $t_H=0$
such that $\rd_{\a_1}\cdots\rd_{\a_n} \bos{F} \big|_{t_H=0}=\bos{\w}_n(e_{\a_1},\cdots, e_{\a_n})$,
i.e., the formal expansion of $\bos{F}$ at $t_H$ is
$$
\bos{F} = t^\a \bos{\w}_1(e_\a) + \Fr{1}{2!} t^\b t^\a \bos{\w}_2(e_\a, e_\b) 
+ \sum_{n=3}^{\infty} \Fr{1}{n!} t^{\a_n}\cdots t^{\a_1}\bos{\w}_n(e_{\a_1},\ldots, e_{\a_n}).
$$
Note that the resulting recursive relations for  the sequence 
$\underline{\bos{\w}}=\bos{\w}_1, \bos{\w}_2,\ldots$  
are to be regarded as definitions of $\bos{\w}_n$ by $\tilde\bos{\iota}=\bos{\w}_1$, 
and $\underline{m}=m_2,m_3,\ldots$ in $H$ and the ordinary  multiplication
of the ground field $\Bbbk$. 

Now we can interpret the  map $\tilde\bos{\iota}: H[[\hbar]]\rightarrow \Bbbk[[\hbar]]$
as a morphism of QFT algebra from the QFT algebra
$\big(H[[\hbar]], 0, \underline{m}\big)$ with zero differential
 to the trivial QFT algebra $(\Bbbk[[\hbar]], 0, \cdot)$ with zero differential.
Both the QFT algebras have trivial quantum descendant algebra and the
sequence  $\underline{\bos{\w}}=\bos{\w}_1, \bos{\w}_2,\ldots$ is the quantum
descendant morphism - a distinguished morphism on $H[[\hbar]]$ into $\Bbbk[[\hbar]]$
as $L_{\infty}$-algebras over $\Bbbk[[\hbar]]$.  

The moral of the  above story is that
we have a confirmation of our metaphor that quantum field theory is a study of morphisms of
QFT algebras, which contains rather complete information on quantum correlation functions.

\newsec{QFT Integral and Quantum WDVV Equation}

In this section we consider a class of BV QFT with an additional datum
called a QFT integral. A QFT integral and QFT cycle have the same classical property
and  could be viewed as two different ways of quantization of classical cycle.
A QFT integral is responsible for formal Frobenius  manifold structure on moduli space
in semi-classical case and gives certain quantum  version of it in general.

\subsection{QFT Integral}

We begin with defining QFT integral and discuss motivations behind the notion
afterward.

\begin{definition}\label{qinta}
A QFT integral $\oint$ of dimension $N$ in BV QFT algebra is a sequence of 
$\Bbbk$-linear maps on $\sC$ into $\Bbbk$, parametrized by $\hbar$,
$$
\oint=\int +\hbar\int^{(1)} +\hbar^2\int^{(2)} +\cdots: \sC\rightarrow \Bbbk[[\hbar]],
$$ 
of ghost number $-N$ satisfying
$$
\oint \bos{K}{a}\cdot {b}=-(-1)^{|\bos{a}|}\oint {a}\cdot \bos{K}{b},
$$
for $\forall a, b \in \sC$.

\end{definition}

\begin{remark}
Note that  QFT integral $\oint$ also define a $\Bbbk[[\hbar]]$-linear map 
on $\sC[[\hbar]]$ into $\Bbbk[[\hbar]]$
such that 
$
\oint \bos{K}\bos{a}\cdot \bos{b}
=-(-1)^{|\bos{a}|}\oint \bos{a}\cdot \bos{K}\bos{b}$
for $\forall \bos{a},\bos{b} \in \sC[[\hbar]]$.
\end{remark}

\begin{corollary}\label{qintc}
A sequence $\oint$ of  $\Bbbk$-linear maps on $\sC$ into $\Bbbk$ parametrized by $\hbar$
is a QFT integral if and only if, for $\forall {a},{b} \in \sC$,
$$
\eqalign{
\oint \bos{K} a &=0,
\cr
\oint (a,b)_{\hbar} &=0,
}
$$
or equivalently $\oint \bos{K} \bos{a} =\oint (\bos{a},\bos{b}\big)_{\hbar} =0$ 
for $\forall \bos{a},\bos{b} \in \sC[[\hbar]]$.
\end{corollary}

\begin{proof}
Set ${b}=1$ in  definition \ref{qinta} and use the property that $1$ is a center of the bracket
to see that $\oint \bos{K} {a} =0$ for $\forall {a} \in \sC$.
Applying $\oint$ to 
the identity
\eqn\dddxx{
\bos{K}\left({a}\cdot{b}\right)=
\bos{K}{a}\cdot{b}+(-1)^{|\bos{a}|}{a}\cdot\bos{K}{b}
-\hbar (-1)^{|\bos{a}|}\big({a},{b}\big)_{\hbar},
}
we have
$$
\eqalign{
0=\oint \bos{K}{a}\cdot{b}+(-1)^{|{a}|}\int {a}\cdot\bos{K}{b}
&=\oint \bos{K}\left({a}\cdot{b}\right)
+\hbar (-1)^{|{a}|}\oint \big({a},{b}\big)_{\hbar}.
}
$$
It follows that $\oint ({a},{b}\big)_{\hbar} =0$ for 
$\forall {a},{b} \in \sC$ 
since $\oint \bos{K}(a\cdot b) =0$.
Conversely assume that $\oint \bos{K} a=\oint (a,b)_{\hbar} =0$ for $\forall \a\b$. 
Then the identity \dddxx\ implies
that $\oint \bos{K}{a}\cdot{b}+(-1)^{|{a}|}\int {a}\cdot\bos{K}{b}=0$.
\qed
\end{proof}

The above  corollary implies that a QFT integral is automatically a QFT cycle, i.e., $\oint \bos{K}=0$.
But the converse is not true in general and QFT integral has the additional
property that $\oint (a,b)_{\hbar} =0$.
A BV QFT algebra with a QFT integral also defines a BV QFT, since a QFT integral is a QFT cycle.
In general we tend to regard a QFT integral an additional structure to a BV QFT.

Now consider
the leading two relations for a QFT integral: 
\eqn\intnc{
\eqalign{
\int Q{a}\cdot {b}+(-1)^{|{a}|}\int {a}\cdot Q{b}&=0
.\cr
\int K^{(1)}{a}\cdot {b}
+(-1)^{|{a}|}\int {a}\cdot K^{(1)}{b}
&=
-\int^{(1)} Q{a}\cdot {b} 
-(-1)^{|{a}|}\int^{(1)}  {a}\cdot Q{b}
.\cr
}
}
We recall that the classical limit of BV QFT algebra $\big(\sC[[\hbar]],\bos{K},\hbox{ }\cdot\hbox{ }\big)$
is a CDGA $\big(\sC,Q,\hbox{ }\cdot\hbox{ }\big)$. Then the $1$st condition \intnc\ means that
the classical limit $\int$ of $\oint$ is a cycle of the CDGA $\big(\sC,Q,\hbox{ }\cdot\hbox{ }\big)$, i.e., 
$\int\sC\rightarrow \Bbbk$ and $\int Q =0$. 
The condition $\int Q=0$ is equivalent the  first condition in \intnc,
since $Q(a\cdot b)= Qa\cdot b +(-1)^{|a|}a\cdot Q b$.
The classical limit $\mc^{(0)}$ of a QFT cycle $\mb{\mc}=\mc^{(0)}+\hbar\mc^{(1)}+\cdots$
also has the same property, while $\mb{\mc}\big(({a},{b})_{\hbar}\big) \neq 0$ in general.
We may view  QFT cycle and QFT integral as two different ways of quantizing cycle of CDGA.

The classical limit $\int$ of a QFT integral $\oint$ induces a unique graded symmetric
$\Bbbk$-bilinear pairing  
$
\left<\hbox{ },\hbox{ }\right>: H^i \otimes H^{N-i}\rightarrow \Bbbk
$
on the cohomology $H$ of
the cochain complex $\big(\sC, Q\big)$ defined by, for $x,y \in H$,
$$
\left<x, y\right>:=\int f(x)\cdot f(y).
$$
The paring is a homotopy invariant since $Q$ is a derivation of the product $\cdot$.
We recall that there is
a unique graded commutative and associative product $m_2:H\otimes H\rightarrow H$
on $H$. Then

\begin{lemma}
The triple $\big(H, m_2, \left<\hbox{ },\hbox{ }\right>\big)$ is a graded commutative Frobenius algebra,
i.e., $m_2$ is a graded commutative and associative product and 
$$
\left<x,y\right>=(-1)^{|x||y|}\left<y,x\right>,\qquad
\big<x,m_2\big(y,z\big)\big> =\big<m_2\big(x,y\big),z\big>
$$

\end{lemma}

\begin{proof}
The graded commutativity of $\big<\hbox{ },\hbox{ }\big>$ follows from the graded commutativity
of the product $\cdot$ on $\sC$.
Recall that that a $\Bbbk$-linear way of choosing representative of every element in $H$
is a cochain map $f:(H,0)\rightarrow (\sC,Q)$, which induces the identity map on $H$
and is a morphism of CDGA  $f:\big(H, 0, m_2\big)\rightarrow \big(\sC, Q,\hbox{ }\cdot\hbox{ }\big)$
up to homotopy: for $\forall x,y \in H$,
$f(x)\cdot f(y) - f\big(m_2(x,y)\big)=Q\l_2(x,y)$
such that $f(x) \in \Ker Q\cap \sC$ and $[f(x)]=x$.
Then, by definition,
$$
\eqalign{
\big<x,m_2\big(y,z\big)\big>
&:= \int f(x)\cdot f\big(m_2(y,z)\big)
\cr
&= \int f(x)\cdot\Big( f(y)\cdot f(z) - Q\l_2(y,z)\Big)
\cr
&=\int f(x)\cdot\big( f(y)\cdot f(z) \big),
}
$$
where we have used $\int \left(\Ker Q\right)\cdot \hbox{Im }Q=0$ for the last equality.
By the similar manipulation we have
$$
\eqalign{
\big<m_2\big(x,y\big),z\big>
&:= \int f\big(m_2(x,y)\big)\cdot f\big(z\big)
\cr
&=\int \big(f(x)\cdot f(y)\big)\cdot f(z).
}
$$
Hence
$$
\eqalign{
\big<x,m_2\big(y,z\big)\big>
-\big<m_2\big(x,y\big),z\big>
&=\int\Big( f(x)\cdot\big( f(y)\cdot f(z) \big)-\big(f(x)\cdot f(y)\big)\cdot f(z)\Big)
\cr
&=0,
}
$$
as is claimed.
\qed
\end{proof}

\begin{remark}
The relation $\Big<x,m_2\big(y,z\big)\Big>=
\Big<m_2\big(x,y\big),z\Big>$
together with the commutativity of $m_2$ is equivalent to
the following relations
\begin{itemize}
\item left-invariance:
$$
\Big<x,m_2\big(y,z\big)\Big>=(-1)^{|x||y|}\Big<y,m_2\big(x,z\big)\Big>,
$$
\item right-invariance:
$$
\Big<m_2\big(x,y\big),z\Big>=(-1)^{|y||z|}\Big<m_2\big(x,z\big),y\Big>,
$$
\end{itemize}
as well as to 
\begin{itemize}
\item left-cyclic:
$$
\Big<x,m_2\big(y,z\big)\Big>=(-1)^{|x|(|y|+|z|)}\Big<y,m_2\big(z,x\big)\Big>,
$$
\item right-cyclic:
$$
\Big<m_2\big(x,y\big),z\Big>=(-1)^{(|x|+|y|)|z|}\Big<m_2\big(z,x\big),y\Big>,
$$
\end{itemize}

\end{remark}

Now we consider the quantum extension map 
$\bos{f}=f+\hbar f^{(1)} +\cdots :H\rightarrow \sC[[\hbar]]$
such that $\bos{K}\bos{f} =\bos{f}\bos{\k}$. We may want to quantize the 
$\Bbbk$-bilinear pairing $
\left<\hbox{ },\hbox{ }\right>: H^i \otimes H^{N-i}\rightarrow \Bbbk$
to certain well-defined sequence $\left<\hbox{ },\hbox{ }\right>_{\hbar}
=\left<\hbox{ },\hbox{ }\right> 
+\hbar \left<\hbox{ },\hbox{ }\right>^{(1)} 
+\hbar^2 \left<\hbox{ },\hbox{ }\right>^{(1)}\cdots$ of  $\Bbbk$-bilinear parings parametrized by $\hbar$.
It is natural to try
$$
\left<x, y\right>_{\hbar} ?= \int \bos{f}(x)\cdot \bos{f}(y),
$$
which does not works, in general, since it depends on quantum homotopy.
The first obstruction is $\bos{\k}$, which is assumed to vanish in this paper.
Hence $\bos{K}\bos{f}=0$ and an arbitrary quantum extension map $\bos{f}^\pr$
homotopic to $\bos{f}$ is given by $\bos{f}^\pr =\bos{f} +\bos{K}\bos{s}$. 
Then
$$
\eqalign{
 \bos{f}^\pr(x)\cdot \bos{f}^\pr(y) 
 -\bos{f}(x)\cdot \bos{f}(y)
 = &\bos{f}(x)\cdot \bos{K}\bos{s}(y)  +\bos{K}\bos{s}(x)\cdot \bos{f}(y) +\bos{K}\bos{s}(x)\cdot\bos{K} \bos{s}(y)
 \cr
 =
 &
 (-1)^{|x|}\bos{K}\left(\bos{f}(x)\cdot \bos{s}(y)\right)
  +\bos{K}\left(\bos{s}(x)\cdot \bos{f}^\pr(y)\right) 
 \cr
 &
 +\hbar \left(\bos{f}(x), \bos{s}(y)\right)_{\hbar}
 +\hbar(-1)^{|x|-1}\left(\bos{s}(x), \bos{f}^\pr(y)\right)_{\hbar}. 
 }
$$
The above computation implies that  the integral $\int$ must be quantized to a QFT integral
$\oint =\int+\hbar \int^{(1)} +\cdots$ such that $\int \bos{f}(x)\cdot \bos{f}(y)$ is quantum homotopy
invariant.

\begin{remark}
In some case $\int$, satisfying 
$\int Q a\cdot b =-(-1)^{|a|}\int a\cdot Qb$,
may be itself be a QFT integral. We call such a QFT integral $\int\equiv \oint$ 
semi-classical and note that it has to satisfy
$$
\int K^{(n)} a\cdot b =-(-1)^{|a|}\int a\cdot K^{(n)}b,
$$
for all $n=1,2,3,\cdots$.
By the way the confusing expression that "something is semi-classical" really means
that ``something classical which is also quantum by itself''.

\end{remark}

\begin{definition} (\hbox{\bf Lemma})
Let $\oint$ be a QFT integral of an anomaly-free BV QFT algebra, that is $\bos{\k}=0$ on $H$.
Then there is a sequence of 
$\left<\hbox{ },\hbox{ }\right>_{\hbar}
=\left<\hbox{ },\hbox{ }\right> +\hbar \left<\hbox{ },\hbox{ }\right>^{(1)} 
+\hbar^2 \left<\hbox{ },\hbox{ }\right>^{(1)}\cdots$ 
of  $\Bbbk$-bilinear parings parametrized by $\hbar$,
$
\left<\hbox{ }, \hbox{ }\right>_{\hbar}:H\otimes H \rightarrow \Bbbk[[\hbar]]
$,
defined by
$$
\left<x, y\right>_{\hbar} = \oint \bos{f}(x)\cdot \bos{f}(y),
$$
which is a quantum homotopy invariant.
\end{definition}

\begin{definition}

A QFT integral $\oint$ is called non-degenerate if the 
$\Bbbk$-bilinear pairing  on the cohomology $H$ defined by
its classical limit $\int$ is non-degenerate.

\end{definition}

\begin{proposition}
$$
\eqalign{
\big<x,m_2(y,z)\big>_{\hbar} -\big<m_2(x,y),z\big>_{\hbar} 
&=
-\hbar\oint \bos{\phi}_1(x)\cdot\bos{\phi}_2(y,z)
-\bos{\phi}_2(x,y)\cdot\bos{\phi}_1(z)
,\cr
\big<x,m_2\big(y,z\big)\big>_{\hbar}-(-1)^{|x||y|}\Big<y,m_2\big(x,z\big)\big>_{\hbar}
&=
-\hbar\oint \bos{\phi}_1(x)\cdot\bos{\phi}_2(y,z)
-(-1)^{|x||y|}\bos{\phi}_1(y)\cdot\bos{\phi}_2(x,z)
,\cr
\big<m_2\big(x,y\big),z\big>_{\hbar}-(-1)^{|y||z|}\big<m_2\big(x,z\big),y\big>_{\hbar}
&=-\hbar \oint \bos{\phi}_2(x,y)\cdot \bos{\phi}_1(z) -(-1)^{|y||z|}\bos{\phi}_2(x,z)\cdot \bos{\phi}_1(y).
}
$$

\end{proposition}

\begin{proof}

Consider the relation
$$
\hbar\bos{\phi}_2(x,y)=\bos{f}(x)\cdot \bos{f}(y) - \bos{f}\big( m_2(x,y) \big)-\bos{K}\bos{\l}_2(x,y).
$$
Then, by definition, we have
$$
\eqalign{
\Big<x,m_2\big(y,z\big)\Big>_{\hbar}
&:= \oint\bos{f}(x)\cdot \bos{f}\big(m_2(y,z)\big)
\cr
&= \oint \bos{f}(x)\cdot\Big( \bos{f}(y)\cdot \bos{f}(z) - \bos{K}\bos{\l}_2(y,z)-\hbar\bos{\phi}_2(y,z)\Big)
\cr
&=\oint \bos{f}(x)\cdot\big( \bos{f}(y)\cdot \bos{f}(z) \big)
-\hbar\oint \bos{\phi}_1(x)\cdot\bos{\phi}_2(y,z)
}
$$
where we have used $\oint \Ker \bos{K}\cdot \hbox{Im }\bos{K}=0$ for the last equality.
After the similar manipulation we also obtain that
$$
\eqalign{
\big< m_2\big( x,y\big),z\big>_{\hbar}
&:= \oint\bos{f} \big( m_2(x,y)\big) \cdot \bos{f} (z)
\cr
&=\oint\big( \bos{f}(x)\cdot \bos{f}(y)\big) \cdot \bos{f} (z)
-\hbar \oint \bos{\phi}_2(x,y)\cdot \bos{\phi}_1(z).
}
$$
Now the associativity of the product $\cdot$ in $\sC$ implies that 
$$
\eqalign{
\Big<x,m_2\big(y,z\big)\Big>_{\hbar}
-\Big<m_2\big(x,y\big),z\Big>_{\hbar}
&=-\hbar\oint\Big( \bos{\phi}_1(x)\cdot\bos{\phi}_2(y,z)
-\bos{\phi}_2(x,y)\cdot\bos{\phi}_1(z)\Big).
}
$$
The remaining two relations can be proved similarly.
\qed. 
\end{proof}

\begin{definition} (Lemma)
Let $\oint$ be a QFT integral of an anomaly-free BV QFT algebra, that is $\bos{\k}=0$ on $H$.
Then there is a sequence 
$\left<\ast,\ast,\ast\right>_{\hbar}
=\left<\ast,\ast,\ast\right> +\hbar \left<\ast,\ast,\ast\right>^{(1)}
+\hbar^2 \left<\ast,\ast,\ast\right>^{(2)}\cdots$  of  $\;\Bbbk$-trilinear parings parametrized by $\hbar$:
$$
\left<\ast,\ast,\ast\right>_{\hbar} :H\otimes H\otimes H \rightarrow \Bbbk[[\hbar]],
$$
which is a quantum homotopy invariant,
where
$$
\eqalign{
\left<x, y,z\right>_{\hbar} 
&:= \oint \Big(\bos{\phi}_2(x,y)\cdot\bos{\phi}_1(z) +(-1)^{|x||y|} \bos{\phi}_1(y)\cdot\bos{\phi}_2(x,z)\Big).
}
$$
such that

(1) $\big<x,y,z\big>_{\hbar} = (-1)^{|y||z|}\big<x,z,y\big>_{\hbar}$,

(2) $\big<e,y,z\big>_{\hbar} =0$,

(3) $
\Big<x,m_2(y,z)\Big>_{\hbar} -\Big<m_2(x,y),z\Big>_{\hbar} 
=\hbar\Big(\big<x,y,z\big>_{\hbar} -(-1)^{|x||z|+|y||z|}\big<z,x,y\big>_{\hbar}\Big)
$

(4)
$
\Big<x,m_2(y,z)\Big>_{\hbar} -(-1)^{|x||y|}\Big<y,m_2(x,z)\Big>_{\hbar} 
=\hbar\Big(\big<x,y,z\big>_{\hbar} -(-1)^{|x||y|}\big<y,x,z\big>_{\hbar}\Big) 
$

(5)
$
\Big<m_2(x,y),z)\Big>_{\hbar} -(-1)^{|y||z|}\Big<m_2(x,z),y\Big>_{\hbar} 
=\hbar\Big((-1)^{|x||z|+|y||z|}\big<z,x,y\big>_{\hbar} -(-1)^{|x||y|}\big<y,x,z\big>_{\hbar}\Big) 
$

\end{definition}

\begin{proof}
Exercise.
\end{proof}

We may go on to define higher pairings and examine their properties. 
Instead we look for generating formula using the solution 
$\bos{\Theta}$ of quantum master equation in the case that $H$ 
is finite dimensional for each ghost number. Let $\{e_\a\}$ be
the basis of $H$ introduced in the beginning of section $3$. Then
$$
\eqalign{
\big<e_\a, e_\b\big>_{\hbar} &=\oint \bos{O}_\a \cdot \bos{O}_\b
,\cr
\big<e_\a, e_\b, e_\g\big>_{\hbar} 
&=\oint\Big( \bos{O}_{\a\b} \cdot \bos{O}_\g +(-1)^{|\a||\b|}\bos{O}_\b\cdot \bos{O}_{\a\g}\Big)
.
}
$$
Define
$$
\eqalign{
\bos{g}_{\b\g} :=
& \oint \rd_\b\bos{\Theta}\cdot \rd_\g\bos{\Theta}
\cr
=& \oint \bos{O}_\a \cdot \bos{O}_\b
+t^\a\oint\Big( \bos{O}_{\a\b} \cdot \bos{O}_\g +(-1)^{|\a||\b|}\bos{O}_\b\cdot \bos{O}_{\a\g}\Big)
+\cdots
}
$$
such that
$$
\bos{g}_{\b\g} = \big<e_\b, e_\g\big>_{\hbar} +t^\a \big<e_\a, e_\b, e_\g\big>_{\hbar} +\cdots.
$$
We then define $n$-nary pairing on $H^{\otimes n}$ into $\Bbbk[[\hbar]]$ by
$$
\big< e_{\a_1}, e_{\a_2}, \cdots, e_{\a_{n-2}}, e_{\b}, e_\g \big>_{\hbar}
:= \rd_{\a_1}\rd_{\a_2}\cdots \rd_{\a_{n-2}} \bos{g}_{\b\g} \big|_{t_H=0}.
$$

\begin{proposition}\label{boma}
Let
$\bos{g}_{\b\g} := \oint \rd_\b\bos{\Theta}\cdot \rd_\g\bos{\Theta} \in \Bbbk[[t_H]][[\hbar]]$.
Then, the following relations are satisfied:

(1) $\bos{g}_{\b\g} =(-1)^{|\b||\g|}\bos{g}_{\g\b}$,

(2) $\rd_0\bos{g}_{\b\g} =0$,

(3) $\hbar \left(\rd_\a\bos{g}_{\b\g} -(-1)^{|\a||\b|}\rd_\b \bos{g}_{\a\g}\right)
= A_{\b\g}{}^\r \bos{g}_{\a\r} -(-1)^{|\a||\b|}  A_{\a\g}{}^\r\bos{g}_{\b\r}$,

(4) $\rd_\a\bos{g}_{\b 0} -(-1)^{|\a||\b|}\rd_\b \bos{g}_{\a 0}=0$.

\end{proposition}

\begin{proof}
The $1$st property is obvious. The $2$nd property follows from the quantum unity that
$\rd_0\bos{\Theta}=1$, which implies that $\rd_0 \rd_\a\bos{\Theta}=0$.
For the $3$rd property, use
the quantum master equation
$
\rd_\b\bos{\Theta}\cdot \rd_\g\bos{\Theta}=\hbar\rd_\b\rd_\g \bos{\Theta} 
+A_{\b\g}{}^\r\rd_\r\bos{\Theta}
+\bos{K}_{\!\bos{\Theta}}\bos{\La}_{\b\g}
$, which implies that
$$
\eqalign{
\bos{g}_{\b\g} 
=& \hbar\oint \rd_\b\rd_\g \bos{\Theta} + A_{\b\g}{}^\r\oint\rd_\r\bos{\Theta}.
+\oint\bos{K}\bos{\La}_{\b\g}
+\oint\big(\bos{\Theta},\bos{\La}_{\b\g}\big)_{\hbar}
\cr
=& \hbar\oint \rd_\b\rd_\g \bos{\Theta} + A_{\b\g}{}^\r\oint\rd_\r\bos{\Theta}.
}
$$
Applying $\hbar\rd_\a$ to the above we have
$$
\eqalign{
\hbar\rd_\a\bos{g}_{\b\g} = 
&
\hbar^2\oint \rd_\a\rd_\b\rd_\g \bos{\Theta} 
+ \hbar\rd_\a A_{\b\g}{}^\r\oint\rd_\r\bos{\Theta}
+\hbar A_{\b\g}{}^\r\oint\rd_{\a}\rd_\r\bos{\Theta}
\cr
=
&\hbar^2\oint \rd_\a\rd_\b\rd_\g \bos{\Theta} 
+ \hbar\rd_\a A_{\b\g}{}^\r\oint\rd_\r\bos{\Theta}
- A_{\b\g}{}^\r A_{\a\r}{}^\s\oint\rd_{\s}\bos{\Theta}
\cr
&
+ A_{\b\g}{}^\r\bos{g}_{\a\r}.
}
$$
Then the $3$rd proposition follows from the potentiality and  the  associativity of $A_{\a\b}{}^\g$.
For the $4$th property, set $\g=0$ to relation $(3)$ and use $A_{\b0}{}^\r= \d_{\b}{}^\g$.
\qed
\end{proof}

%

\subsection{WDVV equation as a semi-classical phenomena}

We call a solution $\bos{\Theta}$ to the quantum master equation semi-classical
if $\bos{\Theta}$ does not depend on $\hbar$, i.e., 
$\bos{\Theta}=\Theta$. Then the quantum master equation is decomposed
into the following form
\eqn\scqme{
\eqalign{
\Theta_\a\cdot \Theta_\b &=A_{\a\b}{}^{\g}\Theta_{\g} 
+ Q\La^{(0)}_{\a\b}
+\left(\Theta,\La^{(0)}_{\a\b}\right)^{(0)},
\cr
\Theta_{\a\b}&=-K^{(1)} \La^{(0)}_{\a\b} -Q \La^{(1)}_{\a\b}-\left(\Theta, \La^{(1)}_{\a\b}\right)^{(0)}
-\left(\Theta,\La^{(0)}_{\a\b}\right)^{(1)},
\cr
 0&= -\sum_{j=1}^n K^{(j)}\La^{(n-j)}_{\a\b} 
 - Q \La^{(n)}_{\a\b} 
 -\sum_{j=0}^n \left(\Theta, \La^{(j)}_{\a\b}\right)^{(n-j)}
  \quad \hbox{for } n\geq 2, 
}
}
where $\Theta_\a :=\rd_\a\Theta$ and $\Theta_{\a\b}:=\rd_\a\rd_\b \Theta$.
The quantum descendant equation is, then, decomposed as
as follows
\eqn\scqde{
\eqalign{
Q\Theta +\Fr{1}{2}\left(\Theta,\Theta)^{(0)}\right) &=0,\cr
K^{(1)}\Theta +\Fr{1}{2}\left(\Theta,\Theta\right)^{(1)} &=0,\cr
K^{(n)}\Theta +\Fr{1}{2}\left(\Theta,\Theta\right)^{(n)} &=0 \quad\hbox{for }n\geq 2.\cr
}
}
We say a BV QFT algebra 
$\big(\sC[[\hbar]], \bos{K}=Q +\hbar K^{(1)} +\hbar^2 K^{(2)} +\cdots, \hbox{ }\cdot\hbox{ }\big)$
is semi-classical if it admit a semi-classical solution to the quantum master equation.
We say a QFT integral $\oint$ semi-classical if $\oint =\int$.

\begin{lemma}
For a semi-classical BV QFT algebra with a semi-classical QFT integral $\oint=\int$,
the metric $\bos{g}_{\a\b}=\oint \Theta_\a \cdot \Theta_\b$ satisfies the following
properties

1. it does not depend on $\hbar$: $\bos{g}_{\a\b}=g_{\a\b}$,

2. it is flat: $\rd_\a g_{\b\g}=0$

3. it is compatible with the $3$-tensor $A_{\a\b}{}^\g$: $A_{\a\b}{}^\r g_{\r\g}= A_{\b\g}{}^\r g_{\a\r}$.

\end{lemma}

\begin{proof}\label{frost}
Note that $\oint Qa\cdot b =-(-1)^{|a|}\oint a\cdot Q b$ and 
$\oint K^{(n)}a\cdot b =-(-1)^|a|\oint a\cdot K^{(n)} b$
for all $a,b \in \sC$ and for all $n\geq 1$, since $\oint$ is semiclassical.
It  also that $\oint (\hbox{ },\hbox{ })^{(\ell)}=0$ for all $\ell=0,1,2,\cdots$ due
to corollary \ref{qintc} in the semi-classical case.

1. Property $1$ is obvious since both $\oint$ and $\Theta_\a$ do not depend on $\hbar$.

2. By definition, we have
$$
\rd_\a g_{\b\g}= \oint \Theta_{\a\b}\cdot \Theta_\g +(-1)^{|\a||\b|}\oint \Theta_\a\cdot \Theta_{\b\g}.
$$
Consider $\oint \Theta_{\a\b}\cdot \Theta_\g$ and substitute $\Theta_{\a\b}$ using the quantum master
equation \scqme\ to have
$$
\eqalign{
\oint \Theta_{\a\b}\cdot \Theta_\g
=
&-\oint\left( K^{(1)} \La^{(0)}_{\a\b}\cdot \Theta_\g 
+\left(\Theta,\La^{(0)}_{\a\b}\right)^{(1)}\cdot \Theta_\g\right)
\cr
&-\oint \left( Q \La^{(1)}_{\a\b}\cdot \Theta_\g
+\left(\Theta, \La^{(1)}_{\a\b}\right)^{(0)}\cdot \Theta_\g\right).
}
$$
Consider two terms in the RHS above separately;

-The $1$st term: We have 
$$
\eqalign{
\oint K^{(1)} \La^{(0)}_{\a\b}\cdot \Theta_\g 
= (-1)^{|\a|+|\b|}\oint\La^{(0)}_{\a\b}\cdot  K^{(1)} \Theta_\g
= -(-1)^{|\a|+|\b|}\oint\La^{(0)}_{\a\b}\cdot \left(\Theta, \Theta_\g\right)^{(1)}
,\cr
}
$$
where we have used $\oint K^{(1)}a\cdot b =-(-1)^{|a|}\oint a\cdot K^{(1)} b$ for the $1$st equality
(note that $|\La^{(\ell)}_{\a\b}|= |e_\a|+|e_\b|-1$ for all $\ell=0,1,2,\cdots$) 
and $K^{(1)}\Theta_\g +\big(\Theta,\Theta_\g)^{(1)}=0$, 
which is a consequence of the quantum descendant equation \scqde,
for the $2$nd equality. Hence, we have
$$
\eqalign{
\oint\left( K^{(1)} \La^{(0)}_{\a\b}\cdot \Theta_\g 
+\left(\Theta,\La^{(0)}_{\a\b}\right)^{(1)}\cdot \Theta_\g\right)
&=\oint\left( -(-1)^{|\a|+|\b|}\La^{(0)}_{\a\b}\cdot \left(\Theta, \Theta_\g\right)^{(1)}
+\left(\Theta,\La^{(0)}_{\a\b}\right)^{(1)}\cdot \Theta_\g\right)
\cr
&=\oint 
\left(\Theta,\La^{(0)}_{\a\b}\cdot \Theta_\g\right)^{(1)}=0,
}
$$
where we have used the Poisson-law \poisson\ for the $2$nd equality and
the property that $\oint \left(\hbox{ }, \hbox{ }\right)^{(\ell)}=0$ for the last equality.

-The $2$nd term:
After the similar manipulations using $\oint Qa\cdot b =-(-1)^{|a|}\oint a\cdot Q b$ 
and $Q\Theta_\g+\big(\Theta,\Theta_\g)^{(0)}=0$, we have
$$
\eqalign{
\oint  Q \La^{(1)}_{\a\b}\cdot \Theta_\g
= (-1)^{|\a|+|\b|}\oint\La^{(1)}_{\a\b}\cdot  Q \Theta_\g
= -(-1)^{|\a|+|\b|}\oint\La^{(1)}_{\a\b}\cdot \left(\Theta, \Theta_\g\right)^{(0)}
,\cr
}
$$
such that
$$
\eqalign{
\oint\left(Q \La^{(1)}_{\a\b}\cdot \Theta_\g 
+\left(\Theta,\La^{(1)}_{\a\b}\right)^{(0)}\cdot \Theta_\g\right)
&=\oint\left( -(-1)^{|\a|+|\b|}\La^{(1)}_{\a\b}\cdot \left(\Theta, \Theta_\g\right)^{(0)}
+\left(\Theta,\La^{(1)}_{\a\b}\right)^{(0)}\cdot \Theta_\g\right)
\cr
&=\oint 
\left(\Theta,\La^{(1)}_{\a\b}\cdot \Theta_\g\right)^{(0)}
=0.
}
$$
where we have used the Poisson-law \poisson\ for the $2$nd equality and
the property that $\oint \left(\hbox{ }, \hbox{ }\right)^{(\ell)}=0$ for the last equality.

Thus $\oint \Theta_{\a\b}\cdot \Theta_\g =0$, which also implies that 
$\oint \Theta_\a\cdot \Theta_{\b\g}=0$. Consequently we have $\rd_\a g_{\b\g}=0$.

(3) From the associativity of the product $\cdot$ we have
\eqn\triaso{
\oint \left(\Theta_\a\cdot \Theta_\b\right)\cdot\Theta_\g
=\oint \Theta_\a\cdot\left( \Theta_\b\cdot\Theta_\g\right).
}
The LHS of the above, after using the quantum master equation \scqme, becomes
$$
\oint \left(\Theta_\a\cdot \Theta_\b\right)\cdot\Theta_\g
= \oint A_{\a\b}{}^\g \Theta_\r\cdot\Theta_\g 
+ \oint \left( Q\La_{\a\b}^{(0)}\cdot \Theta_\g 
+\left(\Theta, \La_{\a\b}^{(0)}\right)^{(0)}\cdot \Theta_\g\right).
$$
Then the first term gives
$$
\oint A_{\a\b}{}^\g \Theta_\r\cdot\Theta_\g 
= A_{\a\b}{}^\g\oint\Theta_\r\cdot\Theta_\g=A_{\a\b}{}^\g g_{\r\g},
$$
while the second term vanishes, since
$$
\eqalign{
\oint \left( Q\La_{\a\b}^{(0)}\cdot \Theta_\g 
+\left(\Theta, \La_{\a\b}^{(0)}\right)^{(0)}\cdot \Theta_\g\right)
&=\oint \left( (-1)^{|\a|+|\b|}\La_{\a\b}^{(0)}\cdot Q\Theta_\g 
+\left(\Theta, \La_{\a\b}^{(0)}\right)^{(0)}\cdot \Theta_\g\right)
\cr
&=\oint \left( -(-1)^{|\a|+|\b|}\La_{\a\b}^{(0)}\cdot \left(\Theta,\Theta_\g\right)^{(0)} 
+\left(\Theta, \La_{\a\b}^{(0)}\right)^{(0)}\cdot \Theta_\g\right)
\cr
&=\oint \left(\Theta, \La_{\a\b}^{(0)}\cdot \Theta_\g\right)^{(0)}=0.
}
$$
Hence $\oint \left(\Theta_\a\cdot \Theta_\b\right)\cdot\Theta_\g= A_{\a\b}{}^\g g_{\r\g}$.
The similar computations for the RHS of \triaso\ gives 
$\oint \Theta_\a\cdot\left( \Theta_\b\cdot\Theta_\g\right)= A_{\b\g}{}^\r g_{\a\r}$.
It follows that
$$
A_{\a\b}{}^\g g_{\r\g}=A_{\b\g}{}^\r g_{\a\r}
$$
as was claimed.
\qed
\end{proof}

\begin{corollary}
Let $A_{\a\b\g}:= A_{\b\g}{}^\r g_{\a\r} \in \Bbbk[[t_H]]$. Then
There exist $\Phi \in \Bbbk[[t_H]]$ such that 
$$
\eqalign{
A_{\a\b\g}&=\Fr{\rd^3 \Phi}{\rd\!t^\a\rd\! t^\b \rd\! t^\g}
,\cr
 g_{\b\g}&=\Fr{\rd^3 \Phi}{\rd\!t^0\rd\! t^\b \rd\! t^\g}.
 }
 $$
\end{corollary}

\begin{proof}
Standard.
\end{proof}

Hence we just have established that a semi-classical BV QFT algebra with a semi-classical QFT integral
induce a structure of Frobenius manifold on $\CM$, albeit the flat metric $g_{\a\b}$ may not be invertible.
We have invertible metric if  the semi-classical QFT integral is non-degenerate, leading to WDDV
equation.

\begin{example}
Consider a BV QFT algebra $(\sC[[\hbar]],\bos{K},\,\cdot\,)$ with the property
that $\bos{K}= Q -\hbar \Delta$, i.e., $K^{(1)}=-\Delta$ and $K^{(n)}=0$ for all $n>1$. 
Then the quadruples $(\sC, Q, \Delta,\,\cdot\,)$ is a differential BV algebra. Assume
that each cohomology class $e_{\a}$ of the complex $(\sC,Q)$ has a representative 
$f([e_{\a}])=O_{\a}$ satisfying $\Delta O_{\a}=0$. 
Hence $\bos{f}(e_\a)=f(e_\a)$ satisfying
$\bos{K}f=0$. Solving the quantum master equation order by order in $t_{H}$, it is obvious
that $\bos{\Theta}=\Theta$. Then the quantum master equation is decomposed
into the following form
$$
\eqalign{
\Theta_{\a}\cdot \Theta_{\b}&=A_{\a\b}{}^{\g}\Theta_{\g} + Q\La_{\a\b}+(\Theta,\La_{\a\b}),
\cr
-\hbar \Theta_{\a\b}&=-\hbar \Delta\La_{\a\b}
}
$$
It follows that our special solution $\Theta$ has the form
$$
\Theta =t^{\a}O_{\a} 
+\sum_{n=2}^{\infty}\Fr{1}{n!}t^{\a_{n}}\cdots t^{\a_{1}} \Delta\, \l_{\a_{1}\cdots\a_{n}},
$$
such that $\Delta \Theta=Q\Theta+(\Theta,\Theta)=0$.
The semi-classical QFT integral
$\oint$ satisfies $\oint Q a\cdot b =-(-1)^{|a|}\oint a\cdot Qb$ and
 $\oint \Delta a\cdot b =-(-1)^{|a|}\oint a\cdot \Delta b$. Hence corresponds to BK integral.\foot{
 Note that the ghost number for $\Delta$ operation is different from the convention in \cite{BK}.}
The above is exactly the special versal solution in \cite{BK} corresponding 
to the flat coordinates on moduli space.
\end{example}

For some semi-classical BV QFT algebra, a semi-classical QFT integral may plays the role of QFT cycle.
Then

\begin{corollary}

For a BV QFT which underlying BV QFT algebra is semi-classical with QFT cycle, which is
a semi-classical QFT integral, the set of $2$-point quantum correlation functions is the flat metric $g_{\a\b}$
and the set of $3$-point quantum correlation functions is the $3$-tensor $A_{\a\b\g}$.

\end{corollary}

\begin{proof}
By the assumption $\left< \bos{\Pi}_{\a\b}\right> =\oint \bos{\Pi}_{\a\b}$, where
$\bos{\Pi}_{\a\b}={\Theta}_\a\cdot \Theta_\b -\hbar \Theta_{\a\b}$.
From the quantum master equation \scqme\ we have
$\oint \Theta_{\a\b} =0$. Hence
$$
\left< \bos{\Pi}_{\a\b}\right> =\oint {\Theta}_\a\cdot \Theta_\b = g_{\a\b}.
$$
Now consider the $3$-point quantum correlators
$$
\bos{\Pi}_{\a\b\g} = {\Theta}_\a\cdot \Theta_\b\cdot\Theta_\g -\hbar \Theta_{\a\b}\cdot {\Theta}_\g
-\hbar \Theta_{\a}\cdot {\Theta}_{\b\g}-\hbar (-1)^{|\a||\b|}\Theta_{\b}\cdot {\Theta}_{\a\g}
+\hbar^2 \Theta_{\a\b\g},
$$
where $\Theta_{\a\b\g}=\rd_\a \Theta_{\b\g}$. Then
$$
\left<\bos{\Pi}_{\a\b\g}\right> = \oint {\Theta}_\a\cdot \Theta_\b\cdot\Theta_\g = A_{\a\b\g},
$$
since we already have shown that $\oint\Theta_{\a\b}\cdot {\Theta}_\g=
\oint \Theta_{\a}\cdot {\Theta}_{\b\g}=\oint\Theta_{\b}\cdot {\Theta}_{\a\g}=0$
while proving lemma \ref{frost} and $\oint \Theta_{\a\b\g}=\rd_\a \oint \Theta_{\b\g}=0$.
\qed.

\end{proof}

\newsec{Proof of the Main Theorem}

The purpose of this section is to prove the main theorem [theorem $3.1$].

Let $\bos{\Theta}^{[1]}:=t^\a\bos{f}(e_\a)=t^\a \bos{O}_\a$ such that
$\bos{K}\bos{\Theta}^{[1]}=0$, since $\bos{K}\bos{O}_\a =0$. 
We set $\bos{P}(1) =\left\{\bos{\Theta}^{[1]}, 0 , 0\right\}$.
For a  natural number $n \geq 2$, we shall build an inductive system $\bos{P}(n)$
$$
\bos{P}(1)\subset\bos{P}(2)\subset \bos{P}(3)\cdots \subset\bos{P}(n-1)\subset \bos{P}(n),
$$
such that $n\rightarrow \infty$ limit implies theorem $3.1$.

\subsection{Setting up $\bos{P}(n)$}

\begin{definition}\label{systemn}
$\bos{P}(n)$, for a fixed $n\geq 2$,  is a system
consist  of 
$$
\eqalign{
\bos{\Theta}&=\bos{\Theta}^{[1]}+\bos{\Theta}^{[2]}+\cdots +\bos{\Theta}^{[n]}
,\cr
\bos{\La}_{\a_{2}\a_{1}}&= \bos{\La}_{\a_{2}\a_{1}}^{[0]}
+\bos{\La}_{\a_{2}\a_{1}}^{[1]}+\cdots +\bos{\La}_{\a_{2}\a_{1}}^{[n-2]}
,\cr
A_{\a_{2}\a_{1}}{}^{\g}&= A^{[0]}_{\a_{2}\a_{1}}{}^{\g}+ A^{[1]}_{\a_{2}\a_{1}}{}^{\g}
+\cdots +  A^{[n-2]}_{\a_{2}\a_{1}}{}^{\g}
,\cr
}
$$
where, for $1\leq k\leq n$, $0\leq j \leq n-2$ and $0\leq\ell\leq n-3$,
$$
\eqalign{
\bos{\Theta}^{[k]}&=
\Fr{1}{k!}t^{\a_{1}}\cdots t^{\a_{k}}\bos{O}_{\a_{k}\cdots \a_{1}} 
\hbox{ where }
\bos{O}_{\a_{k}\cdots \a_{1}} \in\sC[[\hbar]]^{|\a_1|+\cdots+|\a_k|}
,\cr
\bos{\La}_{\a_{2}\a_{1}}^{[j]}&= \Fr{1}{j!}
\bar t^{\r_{1}}\cdots \bar t^{\r_{j}}\bos{\l}_{\r_{j}\cdots\r_{1}\a_{2}\a_{1}}
\hbox{ where }
\bos{\l}_{\a_{k}\cdots \a_{1}} \in\sC[[\hbar]]^{|\a_1|+\cdots+|\a_j|-1}
,\cr
A^{[j]}_{\a_{2}\a_{1}}{}^{\g}&= \Fr{1}{j!}
t^{\r_{1}}\cdots t^{\r_{j}}m_{\r_{j}\cdots\r_{1}\a_{2}\a_{1}}{}^{\g} 
\hbox{ where } m_{\r_{j}\cdots\r_{1}\a_{2}\a_{1}}{}^{\g}  \in \Bbbk
,\cr
}
$$
which satisfy the following properties:

1. graded commutativity:
$
A_{\a_{2}\a_{1}}{}^{\g}-(-1)^{|\a_{2}||\a_{1}|}A_{\a_{1}\a_{}}{}^{\g}=0
\mod t_{H}^{n-1}
$,

2. associativity:
$
A_{\a_{3}\a_{2}}{}^\r A_{\r\a_{1}}{}^\g - A_{\a_{2}\a_{1}}{}^\r A_{\a_{3}\r}{}^\g
=0 
	\mod t_{H}^{n-1}$,
	
3. potentiality: 
$
\rd_{\a_{3}} A_{\a_{2}\a_{1}}{}^\g - (-1)^{|\a_{3}||\a_{2}|} \rd_{\a_{2}} A_{\a_{3}\a_{1}}{}^\g
=0 \mod t_{H}^{n-2}$,

4. unity $A_{0\b}{}^\g =1$

5. quantum master equation:
$$
\eqalign{
\hbar \rd_{\a_2}\rd_{\a_1}\bos{\Theta}
=&\rd_{\a_2}\bos{\Theta}\cdot\rd_{\a_1}\bos{\Theta}
-A_{\a_{2}\a_{1}}{}^{\g}\rd_{\g}\bos{\Theta}
-\bos{K}\bos{\La}_{\a_{2}\a_{1}} - \left(\bos{\Theta}, \bos{\La}_{\a_{2}\a_{1}}\right)_{\hbar}
\cr
&
\mod t_H^{n-1},
}
$$

6, quantum unity: $\rd_0 \bos{\Theta}=1\mod t_H^{n}$.

7. quantum descendant equation:
$
\bos{K}\bos{\Theta} +\Fr{1}{2}\left(\bos{\Theta},\bos{\Theta}\right)_{\hbar} =0
\quad \hbox{\it mod}\quad t_H^{n+1}$.

8. quantum gauge: 
$\bos{\La}_{\a_2\a_1}$
satisfies
$$
\eqalign{
\bos{\La}_{\a_2 0}=&0,
\cr
\bos{\La}_{\a_2\a_1}=&(-1)^{|\a_2||\a_1|}\bos{\La}_{\a_1\a_2}
,\cr
\hbar\rd_{\bar\a}\bos{\La}_{\b\g}-(-1)^{|\a||\b|}\hbar\rd_{\bar\b}\bos{\La}_{\a\g}
=
&-(-1)^{|\a_3|} \rd_{\a_3}\bos{\Theta} \cdot\bos{\La}_{\a_2\a_1}
+(-1)^{|\a_3||\a_2| +|\a_2|}\rd_{\a_2}\bos{\Theta}\cdot \bos{\La}_{\a_3\a_1}
\cr
&
-A_{\a_2\a_1}{}^\r \bos{\La}_{\a_3\r}
+(-1)^{|\a_3||\a_2|}A_{\a_3\a_1}{}^\r \bos{\La}_{\a_2\r} 
\cr
&
-B_{\a_3\a_2\a_1}{}^\r \rd_\r \bos{\Theta}
-\bos{K}{X}_{\a_3\a_2\a_1}-\big(\bos{\Theta},{X}_{\a_3\a_2\a_1}\big)_{\hbar}
\cr
&
\mod t_H^{n-2},
}
$$
where $B_{\a_3\a_2\a_1}{}^\g \in \Bbbk[[t_H]]\mod t_H^{n-2}$ is a certain $4$-tensor 
satisfying
$$
\eqalign{
{B}_{0\a_{2}\a_{1}}{}^{\g}={B}_{\a_{3}0\a_{1}}{}^{\g}={B}_{\a_{3}\a_{2} 0}{}^{\g}=0\mod t_H^{n-2}
,\cr
{B}_{\a_{3}\a_{2}\a_{1}}{}^{\g}+(-1)^{|\a_3||\a_2|}{B}_{\a_{2}\a_{3}\a_{1}}{}^{\g}=0
\mod t_H^{n-2}
,\cr
{B}_{\a_3\a_2\a_1}
-(-1)^{|\a_2||\a_1|}{B}_{\a_3\a_1\a_2}
+(-1)^{|\a_3|(|\a_2|+|\a_1|)}{B}_{\a_2\a_1\a_3}=0
\mod t_H^{n-2},
}
$$
and $X_{\a_3\a_2\a_1} \in \sC[[t_H]]^{|\a_3|+|\a_2| +|\a_1| -2}\mod t_H^{n}$ satisfying
$$
\eqalign{
{X}_{0\a_{2}\a_{1}}={X}_{\a_{3}0\a_{1}}=X_{\a_{3}\a_{2} 0}=0\mod t_H^{n-2}
,\cr
X_{\a_3\a_2\a_1}  +(-1)^{|\a_3||\a_2|} X_{\a_3\a_2\a_1} =0
\mod t_H^{n-2}
,\cr
X_{\a_3\a_2\a_1}
-(-1)^{|\a_2||\a_1|}X_{\a_3\a_1\a_2}
+(-1)^{|\a_3|(|\a_2|+|\a_1|)}{X}_{\a_2\a_1\a_3}=0
\mod t_H^{n-2}
..
}
$$
Remark that the $4$-tensor $B_{\a\b\g}{}^\r$  does not contribute to $\bos{P}(2)$.
\end{definition}

Note that we have already constructed $\bos{P}(2)$ as well as a sketch of constructing $\bos{P}(3)$
out of $\bos{P}(2)$ after the statement of theorem $3.1$.
We shall build $\bos{P}(n+1)$ out of assumed $\bos{P}(n)$.

\begin{remark} The homogeneity 
$ |e_\r| t^\r\rd_\r A_{\a\b}{}^\g =\left(|e_\g|-|e_\b| -|e_\a|\right)A_{\a\b}{}^\g$
follows from the quantum master equation by matching the ghost number.
\end{remark}

\subsubsection{Some consequences of $\bos{P}(n)$.}

Assuming  $\bos{P}(n)$  we examine some of its consequences, which shall be
used to build $\bos{P}(n+1)$ from $\bos{P}(n)$.

It is convenient to introduce the notation that
$\bos{\Theta}_{\a}:=\rd_{\a}\bos{\Theta}$ and $\bos{\Theta}_{\b\a}:=\rd_{\b}\bos{\Theta}_\a$
such that the $\mod t_H^{n-1}$ quantum master equation in $\bos{P}(n)$ 
is rewritten as follows
$$
\hbar \bos{\Theta}_{\a_2\a_1}
=\bos{\Theta}_{\a_2}\cdot\bos{\Theta}_{\a_1}
-A_{\a_{2}\a_{1}}{}^{\g}\bos{\Theta}_\g
-\bos{K}\bos{\La}_{\a_{2}\a_{1}} - \left(\bos{\Theta}, \bos{\La}_{\a_{2}\a_{1}}\right)_{\hbar}
\mod t_H^{n-1}.
$$
Note that $\bos{\Theta}_0=1$.
Then the classical limit of quantum master equation in $\bos{P}(n)$ is
\eqn\clqme{
{\Theta}_{\a_2}\cdot{\Theta}_{\a_1}
=A_{\a_{2}\a_{1}}{}^{\g}{\Theta}_\g
+Q{\La}_{\a_{2}\a_{1}} - \left({\Theta}, {\La}_{\a_{2}\a_{1}}\right)
\mod t_H^{n-1},
}
where ${\Theta}_{\a_1}=\bos{\Theta}_{\a_1}\big|_{\hbar=0}$ and 
${\La}_{\a_{2}\a_{1}}=\bos{\La}_{\a_{2}\a_{1}}\big|_{\hbar=0}$.
We may also use the following decompositions
$$
\bos{\Theta}_{\a} =\bos{\Theta}_{\a}^{[0]} +\bos{\Theta}_{\a}^{[1]}+\cdots +\bos{\Theta}_{\a}^{[n-1]},
$$
where $\bos{\Theta}_{\a}^{[k-1]}=\rd_\a \bos{\Theta}^{[k]}$. 
Note that
$\bos{\Theta}^{[0]}_\a =\bos{O}_\a$
and
$ t^\a \bos{\Theta}_{\a}^{[k-1]}= k  \bos{\Theta}^{[k]}$,
since $\bos{\Theta}^{[k]}$ is a homogeneous polynomial in degree $k$ in $t_H$.

Now we turn to the quantum descendant equation in $\bos{P}(n)$:
$
\bos{K}\bos{\Theta}+\Fr{1}{2}\left(\bos{\Theta},\bos{\Theta}\right)_{\hbar}=0\mod t_H^{n+1}$.
\begin{corollary}
\label{consqa}

Let $\bos{K}_{\bos{\Theta}}:= \bos{K}+\left(\bos{\Theta}, \hbox{ }\right)_{\hbar}\mod t_H^{n+1}$.
On any $\bos{M} \in \sC[[t_H,\hbar]]]\big)[[\hbar]]\mod t_H^{m}$ for $m \leq n+1$,
$$
\bos{K}^{2}_{\bos{\Theta}}\bos{M} =0 \mod t_H^{m}
$$
\end{corollary}
\begin{proof}

Note that  
$\bos{K}_{\bos{\Theta}}\bos{M}
\equiv\bos{K}\bos{M}+\left(\bos{\Theta},\bos{M}\right)_{\hbar}\mod t_H^{m}$,
and
$$
\bos{K}^{2}_{\bos{\Theta}}\bos{M}
=\bos{K}^{2}\bos{X}+\bos{K}\big(\bos{\Theta},\bos{M}\big)_{\hbar}
+\left(\bos{\Theta},\bos{K}\bos{M}\right)_{\hbar}
+\left(\bos{\Theta},\left(\bos{\Theta},\bos{M}\right)_{\hbar}\right)_{\hbar}
\mod t_H^{m}
$$
Using $\bos{K}^{2}=0$, the property that $\bos{K}$ is a derivation of the BV bracket
and the Jacobi identity of the BV bracket, we have
$$
\bos{K}^{2}_{\bos{\Theta}}\bos{M}
=\left(\bos{K}\bos{\Theta}+\Fr{1}{2}\big(\bos{\Theta},\bos{\Theta}\big)_{\hbar},\bos{M}\right)_{\hbar}
\mod t_H^{m}.
$$
Since $\bos{K}\bos{\Theta}+\Fr{1}{2}\left(\bos{\Theta},\bos{\Theta}\right)_{\hbar}=0\mod t^{n+1}$
and $m\leq n+1$, we conclude that
$$
\bos{K}^{2}_{\bos{\Theta}}\bos{M}=0
\mod t_H^{m}.
$$
\qed
\end{proof}
By applying $\rd_{\a_{1}}$ and $\rd_{\a_{2}}$  successively to the quantum descendant equation,
we also have
\begin{corollary}
\label{consqb}

$
\bos{K}\bos{\Theta}_{\a_{1}} +\left(\bos{\Theta},\bos{\Theta}_{\a_{1}}\right)_{\hbar} =0
\mod t_H^{n}.
$
\end{corollary}

Let $\Theta= \bos{\Theta}\bigr|_{\hbar=0}$ and let 
$\Theta_{\g}=\bos{\Theta}_{\g}\bigr|_{\hbar=0}$.
By taking the classical limit of the quantum descendant equation in $\bos{P}(n)$ we have
\begin{corollary}
$Q\Theta +\Fr{1}{2}\left(\Theta,\Theta\right) =0 \mod t_H^{n+1}$
\end{corollary}
By taking the classical limit of  corollary \ref{consqa}, we have
\begin{corollary}
Let ${Q}_{\Theta}:= Q+\left(\Theta, \hbox{ }\right)\mod t_H^{n+1}$.
On any ${M} \in \sC[[t_H]]\mod t_{H}^{m}$ for $m \leq n+1$,
$$
Q^{2}_{\Theta}{M} =0 \mod t_H^{m}.
$$
\end{corollary}
By taking the classical limit of  corollary \ref{consqb}, we have
\begin{corollary}
$Q\Theta_{\g} +\left(\Theta,\Theta_{\g}\right) =0 \mod t_H^{n}$
\end{corollary}
The condition
$Q\Theta_{\g}+ \left(\Theta, \Theta_{\g}\right)=0\mod t_{H}^{n}$
implies that $Q\Theta_{\g}+ \left(\Theta, \Theta_{\g}\right)=0\mod t_{H}^{k}$
for all $1\leq k \leq n$. In particular 
$Q\Theta_{\g}+ \left(\Theta, \Theta_{\g}\right)=0\mod t_{H}^{1}$ is equivalent
to $Q O_{\g}=0$. By our assumption the $Q$-cohomology classes of $\{O_{\g}\}$
form a (linearly independent) basis of $H$, any homogeneous
element $X \in \sC[[t_H]]^{|X|}$ satisfying $Q X =0$ can be expressed as
$X = c^{\g}O_{\g} + Q Y$ for unique set of constants $\{c^{\g}\}$ in $\Bbbk$ and
some $Y \in \sC[[t_H]]^{|X|-1}$ defined modulo $\Ker Q$. Also for any equality in the form
$Q Y = c^{\g}O_{\g}$ implies that $c^{\g}=0$ for all $\g$, since
by taking the $Q$-cohomology class we have $c^{\g}[O_{\g}]\equiv c^{\g}e_{\g}=0$
and $\{e_{\g}\}$ are linearly independent, as well as $Q Y=0$. We shall establish the similar
properties involving $Q +\left(\Theta,\hbox{ }\right)$ and $\{\Theta\}_{\a} \mod t^{n}$
as consequences of the $\bos{P}(n)$.

\begin{proposition}
\label{tech1}

Any  element $M \in \sC[[t_H]]\big)^{|M|}$
defined modulo $t_H^{m}$ for $1\leq m \leq n$ satisfying
$$
Q M +\left(\Theta, M\right)=0\mod t_H^{m}
$$
can be expressed as
$$
M = B^{\g}\Theta_{\g} + Q \La +\left(\Theta,\La\right)_{\hbar} \mod t_H^{m}
$$
where $B^{\g}\in \Bbbk[[t_{H}]]\mod t_H^{m}$ and 
$\La \in\sC[[t_H]]^{|M|-1} \mod t_H^m$ defined modulo 
$\Ker\left( Q+\left(\Theta,\hbox{ }\right)\right)$.
\end{proposition}

\begin{proof}
See Appendix B.\qed
\end{proof}

\begin{corollary}
\label{tech3}

For any  $M \in \sC[[t_H]]^{|M|}$
defined modulo $t_H^{m}$ for $1\leq m \leq n$
for $1\leq m \leq n$ satisfying
$$
Q M +\left(\Theta, M\right)=0\mod t_H^{m},
$$
there is a canonical extension to 
$\bos{M} \in \sC[[t_H,\hbar]]^{|M|} \mod t_H^m$,
which is unique up to $\hbar \hbox{Im } \bos{K}_{\bos{\Theta}}$,
such that $\bos{M} \bigl|_{\hbar=0} =M$ and
$$
\bos{K} \bos{M} +\left(\bos{\Theta}, \bos{M}\right)_{\hbar}=0\mod t_H^{m}.
$$

\end{corollary}

\begin{proof}

From proposition \ref{tech1}, $M$ 
can be expressed as
$$
M = B^{\g}\Theta_{\g} + Q \La +\left(\Theta,\La\right) \mod t_H^{m}
$$
for unique $B^{\g}\in \Bbbk[t_{H}]\mod t_H^m$ and some 
$\La \in \big(\sC\otimes \Bbbk[t_{H}]\big)^{|M|-1}\mod t_H^m$
defined modulo 
$\Ker\left( Q+\left(\Theta,\hbox{ }\right)\right)$.
Let $\bos{\La} \in \sC[[t_{H},\hbar]]^{|M|-1}\mod t_H^m$ such
that $\bos{\La}|_{\hbar=0}=\La$, and let
$$
\bos{M} = B^{\g}\bos{\Theta}_{\g} + \bos{K} \bos{\La} 
+\left(\bos{\Theta},\bos{\La}\right)_{\hbar} \mod t_H^{m},
$$
Then $\bos{M} \bigl|_{\hbar=0} =M$ and 
$\bos{K} \bos{M} +\left(\bos{\Theta}, \bos{M}\right)_{\hbar}=0\mod t_H^{m}$,
 where we have used corollaries \ref{consqa} and \ref{consqb}.

\qed

\end{proof}

\begin{proposition}
\label{tech2}

Let $m\leq n$. An equality in the following form:
$$
Q M +\left(\Theta, M\right)=C^{\g}\Theta_{\g}\mod t_{H}^{m},
$$
where 
$M\in\sC[[t_H]]^{|M|}$
and $C^{\g}\in \Bbbk[[t_{H}]]\mod t_{H}^{m}$,
implies that
$$
Q M +\left(\Theta, M\right)=C^\g =0\mod t_{H}^{m}.
$$

\end{proposition}

\begin{proof}
See Appendix B. \qed
\end{proof}

Now consider
the following two expressions, which are defined modulo $t_{H}^{n-1}$ and built from $\bos{P}(n)$:
\eqn\qtyb{
\eqalign{
\mathbb{M}_{\a_{3}\a_{2}\a_{1}}:=
&
\bos{\Theta}_{\a_{3}\a_{2}}\cdot\bos{\Theta}_{\a_{1}}
+(-1)^{|\a_{3}||\a_{2}|}\bos{\Theta}_{\a_{2}}\cdot\bos{\Theta}_{\a_{3}\a_{1}}
-A_{\a_{2}\a_{1}}{}^{\g}\bos{\Theta}_{\a_{3}\g}
- \left(\bos{\Theta}_{\a_{3}}, \bos{\La}_{\a_{2}\a_{1}}\right)_{\hbar}
\cr
&
\mod t_{H}^{n-1}
,\cr
\mathbb{N}_{\a_{3}\a_{2}\a_{1}} 
:=&
-A_{\a_{2}\a_{1}}{}^{\r}\bos{\La}_{\a_{3}\r}
+(-1)^{|\a_{3}||\a_{2}|} A_{\a_{3}\a_{1}}{}^{\r}\bos{\La}_{\a_{2}\r}
\cr
&
-(-1)^{|\a_{3}|}\bos{\Theta}_{\a_{3}}\cdot\bos{\La}_{\a_{2}\a_{1}}
+(-1)^{|\a_{3}||\a_{2}|} (-1)^{|\a_{2}|}\bos{\Theta}_{\a_{2}}\cdot\bos{\La}_{\a_{3}\a_{1}}
\cr
& \mod t_{H}^{n-1}.
}
}

Then, 
\begin{proposition}
\label{freshtwo}
The two expressions $\mathbb{M}_{\a_{3}\a_{2}\a_{1}}$ and $\mathbb{N}_{\a_{3}\a_{2}\a_{1}}$
satisfy

1. $\mathbb{M}_{\a_{3}\a_{2}\a_{1}}=(-1)^{|\a_2||\a_1|}\mathbb{M}_{\a_{3}\a_{1}\a_{2}}\mod t_H^{n-1}$,

2. $\mathbb{M}_{\a_{3}\a_{2}0}=\mathbb{M}_{\a_{3}0\a_1}=\mathbb{M}_{0\a_{2}\a_{1}}=0\mod t_H^{n-1}$,

3. $\mathbb{N}_{\a_{3}\a_{2}\a_{1}}=-(-1)^{|\a_3||\a_2|}\mathbb{N}_{\a_{2}\a_{3}\a_{1}}\mod t_H^{n-1}$

4. $\mathbb{N}_{\a_{3}\a_{2}0}=\mathbb{N}_{\a_{3}0\a_1}=\mathbb{N}_{0\a_{2}\a_{1}}=0\mod t_H^{n-1}$,

5. $\mathbb{N}_{\a_3\a_2\a_1}
-(-1)^{|\a_2||\a_1|}\mathbb{N}_{\a_3\a_1\a_2}
+(-1)^{|\a_3|(|\a_2|+|\a_1|)}\mathbb{N}_{\a_2\a_1\a_3}=0\mod t_H^{n-1}$,

6. $\bos{K}\mathbb{N}_{\a_{3}\a_{2}\a_{1}} 
+\left(\bos{\Theta},\mathbb{N}_{\a_{3}\a_{2}\a_{1}}\right)_{\hbar}
=\hbar\left(\mathbb{M}_{\a_{3}\a_{2}\a_{1}}-(-1)^{|\a_3||\a_2|}\mathbb{M}_{\a_{2}\a_{3}\a_{1}}
\right)\equiv\hbar\mathbb{M}_{[\a_{3}\a_{2}]\a_{1}}
\mod t_H^{n-1}$
\end{proposition}

\begin{proof} In order.

1. It  follows from the supercommutativity of the product $\cdot$
and the assumptions that $\bos{\Theta}_{\a_3\a_1}= (-1)^{|\a_1||\a_3|}\bos{\Theta}_{\a_1\a_3}\mod t_H^{n-1}$,
$A_{\a_2\a_1}{}^\g= (-1)^{|\a_2||\a_1|}A_{\a_1\a_2}\mod t_H^{n-1}$ and $\bos{\La}_{\a_2\a_1}=
 (-1)^{|\a_2||\a_1|}\bos{\La}_{\a_1\a_2}\mod t_H^{n-1}$ in $\bos{P}(n)$. 
 
2. It follows  from the definition of $\bos{M}_{\a_3\a_2\a_1}$ in \qtyb\ together with the assumptions
 that $\bos{\Theta}_{\b 0}=\bos{\Theta}_{0\b}=0\mod t_H^{n-1}$, $\bos{\Theta}_0=1 \mod t_H^{n-1}$,
 $A_{\b 0}{}^\g =A_{0\b}{}^\g =\d_{\b}{}^\g\mod t_H^{n-1}$
 and $\bos{\La}_{\b 0}=0\mod t_H^{n-1}$ in $\bos{P}(n)$.

3. It is trivial by definition.

4. This can be checked using
the assumptions that $\bos{\Theta}_0=1\mod t_H^{n-1}$,
 $A_{\b 0}{}^\g =A_{0\b}{}^\g =\d_{\b}{}^\g\mod t_H^{n-1}$,
 $\bos{\La}_{\b\a}=(-1)^{|\b||\a|}\bos{\La}_{\a\b}\mod t_H^{n-1}$ 
 and $\bos{\La}_{\b 0}=0\mod t_H^{n-1}$ in $\bos{P}(n)$.

5. This follows directly after comparison using 
$A_{\a_2\a_1}{}^\g= (-1)^{|\a_2||\a_1|}A_{\a_1\a_2}\mod t_H^{n-1}$ and $\bos{\La}_{\a_2\a_1}=
 (-1)^{|\a_2||\a_1|}\bos{\La}_{\a_1\a_2}\mod t_H^{n-1}$ in $\bos{P}(n)$. 

6. For the last property,
multiply $\bos{\Theta}_{\a_{3}}$ 
to the $\mod t_H^{n-1}$ quantum master equation in $\bos{P}(n)$
to obtain that
$$
\eqalign{
\hbar\bos{\Theta}_{\a_{3}}\cdot \bos{\Theta}_{\a_{2}\a_{1}}
=&\bos{\Theta}_{\a_{3}}\cdot\left(\bos{\Theta}_{\a_{2}}\cdot\bos{\Theta}_{\a_{1}}\right)
-A_{\a_{2}\a_{1}}{}^{\g}\bos{\Theta}_{\a_{3}}\cdot\bos{\Theta}_{\g}
-\hbar\left(\bos{\Theta}_{\a_{3}},\bos{\La}_{\a_{2}\a_{1}}\right)_{\hbar}
\cr
&
-\bos{K}\left((-1)^{|\a_{3}|}\bos{\Theta}_{\a_{3}}\cdot\bos{\La}_{\a_{2}\a_{1}}\right)
-\left(\bos{\Theta},(-1)^{|\a_{3}|}\bos{\Theta}_{\a_{3}}\cdot\bos{\La}_{\a_{2}\a_{1}}\right)
\cr
&
\mod t_H^{n-1}.
}
$$
Using  the $\mod t_H^{n-1}$ quantum master equation  again,
we have
$$
\eqalign{
\hbar\bos{\Theta}_{\a_{3}}\cdot \bos{\Theta}_{\a_{2}\a_{1}}
=&\bos{\Theta}_{\a_{3}}\cdot\left(\bos{\Theta}_{\a_{2}}\cdot\bos{\Theta}_{\a_{1}}\right)
-A_{\a_{2}\a_{1}}{}^{\r}A_{\a_{3}\r}{}^{\g}\bos{\Theta}_{\g}
-\hbar A_{\a_{2}\a_{1}}{}^{\r}\bos{\Theta}_{\a_{3}\r}
-\hbar\left(\bos{\Theta}_{\a_{3}},\bos{\La}_{\a_{2}\a_{1}}\right)_{\hbar}
\cr
&
-\bos{K}\left(
A_{\a_{2}\a_{1}}{}^{\r}\bos{\La}_{\a_{3}\r}+
(-1)^{|\a_{3}|}\bos{\Theta}_{\a_{3}}\cdot\bos{\La}_{\a_{2}\a_{1}}
\right)
\cr
&
-\left(\bos{\Theta}, A_{\a_{2}\a_{1}}{}^{\r}\bos{\La}_{\a_{3}\r}+
(-1)^{|\a_{3}|}\bos{\Theta}_{\a_{3}}\cdot\bos{\La}_{\a_{2}\a_{1}}\right)_{\hbar}
\cr
&
\mod t_H^{n-1}.
}
$$
Equivalently
$$
\eqalign{
\bos{\Theta}_{\a_{3}}\cdot\left(\bos{\Theta}_{\a_{2}}\cdot\bos{\Theta}_{\a_{1}}\right)
=&
A_{\a_{2}\a_{1}}{}^{\r}A_{\a_{3}\r}{}^{\g}\bos{\Theta}_{\g}
+\hbar\bos{\Theta}_{\a_{3}}\cdot \bos{\Theta}_{\a_{2}\a_{1}}
+\hbar A_{\a_{2}\a_{1}}{}^{\r}\bos{\Theta}_{\a_{3}\r}
+\hbar\left(\bos{\Theta}_{\a_{3}},\bos{\La}_{\a_{2}\a_{1}}\right)
\cr
&
-\bos{K}\left(
A_{\a_{2}\a_{1}}{}^{\r}\bos{\La}_{\a_{3}\r}
-(-1)^{|\a_{3}|}\bos{\Theta}_{\a_{3}}\cdot\bos{\La}_{\a_{2}\a_{1}}
\right)
\cr
&
-\left(\bos{\Theta}, A_{\a_{2}\a_{1}}{}^{\r}\bos{\La}_{\a_{3}\r}
-(-1)^{|\a_{3}|}\bos{\Theta}_{\a_{3}}\cdot\bos{\La}_{\a_{2}\a_{1}}\right)_{\hbar}
\cr
&
\mod t_H^{n-1}.
}
$$
Then from the identity
$$
\bos{\Theta}_{\a_{3}}\cdot\left(\bos{\Theta}_{\a_{2}}\cdot\bos{\Theta}_{\a_{1}}\right)
-(-1)^{|\a_{3}||\a_{2}|}
\bos{\Theta}_{\a_{2}}\cdot\left(\bos{\Theta}_{\a_{3}}\cdot\bos{\Theta}_{\a_{1}}\right)
=0\mod t_H^{n-1},
$$
we have
$$
\eqalign{
0=&
\left(A_{\a_{2}\a_{1}}{}^{\r}A_{\a_{3}\r}{}^{\g}
-(-1)^{|\a_{3}||\a_{2}|}A_{\a_{3}\a_{1}}{}^{\r}A_{\a_{2}\r}{}^{\g}
\right)\bos{\Theta}_{\g}
\cr
&
+\hbar\left(\bos{\Theta}_{\a_{3}}\cdot\bos{\Theta}_{\a_{2}\a_{1}}
-(-1)^{|\a_{3}||\a_{2}|}\bos{\Theta}_{\a_{2}}\cdot\bos{\Theta}_{\a_{3}\a_{1}}\right)
\cr
&
+\hbar\left(A_{\a_{2}\a_{1}}{}^{\g}\bos{\Theta}_{\a_{3}\g}
-(-1)^{|\a_{3}||\a_{2}|}A_{\a_{3}\a_{1}}{}^{\g}\bos{\Theta}_{\a_{2}\g}\right)
\cr
&
+\hbar\left( \left(\bos{\Theta}_{\a_{3}}, \bos{\La}_{\a_{2}\a_{1}}\right)_{\hbar}
-(-1)^{|\a_{3}||\a_{2}|} \left(\bos{\Theta}_{\a_{2}}, \bos{\La}_{\a_{3}\a_{1}}\right)_{\hbar}\right)
\cr
&
-\bos{K}\left(
-A_{\a_{2}\a_{1}}{}^{\r}\bos{\La}_{\a_{3}\r}
+(-1)^{|\a_{3}||\a_{2}|} A_{\a_{3}\a_{1}}{}^{\r}\bos{\La}_{\a_{2}\r}
-(-1)^{|\a_{3}|}\bos{\Theta}_{\a_{3}}\cdot\bos{\La}_{\a_{2}\a_{1}}
+(-1)^{|\a_{3}||\a_{2}|} (-1)^{|\a_{2}|}\bos{\Theta}_{\a_{2}}\cdot\bos{\La}_{\a_{3}\a_{1}}
\right)
\cr
&
-\left(\bos{\Theta},
-A_{\a_{2}\a_{1}}{}^{\r}\bos{\La}_{\a_{3}\r}
+(-1)^{|\a_{3}||\a_{2}|} A_{\a_{3}\a_{1}}{}^{\r}\bos{\La}_{\a_{2}\r}
-(-1)^{|\a_{3}|}\bos{\Theta}_{\a_{3}}\cdot\bos{\La}_{\a_{2}\a_{1}}
+(-1)^{|\a_{3}||\a_{2}|} (-1)^{|\a_{2}|}\bos{\Theta}_{\a_{2}}\cdot\bos{\La}_{\a_{3}\a_{1}}
\right)_{\hbar}
\cr
&\mod t_H^{n-1}.
}
$$
Now we use that $A_{\a_{2}\a_{1}}{}^{\r}A_{\a_{3}\r}{}^{\g}
-(-1)^{|\a_{3}||\a_{2}|}A_{\a_{3}\a_{1}}{}^{\r}A_{\a_{2}\r}{}^{\g}=0 \mod t_{H}^{n-1}
$
and compare the remaining terms with the definitions 
 of $\mathbb{N}_{\a_{3}\a_{2}\a_{1}}\mod
t_{H}^{n-1}$ and $\mathbb{M}_{\a_{3}\a_{2}\a_{1}}\mod
t_{H}^{n-1}$ in  the equation \qtyb\ to establish  the property.
\qed
\end{proof}

Finally, we consider the quantum gauge 
the $3$rd condition for the quantum gauge (property $8$ of $\bos{P}(n)$),
which can be written as follows:
\eqn\qgaugex{
\eqalign{
\hbar \rd_{[\bar\a_3}\bos{\La}_{\a_2]\a_1}
=
\mathbb{N}_{\a_3\a_2\a_1}
-B_{\a_3\a_2\a_1}{}^\r  \bos{\Theta}_\r
-\bos{K}{X}_{\a_3\a_2\a_1} -\big(\bos{\Theta}, {X}_{\a_3\a_2\a_1}\big)_{\hbar}
\mod t_H^{n-2},
}
}
where $\mathbb{N}_{\a_3\a_2\a_1}$ contributes to the above modulo $t_H^{n-2}$ only.

\subsection{Building $\bos{P}(n+1)$ out of $\bos{P}(n)$}

Our goal in this subsection is to build $\bos{P}(n+1)$ out of assumed $\bos{P}(n)$.
We are going to proceed in the following orders: (1) define $\tilde{A}_{\a\b}{}^\g$
such that $\tilde{A}_{0\b}{}^\g=\d_\b{}^\g$ and 
$\tilde{A}_{\a\b}{}^\g= {A}_{\a\b}{}^\g \mod t_H^{n-1}$, (2) establish
the $\mod t_H^n$ graded commutativity and associativity of $\tilde{A}_{\a\b}{}^\g$,
(3) establish the $\mod t_H^{n-1}$ potentiality of $\tilde{A}_{\a\b}{}^\g$, (4) define
$\tilde\bos{\Theta}$ such that $\rd_0\tilde\bos{\Theta}=1$ and 
$\tilde\bos{\Theta}=\bos{\Theta} \mod t_H^{n+1}$
and show that $\tilde\bos{\Theta}$ satisfy the $\mod t_H^n$ quantum master
equation and  the $\mod t_H^{n+2}$ 
quantum descendant equation.

\subsubsection{Definition  of $\tilde A_{\a\b}{}^\g$ and its immediate properties.}

We first consider
the following expression, {\it built from} $\bos{P}(n)$, 
\eqn\qtya{
\mathbb{M}_{\a_{2}\a_{1}}:=
\bos{\Theta}_{\a_{2}}\cdot\bos{\Theta}_{\a_{1}}
\mod t_{H}^{n},
}
which is  defined modulo $t_{H}^{n}$
since $\bos{\Theta}_{\a}$ are defined modulo $t_{H}^{n}$ in $\bos{P}(n)$.
Then, we obtain that
\eqn\qtyab{
\bos{K}\mathbb{M}_{\a_{2}\a_{1}}
+\left(\bos{\Theta},\mathbb{M}_{\a_{2}\a_{1}}\right)
=-\hbar (-1)^{|\a_2|}\big(\bos{\Theta}_{\a_{2}}, \bos{\Theta}_{\a_{1}}\big)_{\hbar} 
\mod t_{H}^{n}
}
by a direct computation.

\begin{lemma}\label{freshone}
There are unique $\tilde A_{\a_2\a_1}{}^\g \in \Bbbk[[t_{H}]]\mod t_{H}^{n}$
and some $\tilde\La_{\a_2\a_1} \in \sC[[t_H]]^{|\a|_2+|\a_1|-1}\mod t_H^n$ 
satisfying
\eqn\beba{
\Theta_{\a_{2}}\cdot\Theta_{\a_{1}}= \tilde{A}_{\a_{2}\a_{1}}{}^{\g}\Theta_{\g} + 
Q{\tilde \La}_{\a_{2}\a_{1}}
+\left({\Theta},\tilde{\La}_{\a_{2}\a_{1}}\right)\mod t_{H}^{n},
}
with the following properties

1. $\tilde A_{\a_{2}\a_{1}}{}^{\g}
-(-1)^{|\a|_{2}|\a_{1}|}\tilde A_{\a_{1}\a_{2}}{}^{\g}=0 \mod t_H^n$,

2. $\tilde{A}_{0\a_1}{}^{\g}=\d_{\a_1}{}^\g \mod t_H^n$,

3. $\tilde A_{\a_{2}\a_{1}}{}^{\r}\tilde A_{\a_{3}\r}{}^{\g}
-(-1)^{|\a_{3}||\a_{2}|}
\tilde A_{\a_{3}\a_{1}}{}^{\r} \tilde A_{\a_{2}\r}{}^{\g}=0 \mod t^{n}_{H}$,
 
4. $\tilde\La_{\a_2\a_1} = (-1)^{|\a_2||\a_1|}\tilde\La_{\a_1\a_2}\mod t_H^n$,

5. $\tilde \La_{0\a_1}=0 \mod t_H^n$.

\end{lemma}

\begin{proof}

By taking classical limit of \qtyb, we obtain that
$Q{\CM}_{\a_{2}\a_{1}}
+\left({\Theta},{\CM}_{\a_{2}\a_{1}}\right)
=0\mod t_{H}^{n}$,
where $\CM_{\a_{2}\a_{1}}:=\mathbb{M}_{\a_{2}\a_{1}}\bigr|_{t_{H}=0}
=\Theta_{\a_{2}}\cdot \Theta_{\a_{1}}\in \sC[[t_{H}]]^{|\a_{2}|+|\a_{1}|}\mod t_H^n$.
Applying  proposition \ref{tech1}, we conclude that
there are unique $3$-tensor $\tilde A_{\a_{2}\a_{1}}{}^{\g}\in \Bbbk[[t_{H}]]\mod t_{H}^{n}$
and some $\tilde{X}_{\a_{2}\a_{1}}\in \sC[[t_H]]^{|\a_{2}|+|\a_{1}|} \mod t_H^n$,
defined modulo $\Ker Q_\Theta$,
such that
\eqn\begba{
\Theta_{\a_{2}}\cdot\Theta_{\a_{1}}= \tilde{A}_{\a_{2}\a_{1}}{}^{\g}\Theta_{\g} + 
Q{\tilde X}_{\a_{2}\a_{1}}
+\left({\Theta},\tilde{X}_{\a_{2}\a_{1}}\right)\mod t_{H}^{n}.
}
From the super-commutativity $\Theta_{\a_{2}}\cdot\Theta_{\a_{1}}-
(-1)^{|\a|_{2}|\a_{1}|}\Theta_{\a_{1}}\cdot\Theta_{\a_{2}}=0 \mod t_{H}^{n}$,
the  equation \begba\ implies that
$$
\left(\tilde A_{\a_{2}\a_{1}}{}^{\g}
-(-1)^{|\a|_{2}|\a_{1}|}\tilde A_{\a_{1}\a_{2}}{}^{\g}
\right)\Theta_{\g}
=-Q\tilde X_{[\a_{2}\a_{1}]}
-\left({\Theta},\tilde{X}_{[\a_{2}\a_{1}]}\right)\mod t_{H}^{n}.
$$
where $\tilde X_{[\a_{2}\a_{1}]}:=
\tilde X_{\a_{2}\a_{1}}-(-1)^{|\a|_{2}|\a_{1}|}\tilde X_{\a_{1}\a_{2}}\mod t_{H}^{n}$.
We now apply  proposition \ref{tech2}  to conclude that
$$
\tilde A_{\a_{2}\a_{1}}{}^{\g}
-(-1)^{|\a|_{2}|\a_{1}|}\tilde A_{\a_{1}\a_{2}}{}^{\g}=0 \mod t_H^n
$$
and $Q\tilde X_{[\a_{2}\a_{1}]}+\left({\Theta},\tilde{X}_{[\a_{2}\a_{1}]}\right)=0$.
Let
$$
\tilde\La_{\a_2\a_1} 
:= \Fr{1}{2}\left(\tilde X_{\a_2\a_1}+(-1)^{|\a_2||\a_1|}\tilde X_{\a_2\a_1} \right)\mod t_H^n,
$$
such that $\tilde\La_{\a_2\a_1} = (-1)^{|\a_2||\a_1|}\tilde\La_{\a_1\a_2}\mod t_H^n$. 
Then \begba\ is rewritten as follows
$$
\Theta_{\a_{2}}\cdot\Theta_{\a_{1}}= \tilde{A}_{\a_{2}\a_{1}}{}^{\g}\Theta_{\g} + 
Q{\tilde \La}_{\a_{2}\a_{1}}
+\left({\Theta},\tilde{\La}_{\a_{2}\a_{1}}\right)\mod t_{H}^{n}.
$$
Hence we have established the relation \beba\ as well as the properties $1$ and $4$.
For the properties $2$ and $5$,
set $\a_2=0$ in the above equation and use $\Theta_0=1$ to deduce that
$\tilde{A}_{0\b}{}^{\g}=\d_\b{}^\g$ and $\tilde \La_{0\a_1}=0 \mod t_H^n$.

For the last remaining property $3$, 
consider the following identity 
\eqn\bebb{
\Theta_{\a_{3}}\cdot\left(\Theta_{\a_{2}}\cdot\Theta_{\a_{1}}\right)
-(-1)^{|\a_{3}||\a_{2}|}\Theta_{\a_{2}}\cdot\left(\Theta_{\a_{3}}\cdot\Theta_{\a_{1}}\right)
=0,
\mod t_{H}^{n}
}
which is 
due to the associativity and the super-commutativity of the product.
Using the relation \beba, we have
$$
\eqalign{
\Theta_{\a_{3}}\cdot\left(\Theta_{\a_{2}}\cdot\Theta_{\a_{1}}\right)
=&
\tilde  A_{\a_{2}\a_{1}}{}^{\r}\Theta_{\a_{3}}\cdot\Theta_{\r} + 
Q_{\Theta}\left((-1)^{|a_{3}|}\Theta_{\a_{3}}\cdot\tilde{\La}_{\a_{2}\a_{1}}\right)
\mod t_{H}^{n},
}
$$
where we used the property that $Q_{\Theta} \mod t_{H}^{n+1}$ 
is a derivation of the product and $Q_{\Theta}\Theta_{\a}=0\mod t_{H}^{n}$.
Using the relation \beba\ again for $\Theta_{\a_{3}}\cdot\Theta_{\r}$, we have
$$
\eqalign{
\Theta_{\a_{3}}\cdot\left(\Theta_{\a_{2}}\cdot\Theta_{\a_{1}}\right)
=&
 \tilde A_{\a_{2}\a_{1}}{}^{\r}\tilde A_{\a_{3}\r}{}^{\g}\Theta_{\g} + 
Q_{\Theta}\left(\tilde A_{\a_{2}\a_{1}}{}^{\r}\tilde \La_{\r \a_{3}}
+(-1)^{|a_{3}|}\Theta_{\a_{3}}\cdot\tilde \La_{\a_{2}\a_{1}}\right)
\mod t_{H}^{n}.
}
$$
Then the identity \bebb\ imply that
\eqn\bebc{
\left(
\tilde A_{\a_{2}\a_{1}}{}^{\r}\tilde A_{\a_{3}\r}{}^{\g}
-(-1)^{|\a_{3}||\a_{2}|}
\tilde A_{\a_{3}\a_{1}}{}^{\r} \tilde A_{\a_{2}\r}{}^{\g}\right)\Theta_{\g}
= -Q \tilde\CN_{\a_{3}\a_{2}\a_{1}}- \left(\Theta, \tilde\CN_{\a_{3}\a_{2}\a_{1}}\right)
\mod t_{H}^{n},
}
where
$$
\eqalign{
 \tilde\CN_{\a_{3}\a_{2}\a_{1}}
 :=&
\tilde A_{\a_{2}\a_{1}}{}^{\r}\tilde{\La}_{\a_{3}\r}
-(-1)^{|\a_{3}||\a_{2}|} \tilde A_{\a_{3}\a_{1}}{}^{\r}\tilde{X}_{\a_{2}\r}
\cr
&
-(-1)^{|\a_{3}|}{\Theta}_{\a_{3}}\cdot\tilde{\La}_{\a_{2}\a_{1}}
+(-1)^{|\a_{3}||\a_{2}|} (-1)^{|\a_{2}|}{\Theta}_{\a_{2}}\cdot\tilde{\La}_{\a_{3}\a_{1}}
\mod t_{H}^{n}.
}
$$
Applying  proposition \ref{tech2} in Appendix $A$ to the equation \bebc\ to conclude
that
$$
\tilde A_{\a_{2}\a_{1}}{}^{\r}\tilde A_{\a_{3}\r}{}^{\g}
-(-1)^{|\a_{3}||\a_{2}|}
\tilde A_{\a_{3}\a_{1}}{}^{\r} \tilde A_{\a_{2}\r}{}^{\g}=0 \mod t^{n}_{H}.
$$
\qed
\end{proof}
\begin{corollary}
Associativity modulo $t^{n}_{H}$
$$
 \tilde A_{\a_{3}\a_{2}}{}^{\r} \tilde A_{\r\a_{1}}{}^{\g}
 - \tilde A_{\a_{2}\a_{1}}{}^{\r} \tilde A_{\a_{3}\r}{}^{\g}=0 \mod t^{n}_{H}.
$$
\end{corollary}
\begin{proof}
Combine properties $1$ and $3$ in lemma \ref{freshone}.
\qed
\end{proof}

\begin{corollary} Reduction modulo $t_H^{n-1}$:
$\tilde A_{\a_2\a_1}{}^\g = A_{\a_2\a_1}{}^\g \mod t_H^{n-1}$ and
$\tilde \La_{\a_2\a_1}=\La_{\a_2\a_1} \mod t_H^{n-1}$.
\end{corollary}

\begin{proof}
Compare \beba\ with the classical limit  \clqme\ of quantum master equation in $\bos{P}(n)$.
\end{proof}

\subsubsection{Potentiality  $\tilde A_{\a\b}{}^\g$.}

Here is the idea of our proof.
Recall the defining equation \beba\ of $\tilde A_{\a\b}{}^\g$ in lemma \ref{freshone}:
$$
\eqalign{
\Theta_{\a_{2}}\cdot\Theta_{\a_{1}}
&= \tilde A_{\a_{2}\a_{1}}{}^{\g}\Theta_{\g} + 
Q\tilde {\La}_{\a_{2}\a_{1}}
+\left({\Theta},\tilde{\La}_{\a_{2}\a_{1}}\right)
\mod t_{H}^{n},
}
$$
where 
$\tilde A_{\a_{2}\a_{1}}{}^{\g}= A_{\a_{2}\a_{1}}{}^{\g}\mod t_H^{n-1}$ and
$\tilde{\La}_{\a_2\a_1} = \La_{\a_2\a_1} \mod t_H^{n-1}$.
Apply $\rd_{\a_3}$ to the above equation to have
$$
\eqalign{
\Theta_{\a_{3}\a_{2}}\cdot\Theta_{\a_{1}}
+(-1)^{|\a_3||\a_2|}\Theta_{\a_{2}}\cdot\Theta_{\a_3\a_{1}}
&-\tilde A_{\a_{2}\a_{1}}{}^{\g}\Theta_{\a_3\g}
-\left({\Theta}_{\a_3},\tilde{\La}_{\a_{2}\a_{1}}\right)
\cr
&
=\rd_{\a_3}\tilde A_{\a_{2}\a_{1}}{}^{\g}\Theta_{\g} 
+Q_{\!\Theta}(-1)^{|\a_3|}\rd_{\a_3}\tilde\La_{\a_2\a_1}
\mod t_H^{n-1}.
}
$$
Denote the LHS of above by
$\CM_{\a_3\a_2\a_1}$:
$$
\CM_{\a_3\a_2\a_1}
=\Theta_{\a_{3}\a_{2}}\cdot\Theta_{\a_{1}}
+(-1)^{|\a_3||\a_2|}\Theta_{\a_{2}}\cdot\Theta_{\a_3\a_{1}}
- A_{\a_{2}\a_{1}}{}^{\g}\Theta_{\a_3\g}
-\left({\Theta}_{\a_3},{\La}_{\a_{2}\a_{1}}\right)
\mod t_H^{n-1},
$$
since
$\tilde A_{\a_{2}\a_{1}}{}^{\g}= A_{\a_{2}\a_{1}}{}^{\g}\mod t_H^{n-1}$ and
$\tilde{\La}_{\a_2\a_1} = \La_{\a_2\a_1} \mod t_H^{n-1}$.
Hence we have
\eqn\franx{
\eqalign{
\CM_{[\a_3\a_2]\a_1}
=
&\left(\rd_{\a_3}\tilde A_{\a_{2}\a_{1}}{}^{\g}-(-1)^{|\a_3||\a_2|}
\rd_{\a_2}\tilde A_{\a_{3}\a_{1}}{}^{\g}\right)
\Theta_{\g} 
\cr
&
+Q_{\Theta}\left((-1)^{|\a_3|}\rd_{\a_3}\tilde\La_{\a_2\a_1}
-(-1)^{|\a_2|+|\a_3||\a_2|}\rd_{\a_3}\tilde\La_{\a_2\a_1}\right)
\mod t_H^{n-1}.
}
}
Thus we need to show that $\CM_{[\a_3\a_2]\a_1}\in \hbox{Im }Q_{\Theta}$ to
establish that
$$
\rd_{\a_{3}} \tilde A_{\a_{2}\a_{1}}{}^\g - (-1)^{|\a_{3}||\a_{2}|} \rd_{\a_{2}} \tilde A_{\a_{3}\a_{1}}{}^\g
=0 \mod t_{H}^{n-1}.
$$

Note that $\CM_{\a_3\a_2\a_1}$ is exactly the classical limit of $\mathbb{M}_{\a_3\a_2\a_1}$,
the properties of which  are listed in proposition \ref{freshtwo}. We shall need a stronger proposition.

\begin{proposition}\label{freshthree}
There is certain 
$\tilde\mathbb{Y}_{\a_{3}\a_{2}\a_{1}}\in \sC[[t_H,\hbar]]^{|\a_3|+|\a_2|+|\a_1|-1}\mod t_H^{n-1}$
such that
$$
\tilde\mathbb{Y}_{\a_3\a_2\a_1} = \rd_{[\bar\a_3}\bos{\La}_{\a_2]\a_1} \mod t_H^{n-2},
$$
and

1.
$
\bos{K} \tilde\mathbb{Y}_{\a_{3}\a_{2}\a_{1}}
+\left(\bos{\Theta}, \tilde\mathbb{Y}_{\a_{3}\a_{2}\a_{1}}\right)_{\hbar}
= \mathbb{M}_{[\a_{3}\a_{2}]\a_{1}}\mod t_{H}^{n-1}$,

2.
$\tilde\mathbb{Y}_{\a_3\a_2\a_1} =-(-1)^{|\a_3||\a_2|}\tilde\mathbb{Y}_{\a_2\a_3\a_1}\mod t_{H}^{n-1}$,

3. 
$\tilde\mathbb{Y}_{0\a_2\a_1} =\tilde\mathbb{Y}_{\a_3 0\a_1} 
=\tilde\mathbb{Y}_{\a_3\a_2 0} =0\mod t_{H}^{n-1}$,

4.
$
\tilde\mathbb{Y}_{\a_3\a_2\a_1}
-(-1)^{|\a_2||\a_1|}\tilde\mathbb{Y}_{\a_3\a_1\a_2}
+(-1)^{|\a_3|(|\a_2|+|\a_1|)}\tilde\mathbb{Y}_{\a_2\a_1\a_3}
=0 \mod t_{H}^{n-1}$.

\end{proposition}

\begin{proof}
Let ${\CN}_{\a_{3}\a_{2}\a_{1}} \mod t_{H}^{n-1}$ be the classical
limit of $\mathbb{N}_{\a_{3}\a_{2}\a_{1}} \mod t_{H}^{n-1}$.
Then  proposition \ref{freshtwo} in the classical limit $\hbar=0$ 
becomes
$$
Q{\CN}_{\a_{3}\a_{2}\a_{1}} +\left({\Theta},{\CN}_{\a_{3}\a_{2}\a_{1}}\right)
=0\mod t_H^{n-1}.
$$
Note also that 
$$
\eqalign{
{\CN}_{\a_{3}\a_{2}\a_{1}}=-(-1)^{|\a_3||\a_2|}\CN_{\a_{2}\a_{3}\a_{1}}
\mod t_H^{n-1}
,\cr
{\CN}_{\a_{3}\a_{2}0}={\CN}_{\a_{3}0\a_1}={\CN}_{0\a_{2}\a_{1}}=0
\mod t_H^{n-1}
,\cr
{\CN}_{\a_3\a_2\a_1}
-(-1)^{|\a_2||\a_1|}{\CN}_{\a_3\a_1\a_2}
+(-1)^{|\a_3|(|\a_2|+|\a_1|)}{\CN}_{\a_2\a_1\a_3}=0\mod t_H^{n-1},
}
$$
which are the classical limits of properties $3,4$ and $5$ in proposition \ref{freshtwo}.
We also recall  
the $3$rd condition for the quantum gauge (property $8$ of $\bos{P}(n)$ as is written in \qgaugex):
\eqn\frany{
\hbar \rd_{[\bar\a_3}\bos{\La}_{\a_2]\a_1}
=
\mathbb{N}_{\a_3\a_2\a_1}
-B_{\a_3\a_2\a_1}{}^\g  \bos{\Theta}_\g
-\bos{K}{X}_{\a_3\a_2\a_1} -\big(\bos{\Theta}, {X}_{\a_3\a_2\a_1}\big)_{\hbar}
\mod t_H^{n-2},
}
which classical limit is
$$
{\CN}_{\a_3\a_2\a_1}
=B_{\a_3\a_2\a_1}{}^\g  {\Theta}_\g
-Q{X}_{\a_3\a_2\a_1} -\big({\Theta}, {X}_{\a_3\a_2\a_1}\big)
\mod t_H^{n-2}.
$$
Hence, by proposition \ref{tech1}, 
there are unique $\tilde{B}_{\a_{3}\a_{2}\a_{1}} \in \Bbbk[[t_{H}]] \mod t_{H}^{n-1}$
and some $\tilde X_{\a_{3}\a_{2}\a_{1}} \in \sC[[t_{H}]]\mod t_H^{n-1}$ defined modulo
$\Ker Q_{\Theta}$  satisfying
\eqn\frand{
{\CN}_{\a_{3}\a_{2}\a_{1}}=\tilde{B}_{\a_{3}\a_{2}\a_{1}}{}^{\g}\Theta_{\g}
+Q \tilde X_{\a_{3}\a_{2}\a_{1}} +\left(\Theta, \tilde X_{\a_{3}\a_{2}\a_{1}}\right)\mod t_{H}^{n-1},
}
such that $\tilde B_{\a_3\a_2\a_1}{}^\g= B_{\a_3\a_2\a_1}\mod t_H^{n-2}$,
$\tilde \La_{\a_3\a_2\a_1}{}^\g= \La_{\a_3\a_2\a_1}\mod t_H^{n-2}$ and
\eqn\frana{
\eqalign{
\tilde{B}_{\a_{3}\a_{2}\a_{1}}{}^{\g}=-(-1)^{|\a_3||\a_2|}\tilde{B}_{\a_{2}\a_{3}\a_{1}}{}^{\g}
\mod t_H^{n-1}
,\cr
\tilde{B}_{0\a_{2}\a_{1}}{}^{\g}=\tilde{B}_{\a_{3}0\a_{1}}{}^{\g}=\tilde{B}_{\a_{3}\a_{2} 0}{}^{\g}=0\mod t_H^{n-1}
,\cr
\tilde{B}_{\a_3\a_2\a_1}
-(-1)^{|\a_2||\a_1|}\tilde{B}_{\a_3\a_1\a_2}
+(-1)^{|\a_3|(|\a_2|+|\a_1|)}\tilde{B}_{\a_2\a_1\a_3}=0
\mod t_H^{n-1}
,\cr
\tilde X_{\a_3\a_2\a_1} = -(-1)^{|\a_3||\a_2|} \tilde X_{\a_3\a_2\a_1} 
\mod t_H^{n-1}
,\cr
\tilde{X}_{0\a_{2}\a_{1}}=\tilde{X}_{\a_{3}0\a_{1}}=\tilde{X}_{\a_{3}\a_{2} 0}=0\mod t_H^{n-1}
,\cr
\tilde{X}_{\a_3\a_2\a_1}
-(-1)^{|\a_2||\a_1|}\tilde{X}_{\a_3\a_1\a_2}
+(-1)^{|\a_3|(|\a_2|+|\a_1|)}\tilde{X}_{\a_2\a_1\a_3}=0
\mod t_H^{n-1}.
}
}

It follows, from \frand, that
the following  expression
$$
\mathbb{N}_{\a_{3}\a_{2}\a_{1}}-\tilde{B}_{\a_{3}\a_{2}\a_{1}}{}^{\g}\bos{\Theta}_{\g}
-\bos{K}\tilde{X}_{\a_{3}\a_{2}\a_{1}} 
-\left(\bos{\Theta},\tilde{X}_{\a_{3}\a_{2}\a_{1}}\right)_{\hbar}
\mod t_{H}^{n-1},
$$
is  divisible by $\hbar$. Hence we can define
$\tilde\mathbb{Y}_{\a_{3}\a_{2}\a_{1}}$ by the following formula:
\eqn\franc{
\hbar \tilde\mathbb{Y}_{\a_{3}\a_{2}\a_{1}}
:=
\mathbb{N}_{\a_{3}\a_{2}\a_{1}}-\tilde{B}_{\a_{3}\a_{2}\a_{1}}{}^{\g}\bos{\Theta}_{\g}
-\bos{K}\tilde{X}_{\a_{3}\a_{2}\a_{1}} 
-\left(\bos{\Theta},\tilde{X}_{\a_{3}\a_{2}\a_{1}}\right)_{\hbar}
\mod t_{H}^{n-1},
}
such that $\tilde\mathbb{Y}_{\a_3\a_2\a_1} = \rd_{[\bar\a_3}\bos{\La}_{\a_2]\a_1} \mod t_H^{n-2}$.

Now we are ready to check all the properties in order.

1.
Apply $\bos{K}_{\bos{\Theta}}$ to \franc\
to obtain that
$$
\hbar\bos{K} \tilde\mathbb{Y}_{\a_{3}\a_{2}\a_{1}}
+\hbar\left(\bos{\Theta}, \tilde\mathbb{Y}_{\a_{3}\a_{2}\a_{1}}\right)_{\hbar}
=\bos{K} \mathbb{N}_{\a_{3}\a_{2}\a_{1}}+\left(\bos{\Theta}, \mathbb{N}_{\a_{3}\a_{2}\a_{1}}\right)_{\hbar}
\mod t_{H}^{n-1}.
$$
Then, 
from  property $6$ in proposition \ref{freshtwo}, 
we conclude that
$$
\bos{K} \tilde\mathbb{Y}_{\a_{3}\a_{2}\a_{1}}+\left(\bos{\Theta}, \tilde\mathbb{Y}_{\a_{3}\a_{2}\a_{1}}\right)_{\hbar}
= \mathbb{M}_{[\a_{3}\a_{2}]\a_{1}}\mod t_{H}^{n-1}.
$$

2. 
From  property $3$ in proposition \ref{freshtwo} and the relations in \frana,
we conclude that 
$$
\tilde\mathbb{Y}_{\a_3\a_2\a_1} =-(-1)^{|\a_3||\a_2|}\tilde\mathbb{Y}_{\a_2\a_3\a_1}\mod t_{H}^{n-1}.
$$

3.
From property $4$ in proposition \ref{freshtwo} and the relations in \frana,
we conclude that 
$$
\tilde\mathbb{Y}_{0\a_2\a_1} =\tilde\mathbb{Y}_{\a_3 0\a_1} =\tilde\mathbb{Y}_{\a_3\a_2 0} =0\mod t_{H}^{n-1}.
$$

4.
From property $5$ in proposition \ref{freshtwo} and the relations in \frana,
we conclude that 
$$
\tilde\mathbb{Y}_{\a_3\a_2\a_1}
-(-1)^{|\a_2||\a_1|}\tilde\mathbb{Y}_{\a_3\a_1\a_2}
+(-1)^{|\a_3|(|\a_2|+|\a_1|)}\tilde\mathbb{Y}_{\a_2\a_1\a_3}
=0 
\mod t_{H}^{n-1}
.
$$
 \qed
 \end{proof}

Now we are ready to establish the potentiality of $\tilde A_{\a_2\a_1}{}^\g$.

\begin{lemma} 
\label{freshfour}

Potentiality: 
$\rd_{\a_{3}}\tilde A_{\a_{2}\a_{1}}{}^{\g}-(-1)^{|\a_{3}||\a_{2}}
\rd_{\a_{2}}\tilde A_{\a_{3}\a_{1}}{}^{\g}=0
\mod t_{H}^{n-1}$.
\end{lemma}

\begin{proof}
Recall the the relation \franx:
$$
\eqalign{
\CM_{[\a_3\a_2]\a_1}
=
&\left(\rd_{\a_3}\tilde A_{\a_{2}\a_{1}}{}^{\g}-(-1)^{|\a_3||\a_2|}
\rd_{\a_2}\tilde A_{\a_{3}\a_{1}}{}^{\g}\right)
\Theta_{\g} 
\cr
&
+Q_{\Theta}\left((-1)^{|\a_3|}\rd_{\a_3}\tilde\La_{\a_2\a_1}
-(-1)^{|\a_2|+|\a_3||\a_2|}\rd_{\a_3}\tilde\La_{\a_2\a_1}\right)
\mod t_H^{n-1}.
}
$$
Now the classical limit of property $1$ in proposition \ref{freshthree}
gives
\eqn\frany{
\CM_{[\a_3\a_2]\a_1}= Q_{\Theta} \tilde Y_{\a_3\a_2\a_1} \mod t_H^{n-1}
}
where $Y_{\a_3\a_2\a_1}= \mathbb{Y}_{\a_3\a_2\a_1}\big|_{\hbar=0}$.
Comparing \franx\ with \frany, we have
$$
\eqalign{
\Big(\rd_{\a_3}\tilde A_{\a_{2}\a_{1}}{}^{\g}-&(-1)^{|\a_3||\a_2|}
\rd_{\a_2}\tilde A_{\a_{3}\a_{1}}{}^{\g}\Big)
\Theta_{\g} 
\cr
&
=
Q_\Theta\left(
\tilde Y_{\a_3\a_2\a_1} -(-1)^{|\a_3|}\rd_{\a_3}\tilde\La_{\a_2\a_1}
+(-1)^{|\a_2|+|\a_3||\a_2|}\rd_{\a_3}\tilde\La_{\a_2\a_1}\right)
\mod t_H^{n-1}.
}
$$
Applying proposition \ref{tech1}, we conclude that
$$
\rd_{\a_3}\tilde A_{\a_{2}\a_{1}}{}^{\g}-(-1)^{|\a_3||\a_2|}
\rd_{\a_2}\tilde A_{\a_{3}\a_{1}}{}^{\g}=0\mod t_H^{n-1}.
$$
\qed
\end{proof}

\begin{corollary} \label{babystep}

There is a choice such that
$$
\tilde Y_{\a_3\a_2\a_1} =(-1)^{|\a_3|}\rd_{\a_3}\tilde\La_{\a_2\a_1}
-(-1)^{|\a_2|+|\a_3||\a_2|}\rd_{\a_3}\tilde\La_{\a_2\a_1}
\mod t_H^{n-1}.
$$
\end{corollary}

\subsubsection{Definition of $\;\tilde\bos{\Theta} \mod t_H^{n+2}$.}

Now we are about to take the final step, which requires
one more proposition, corresponding to the quantum gauge in $\bos{P}(n+1)$: 

\begin{proposition}\label{painful}
There is a $\tilde \bos{\La}_{\a_2\a_1}\in \sC[[t_H]]^{|\a_2|+|\a_1|-1}\mod t_H^n$ such that

1.  $\tilde \bos{\La}_{\a_2\a_1}\big|_{\hbar=0}=\tilde\La_{\a_2\a_1}
$,

2. $\tilde \bos{\La}_{\a_2\a_1}=\bos{\La}_{\a_2\a_1}\mod t_H^{n-1}$,

3. $\tilde \bos{\La}_{\a_2\a_1}=(-1)^{|\a_2||\a_1|}\tilde\bos{\La}_{\a_1\a_2}
$,

4. $\tilde \bos{\La}_{\a_2 0}=0
$,

5. $\tilde\mathbb{Y}_{\a_3\a_2\a_1} =\rd_{[\a_3}\tilde\bos{\La}_{\a_2]\a_1}
\mod t_H^{n-1}$.
\end{proposition}

\begin{proof}

From the condition in proposition \ref{freshthree} that
$$
\tilde\mathbb{Y}_{\a_3\a_2\a_1} =\rd_{[\a_3}\bos{\La}_{\a_2]\a_1}
\mod t_H^{n-2},
$$
we are  looking for a $\tilde \bos{\La}^{[n-1]}_{\a_2\a_1}$ satisfying

$(a)$.
$\tilde \bos{\La}^{[n-1]}_{\a_2\a_1}\big|_{\hbar}=\tilde {\La}^{[n-1]}_{\a_2\a_1}$

$(b)$.
 $\tilde \bos{\La}_{\a_2\a_1}^{[n-1]}=(-1)^{|\a_2||\a_1|}\tilde\bos{\La}_{\a_1\a_2}^{[n-1]}
$,

$(c)$. 
$\tilde \bos{\La}_{\a_2 0}^{[n-1]}=0
$,

$(d)$.
$\tilde\mathbb{Y}_{\a_3\a_2\a_1}^{[n-2]} = \rd_{\a_3}\tilde\bos{\La}_{\a_2\a_1}^{[n-1]}
$.

Assuming the above conditions, we set 
$\tilde\bos{\La}_{\a_2\a_1}= \bos{\La}_{\a_2\a_1} + \tilde \bos{\La}^{[n-1]}_{\a_2\a_1}$.
Then all the properties in the proposition are satisfied.

Consider
\eqn\relv{
\mathbb{C}_{\a_2\a_1}:= \bos{\Theta}_{\a_2}\cdot\bos{\Theta}_{\a_1} - A_{\a_2\a_1}{}^\g\bos{\Theta}_\g
-\big(\bos{\Theta},\bos{\La}_{\a_2\a_1}\big)_{\hbar} \mod t_H^n.
}
Note that the classical limit of $\mathbb{C}_{\a_2\a_1}$ is
$$
\mathbb{C}_{\a_2\a_1}\big|_{\hbar=0}=
\Theta_{\a_2}\cdot \Theta_{\a_1}- A_{\a_2\a_1}{}^\g \Theta_\g
-\big(\Theta, \La_{\a_2\a_1}\big) \mod t_H^n.
$$
Comparing above with
$
\Theta_{\a_2}\cdot \Theta_{\a_1}= \tilde A_{\a_2\a_1}{}^\g \Theta_\g
+Q \tilde \La_{\a_2\a}+\big(\Theta, \La_{\a_2\a_1} \big) \mod t_H^n
$,
we have
\eqn\relvxa{
\mathbb{C}_{\a_2\a_1}\big|_{\hbar=0} = \tilde A_{\a_2\a_1}^{[n-1]\g} O_\g + Q \tilde\La_{\a_2\a_1}
\mod t_H^n
.
}
Note also that
$$
\eqalign{
\rd_{[\a_3}\mathbb{C}_{\a_2]\a_1} 
=& \mathbb{M}_{[a_3\a_2]\a_1} -\left(\bos{\Theta}, \rd_{\bar\a_3}\bos{\La}_{\a_2]\a_1}\right)_{\hbar}
\cr
=&
\bos{K}\tilde\mathbb{Y}_{\a_3\a_2\a_1}
+\left(\bos{\Theta}, \tilde\mathbb{Y}_{\a_3\a_2\a_1}-\rd_{\bar\a_3}\bos{\La}_{\a_2]\a_1}\right)_{\hbar}
\cr
&\mod t_H^{n-1}
}
$$
where we have used 
$ \mathbb{M}_{[a_3\a_2]\a_1} = \bos{K}\tilde\mathbb{Y}_{\a_3\a_2\a_1}
+\left(\bos{\Theta}, \tilde\mathbb{Y}_{\a_3\a_2\a_1}\right)\mod t_H^{n-1}$ 
(property $1$ in proposition \ref{freshthree}).
It follows that
\eqn\relvxb{
\rd_{[\a_3}\mathbb{C}_{\a_2]\a_1} = \bos{K}\tilde\mathbb{Y}_{\a_3\a_2\a_1}\mod t_H^{n-1},
}
since
$\tilde\mathbb{Y}_{\a_3\a_2\a_1} -\rd_{[\bar\a_3}\bos{\La}_{\a_2]\a_1}=0\mod t_H^{n-2}$
and $\bos{\Theta}\big|_{t_H=0}=0$.

Now consider the word-length $(n-1)$ part $\mathbb{C}^{[n-1]}_{\a_{2}\a_{1}}$ of $\mathbb{C}_{\a_{2}\a_{1}}$: 
\eqn\relva{
\eqalign{
\mathbb{C}^{[n-1]}_{\a_{2}\a_{1}}=
&
\sum_{k=0}^{n-1}
\bos{\Theta}^{[k]}_{\a_{2}}\cdot\bos{\Theta}^{[n-1-k]}_{\a_{1}}
-\sum_{k=0}^{n-2}  A_{\a_2\a_1}^{[k]\g} \bos{\Theta}_\g^{[n-1-k]}
-\sum_{k=0}^{n-2} \left(\bos{\Theta}^{[n-1-k]}, \bos{\La}^{[k]}_{\a_{2}\a_{1}}\right)_{\hbar}.
}
}
We note, from \relvxa\ and \relvxb, that
\eqn\relvb{
\eqalign{
\mathbb{C}^{[n-1]}_{\a_{2}\a_{1}}\big|_{\hbar=0}
&=\tilde A^{[n-1]}_{\a_2\a_1}{}^\g O_\g +Q \tilde\La^{[n-1]}_{\a_2\a_1}
,\cr
\rd_{[\a_3}\mathbb{C}^{[n-1]}_{\a_2]\a_1} &= \bos{K}\tilde\mathbb{Y}^{[n-2]}_{\a_3\a_2\a_1}
.
}
}
Hence $\tilde\mathbb{Y}^{[n-2]}_{\a_3\a_2\a_1}
-\rd^{\phantom{[]}}_{[\bar\a_3}\tilde\La^{[n-1]}_{\a_2]\a_1}$ is divisible by $\hbar$.

Now we consider the following expansions:
$$
\eqalign{
\tilde\mathbb{Y}^{[n-2]}_{\a_3\a_2\a_1}
&=\Fr{1}{(n-2)!}t^{\bar\a_4}\cdots t^{\bar\a_{n+1}}
\tilde\bos{\xi}_{\a_{n+1}\cdots \a_4\a_3\a_2\a_1}
\hbox{ where } \bos{\xi}_{\a_{n+1}\cdots \a_1}\in \sC[[\hbar]]^{|\a_1|+\cdots +|\a_{n+1}|-1}
,\cr
\tilde\La^{[n-1]}_{\a_2\a_1}
&=\Fr{1}{(n-1)!}t^{\bar\a_3}\cdots t^{\bar\a_{n+1}}\tilde\l_{\a_{n+1}\a_n\cdots \a_3\a_2\a_1}{}^\g
\hbox{ where } \tilde{\l}_{\a_{n+1}\cdots \a_1}\in \sC^{|\a_1|+\cdots +|\a_{n+1}|-1}
,
\cr
}
$$
such that
$$
\rd^{\phantom{n}}_{[\bar\a_3}\tilde\La^{[n-1]}_{\a_2]\a_1}
=\Fr{1}{(n-2)}t^{\bar\a_4}\cdots t^{\bar\a_{n+1}}
\left(
\tilde\l_{\a_{n+1}\a_n\cdots\a_4 \a_3\a_2\a_1}{}^\g
-(-1)^{|\a_3||\a_2|}\tilde\l_{\a_{n+1}\a_n\cdots\a_4 \a_2\a_3\a_1}{}^\g
\right).
$$
We can define 
$\tilde\bos{\eta}_{\a_{n+1}\a_n\cdots\a_4 \a_3\a_2\a_1}\in \sC[[\hbar]]^{|\a_1|+\cdots +|\a_{n+1}|-1}$ 
as follows:
$$
\hbar \tilde\bos{\eta}_{\a_{n+1}\a_n\cdots\a_4 \a_3\a_2\a_1}
=
\tilde\bos{\xi}_{\a_{n+1}\a_n\cdots\a_4 \a_3\a_2\a_1} - 
\left(\l_{\a_{n+1}\a_n\cdots\a_4 \a_3\a_2\a_1}
-(-1)^{|\a_3||\a_2|}\tilde\l_{\a_{n+1}\a_n\cdots\a_4 \a_2\a_3\a_1}{}^\g
\right),
$$
since $\tilde\mathbb{Y}^{[n-2]}_{\a_3\a_2\a_1}
-\rd^{\phantom{[]}}_{[\bar\a_3}\tilde\La^{[n-1]}_{\a_2]\a_1}$ is divisible by $\hbar$.
Then
$$
\tilde\mathbb{Y}^{[n-2]}_{\g\b\a}=
\rd^{\phantom{[]}}_{[\bar\g}\tilde{\La}^{[n-1]}_{\b]\a} 
-\Fr{\hbar}{(n-2)!}t^{\bar\r_2}\cdots t^{\bar\r_{n-1}}
\tilde\bos{\eta}_{\r_{n-1}\cdots\r_2\g \b \a}.
$$
Note also that
$$
\eqalign{
\tilde\bos{\eta}_{\r_{n-1}\cdots\r_2\g \b \a}-(-1)^{|\g||\b|}\tilde\bos{\eta}_{\r_{n-1}\cdots\r_2\b\g \a}
&=2\tilde\bos{\eta}_{\r_{n-1}\cdots\r_2\g \b \a},
\cr
(-1)^{|\b||\a|}\tilde\bos{\eta}_{\r_{n-1}\cdots\r_2 \g\a\b}
-(-1)^{|\g|(|\b|+|\a|)}\tilde\bos{\eta}_{\r_{n-1}\cdots\r_2\b \a \g}
&=\tilde\bos{\eta}_{\r_{n-1}\cdots\r_2\g \b \a}.
}
$$

Finally, we define
$$
\tilde\bos{\La}^{[n-1]}_{\b\a} := \tilde \La^{n-1}_{\b\a} -\Fr{\hbar}{3}\Fr{1}{(n-1)!}t^{\bar\r_1}\cdots t^{\bar\r_{n-1}}
\left(\tilde\bos{\eta}_{\r_{n-1}\cdots\r_1 \b \a}+(-1)^{|\b||\a|}\tilde\bos{\eta}_{\r_{n-1}\cdots\r_1 \a \b}\right)
$$
Then, we have

$(a)$
$\tilde \bos{\La}^{[n-1]}_{\a_2\a_1}\big|_{\hbar}=\tilde {\La}^{[n-1]}_{\a_2\a_1}$,

$(b)$
 $\tilde \bos{\La}_{\a_2\a_1}^{[n-1]}=(-1)^{|\a_2||\a_1|}\tilde\bos{\La}_{\a_1\a_2}^{[n-1]}$,

 $(c)$
$\tilde \bos{\La}_{\a_2 0}^{[n-1]}=0$.

Hence all that is remain is to check
that $\tilde\mathbb{Y}_{\a_3\a_2\a_1}^{[n-2]} = 
\rd^{\phantom{[]}}_{[\a_3}\tilde\bos{\La}_{\a_2]\a_1}^{[n-1]}$:
$$
\eqalign{
\rd^{\phantom{[]}}_{[\bar\g}\tilde\bos{\La}^{[n-1]}_{\b]\a} 
=&\rd^{\phantom{[]}}_{[\bar\g}\tilde{\La}^{[n-1]}_{\b]\a} 
-\Fr{\hbar}{3}\Fr{1}{(n-2)!}t^{\bar\r_2}\cdots t^{\bar\r_{n-1}}
\Big(\tilde\bos{\eta}_{\r_{n-1}\cdots\r_2\g \b \a}-(-1)^{|\g||\b|}\tilde\bos{\eta}_{\r_{n-1}\cdots\r_2\b\g \a}
\Big)
\cr
&
-\Fr{\hbar}{3}\Fr{1}{(n-2)!}t^{\bar\r_2}\cdots t^{\bar\r_{n-1}}
\Big((-1)^{|\b||\a|}\tilde\bos{\eta}_{\r_{n-1}\cdots\r_2 \g\a\b}
-(-1)^{|\g|(|\b|+|\a|)}\tilde\bos{\eta}_{\r_{n-1}\cdots\r_2\b \a \g}
\Big)
\cr
=&\rd^{\phantom{[]}}_{[\bar\g}\tilde{\La}^{[n-1]}_{\b]\a} 
-\Fr{2\hbar}{3}\Fr{1}{(n-2)!}t^{\bar\r_2}\cdots t^{\bar\r_{n-1}}
\tilde\bos{\eta}_{\r_{n-1}\cdots\r_2\g \b \a}
\cr
&
-\Fr{\hbar}{3}\Fr{1}{(n-2)!}t^{\bar\r_2}\cdots t^{\bar\r_{n-1}}
\tilde\bos{\eta}_{\r_{n-1}\cdots\r_2\g \b \a}
\cr
=&
\rd^{\phantom{[]}}_{[\bar\g}\tilde{\La}^{[n-1]}_{\b]\a} 
-\Fr{\hbar}{(n-2)!}t^{\bar\r_2}\cdots t^{\bar\r_{n-1}}
\tilde\bos{\eta}_{\r_{n-1}\cdots\r_2\g \b \a}
\cr
=&\tilde\mathbb{Y}_{\g\b\a}^{[n-1]}.
}
$$
\qed.
\end{proof}

\begin{corollary}
Let
$$
\tilde\bos{L}_{\a_{2}\a_{1}}:=
\bos{\Theta}_{\a_{2}}\cdot\bos{\Theta}_{\a_{1}}
-\tilde A_{\a_{2}\a_{1}}{}^{\g}\bos{\Theta}_{\g}
-\bos{K} \tilde\bos{\La}_{\a_{2}\a_{1}}
- \left(\bos{\Theta}, \tilde\bos{\La}_{\a_{2}\a_{1}}\right)_{\hbar}
\mod t_{H}^{n}
.
$$ 
Then

1. $\tilde\bos{L}_{\a_{2}\a_{1}}$ is divisible by $\hbar$,

2. $\tilde\bos{L}_{\a_{2}\a_{1}}=(-1)^{|\a_2||\a_1|}\tilde\bos{L}_{\a_{1}\a_{2}}\mod t_H^{n}$,

3. $\tilde\bos{L}_{\a_{2} 0}=0\mod t_H^n$,

4. $\rd_{\a_3}\tilde\bos{L}_{\a_{2}\a_{1}}=(-1)^{|\a_3||\a_2|}\rd_{\a_2}\tilde\bos{L}_{\a_{3}\a_{1}}\mod t_H^{n-1}$

\end{corollary}

\begin{proof}  
Straightforward to check in order:

1. The classical limit of $\tilde\bos{L}_{\a_2\a_1}$ vanishes due to the relation \beba.

2.  Use the relations $\tilde A_{\a_2\a_1}{}^\g
=(-1)^{|a_2||\a_1|}\tilde A_{\a_1\a_2} \mod t_H^n$  and 
$\tilde \bos{\La}_{\a_2\a_1}=(-1)^{|\a_2||\a_1|}\tilde\bos{\La}_{\a_1\a_2}\mod t_H^{n}$,
as well as supercommutativity of the product $\cdot$. 

3. Use the unities that $\bos{\Theta}_0=1$ and $\tilde A_{0\b}{}^\g=\d_{\b}{}^\g \mod t_H^n$
as well as the relation $\tilde\bos{\La}_{\a_{2} 0}=0$.

4. By applying $\rd_{\a_3}$ we have
$$
\rd_{\a_3}\tilde\bos{L}_{\a_{2}\a_{1}}=
\mathbb{M}_{\a_3\a_2\a_1}
-\rd_{\a_3}\tilde A_{\a_{2}\a_{1}}{}^{\g}\bos{\Theta}_{\g}
-\bos{K}_{\bos{\Theta}} \rd_{\bar\a_3}\tilde\bos{\La}_{\a_{2}\a_{1}}
\mod t_{H}^{n-1}.
$$
Hence
$$
\eqalign{
\rd_{\a_3}\tilde\bos{L}_{\a_{2}\a_{1}}-&(-1)^{|\a_3||\a_2|}\rd_{\a_2}\tilde\bos{L}_{\a_{3}\a_{1}}
\cr
&=\mathbb{M}_{[\a_3\a_2]\a_1}
-\bos{K}_{\bos{\Theta}}\rd_{[\bar\a_3}\tilde\bos{\La}_{\a_{2}]\a_{1}}
\cr
&=
-\bos{K}_{\bos{\Theta}}\left(\bos{Y}_{\a_3\a_2\a_1}-\rd_{[\bar\a_3}\tilde\bos{\La}_{\a_{2}]\a_{1}}\right)
\cr
&=0,
}
$$
where we have used the potentiality 
$\rd_{\a_3}\tilde A_{\a_2\a_1}{}^\g - (-1)^{|\a_3||\a_2|}\rd_{\a_2}\tilde A_{\a_3\a_1}{}^\g=0 \mod t_H^{n-1}$ 
for the $1$st equality and property $5$ in proposition \ref{painful}.
\qed
\end{proof}

{}From the quantum master equation in $\bos{P}(n)$, we already know that
$$
\tilde\bos{L}_{\a_{2}\a_{1}}=\hbar \rd_{\a_2}\rd_{\a_1}\bos{\Theta} \mod t_H^{n-1},
$$
so that the only new piece of information in the expression $\tilde\bos{L}_{\a_2\a_1}$
is its component $\tilde\bos{L}^{[n-1]}_{\a_{2}\a_{1}}$ with the word-length $n-1$ in $t_{H}$:
\eqn\bebf{
\eqalign{
\tilde\bos{L}^{[n-1]}_{\a_{2}\a_{1}}=
&
\sum_{k=0}^{n-1}
\bos{\Theta}^{[k]}_{\a_{2}}\cdot\bos{\Theta}^{[n-1-k]}_{\a_{1}}
-\sum_{k=0}^{n-2} 
\Big( A_{\a\b}^{[k]\g} \bos{\Theta}_\g^{[n-1-k]}
+\left(\bos{\Theta}^{[n-1-k]}, \bos{\La}^{[k]}_{\a_{2}\a_{1}}\right)_{\hbar}
\Big)
\cr
&
-\tilde A_{\a_{2}\a_{1}}^{[n-1]\g}\bos{\Theta}_{\g}^{[0]}
-\bos{K}\tilde\bos{\La}^{[n-1]}_{\a_{2}\a_{1}}.
}
}
Set $\hbar\tilde\bos{\Theta}^{[n+1]} = \Fr{1}{n(n+1)}t^{\a_1}t^{\a_2} \tilde\bos{L}^{[n-1]}_{\a_2\a_1}$
and define 
$$
\eqalign{
\tilde\bos{\Theta} :=&\bos{\Theta} + \tilde\bos{\Theta}^{[n+1]}
=\bos{\Theta}^{[1]} + \cdots +\bos{\Theta}^{[n]} +\tilde\bos{\Theta}^{[n+1]}
}
$$
so that
$\tilde\bos{\Theta} =\bos{\Theta}\mod t_H^{n+1}$
and $\rd_\a\tilde\bos{\Theta} =\bos{\Theta}_\a\mod t_H^{n}$.
Then the following lemma finishes our construction $\bos{P}(n+1)$ out of the assumed $\bos{P}(n)$.

\begin{lemma}
$\tilde \bos{\Theta}$ satisfies

1. the $\mod t_H^{n}$ quantum master equation:
$$
\hbar\rd_{\a_{2}}\rd_{\a_1}\tilde\bos{\Theta}=
\rd_{\a_{2}}\tilde\bos{\Theta}\cdot\rd_{\a_{1}}\tilde\bos{\Theta}
-\tilde A_{\a_{2}\a_{1}}{}^{\g}\rd_{\g}\tilde\bos{\Theta}
-\bos{K}\tilde\bos{\La}_{\a_{2}\a_{1}} - \left(\tilde\bos{\Theta}, \tilde\bos{\La}_{\a_{2}\a_{1}}\right)_{\hbar}
\mod t_H^{n},
$$

2.  the quantum unity $\rd_0 \tilde\bos{\Theta}=1 \mod t_H^{n+1}$,

3. the $\mod t_H^{n+2}$ quantum descendant equation:
$\bos{K}\tilde\bos{\Theta} +\Fr{1}{2}\big(\tilde\bos{\Theta},\tilde\bos{\Theta}\big)_{\hbar}
=0\mod t_H^{n+2}$.

\end{lemma}

\begin{proof} In order.

1.
It is sufficient to show that $\hbar \rd_{\a_2}\rd_{\a_1}\tilde\bos{\Theta}=\tilde\bos{L}_{\a_2\a_1}^{[n-1]}$,
which is combined  with the quantum master equation $\mod t_H^{n-1}$ in $\bos{P}(n)$ to have
the $\mod t_H^{n}$ equation. 
From $\tilde\bos{\Theta}^{[n+1]} = \Fr{1}{n(n+1)}t^{\b}t^{\g} \tilde\bos{L}^{[n-1]}_{\g\b}$,
we have
$$
\eqalign{
\hbar\rd_{\a_1}&\tilde\bos{\Theta}^{[n+1]} 
\cr
= 
&
\Fr{2}{n(n+1)}t^{\g} \tilde\bos{L}^{[n-1]}_{\g\a_1}
+\Fr{1}{n(n+1)}(-1)^{(|\m|+|\n|)|\a_1|}t^{\n}t^{\m} \rd_{\a_1}\tilde\bos{L}^{[n-1]}_{\m\n}
\cr
= 
&
\Fr{2}{n(n+1)}t^{\g} \tilde\bos{L}^{[n-1]}_{\g\a_1}
+\Fr{1}{n(n+1)}t^{\n}t^{\m}\left((-1)^{|\n||\a_1|} \rd_\m\tilde\bos{L}^{[n-1]}_{\a_1\n}
+(-1)^{(|\m|+|\n|)|\a_1|} \rd_{[\a_1}\tilde\bos{L}^{[n-1]}_{\m]\n}\right)
\cr
= 
&
\Fr{2}{n(n+1)}t^{\g} \tilde\bos{L}^{[n-1]}_{\g\a_1}
+\Fr{1}{n(n+1)}t^{\n}t^{\m} \rd_\m\tilde\bos{L}^{[n-1]}_{\n\a_1}
+\Fr{1}{n(n+1)}(-1)^{(|\m|+|\n|)|\a_1|}t^{\n}t^{\m} \rd_{[\a_1}\tilde\bos{L}^{[n-1]}_{\m]\n}
\cr
= 
&
\Fr{1}{n}t^{\g} \tilde\bos{L}^{[n-1]}_{\g\b}
+\Fr{1}{n(n+1)}(-1)^{(|\m|+|\n||\a_1|}t^{\n}t^{\m} \rd_{[\a_1}\tilde\bos{L}^{[n-1]}_{\m]\n},
}
$$
where we have used $\tilde\bos{L}^{[n-1]}_{\a_2\a_1}=(-1)^{|\a_2||\a_1|}\tilde\bos{L}_{\a_1\a_2}^{[n-1]}$
for the $1$st and the $3$rd equalities and 
$t^\m\rd_\m \bos{L}^{[n-1]}_{\n\a_1}=(n-1)\bos{L}^{[n-1]}_{\n\a_1}$
for the last equality. It follows that
$\rd_{\a_1}\tilde\bos{\Theta}^{[n+1]} =\Fr{1}{n}t^{\g} \bos{L}^{[n-1]}_{\g\a_1}$
since  $\rd_{[\a_1}\bos{L}^{[n-1]}_{\m]\n}=0$.
The similar computation
show that $\hbar \rd_{\a_2}\rd_{\a_1}\tilde\bos{\Theta}=\tilde\bos{L}_{\a_2\a_1}^{[n-1]}$..

2.
The quantum unity $\rd_0\tilde\bos{\Theta}=1\mod t_H^{n+1}$ follows from
$\rd_0\tilde\bos{\Theta}=\rd_0\bos{\Theta}=1\mod t_H^{n}$ 
and $\hbar\rd_0 \tilde\bos{\Theta}^{[n+1]}=\Fr{1}{n}t^\g\tilde\bos{L}^{[n-1]}_{\g 0}=0$.

3.
In $\bos{P}(n)$ we have assumed that
$
\bos{K}\bos{\Theta} +\Fr{1}{2}\big(\bos{\Theta},\bos{\Theta}\big)_{\hbar}
=\bos{K}\tilde\bos{\Theta} +\Fr{1}{2}\big(\tilde\bos{\Theta},\tilde\bos{\Theta}\big)_{\hbar}
 =0\mod t_H^{n+1}$.
So it is suffice to show that
\eqn\targetx{
\bos{K}\tilde\bos{\Theta}^{[n+1]}
+\Fr{1}{2}\sum_{k=1}^{n}\Big(\bos{\Theta}^{[n+1-k]},\bos{\Theta}^{[k]}\Big)_{\hbar} =0.
}
Note that the quantum master equation in $1$ can be written as follows:
$$
\hbar\rd_{\a_{2}}\rd_{\a_1}\tilde\bos{\Theta}=
\rd_{\a_{2}}\bos{\Theta}\cdot\rd_{\a_{1}}\bos{\Theta}
-\tilde A_{\a_{2}\a_{1}}{}^{\g}\rd_{\g}\bos{\Theta}
-\bos{K}\tilde\bos{\La}_{\a_{2}\a_{1}} - \left(\bos{\Theta}, \tilde\bos{\La}_{\a_{2}\a_{1}}\right)_{\hbar}
\mod t_H^{n}.
$$
Applying $\bos{K}_{\bos{\Theta}}$ to the above, we have
$$
\bos{K}\rd_{\a_{2}}\rd_{\a_1}\tilde\bos{\Theta}
+\left(\bos{\Theta}, \rd_{\a_{2}}\rd_{\a_1}\tilde\bos{\Theta}\right)_{\hbar}
=
-\hbar(-1)^{|\a_2|}\Big(\rd_{\a_{2}}\bos{\Theta},\rd_{\a_{1}}\bos{\Theta}\Big)_{\hbar}
\mod t_H^{n}.
$$
Now consider  the word length $(n-1)$ part of  the above identity:
$$
\bos{K} \left(\rd_{\a_{2}}\rd_{\a_{1}}\tilde\bos{\Theta}^{[n+1]}\right)
+\sum_{k=1}^{n}\left(\bos{\Theta}^{[n+1-k]}, \rd_{\a_{2}}\rd_{\a_{1}}\bos{\Theta}^{[k]}\right)_{\hbar}
=
-(-1)^{|\a_{2}|}\sum_{k=1}^{n}\left(\rd_{\a_{2}}\bos{\Theta}^{[n+1-k]}
\rd_{\a_{1}}\bos{\Theta}^{[k]}\right)_{\hbar},
$$
and multiply $(-1)^{|\a_{2}|+|\a_{1}|}t^{\a_{1}}t^{\a_{2}}$ to sum over the repeated indices
such that
$$
\eqalign{
\bos{K} \left(t^{\a_{1}}t^{\a_{2}}\rd_{\a_{2}}\rd_{\a_{1}}\tilde\bos{\Theta}^{[n+1]}\right)
+&\sum_{k=1}^{n}
\left(\bos{\Theta}^{[n+1-k]}, t^{\a_{1}}t^{\a_{2}}\rd_{\a_{2}}\rd_{\a_{1}}\bos{\Theta}^{[k]}\right)_{\hbar}
\cr
&
=-\sum_{k=1}^{n}\left(t^{\a_{2}}\rd_{\a_{2}}\bos{\Theta}^{[n+1-k]}, 
t^{\a_{1}}\rd_{\a_{1}}\bos{\Theta}^{[k]}\right)_{\hbar}.
}
$$
Then, from the homogeneity of $\bos{\Theta}^{[k]}$ 
such that $t^\a \rd_\a \bos{\Theta}^{[k]}=k\bos{\Theta}^{[k]}$,
we obtain that
$$
n(n+1)\bos{K} \tilde\bos{\Theta}^{[n+1]}
+\sum_{k=1}^{n}k(k-1)\left(\bos{\Theta}^{[n+1-k]}, \bos{\Theta}^{[k]}\right)_{\hbar}
=-\sum_{k=1}^{n}(n+1-k)k\left(\bos{\Theta}^{[n+1-k]}, \bos{\Theta}^{[k]}\right)_{\hbar},
$$
which is simplified as follows:
$$
n(n+1)\bos{K} \tilde\bos{\Theta}^{[n+1]}
+\sum_{k=1}^{n}k\left(\bos{\Theta}^{[n+1-k]}, \bos{\Theta}^{[k]}\right)_{\hbar}
=0.
$$
Using the following re-summation:
$$
\eqalign{
\sum_{k=1}^{n}k\left(\bos{\Theta}^{[n+1-k]}, \bos{\Theta}^{[k]}\right)_{\hbar}
&=\sum_{k=1}^{n}(n+1-k)\left(\bos{\Theta}^{k},\bos{\Theta}^{[n+1-k]}\right)_{\hbar}
\cr
&=\sum_{k=1}^{n}(n+1-k)\left(\bos{\Theta}^{[n+1-k]}, \bos{\Theta}^{[k]}\right)_{\hbar},
}
$$
we have 
$$
(n+1)\bos{K} \tilde\bos{\Theta}^{[n+1]}
+\Fr{(n+1)}{2}\sum_{k=1}^{n}\left(\bos{\Theta}^{[n+1-k]}, \bos{\Theta}^{[k]}\right)_{\hbar}
=0,
$$
which is equivalent the relation \targetx.
\qed
\end{proof}

Finally take $n\rightarrow \infty$ and we are done.

\appendix
\section{Appendix}
\xdef\nsecsym{A.}

The purpose of this appendix is to prove proposition \ref{flemma} and \ref{flemmb},
which compare
 $\big(\sC[[\hbar]], \bos{K}\big)$
and $\big(H[[\hbar]], \bos{\k}\big)$ both as cochain complexes over $\Bbbk[[\hbar]]$.

Recall that a quantization map $\bos{f}=f +\hbar f^{(1)}+\cdots$ 
is a ghost number preserving $\Bbbk[[\hbar]]$-linear map on $H[[\hbar]]$
to $\sC[[\hbar]]$ satisfying $\bos{K}\bos{f}= \bos{f}\bos{\k}$ such that $Qf=0$, $f(0)=0$, and $f$ induces
 the identity map on $H$, i.e, the $Q$-cohomology class of $f(x)$ is $x$ for all $x \in H$.
Let $\g \in \Ker Q \cap \sC$. We denote the $Q$-cohomology class of $\g$ by $[\g]\in H$.
Note that $f\left([\g]\right) = \g  \mod \hbox{Im }Q$ since $f$ induces the identity map on $H$.
Note also that an equality of the type $f(x) = Q \l$, where $x\in H^{|x|}$ and $\l \in \sC^{|x|-1}$
implies that (i) $x=0$ since $\left[f(x)\right] = \left[Q\l\right]=0$ and $\left[f(x)\right]=x$,
and (ii) $Q\l=0$ since $f(0)=0$.

We begin with proposition \ref{flemma}:

\begin{proposition}
Any homogeneous $\bos{\eta}\in \sC[[\hbar]]^{|\bos{\eta}|}$ satisfying $\bos{K}\bos{\eta}=0$,
can be expressed as
$$
\bos{\eta} =\bos{f}(\bos{x}) + \bos{K} \bos{\l},
$$
for certain pair $\left(\bos{x}, \bos{\l}\right) \in H[[\hbar]]^{|\bos{\eta}|}\oplus \sC[[\hbar]]^{|\bos{\eta}|-1}$
such that $\bos{\k}\bos{x}=0$.
Let $\left(\bos{x}^\pr, \bos{\l}^\pr\right) \in H[[\hbar]]^{|\bos{\eta}|}\oplus \sC[[\hbar]]^{|\bos{\eta}|-1}$
be any other pair satisfying 
$$
\bos{\eta} =\bos{f}(\bos{x}^\pr) + \bos{K} \bos{\l}^\pr.
$$ 
Then there is
certain pair $\left(\bos{y}, \bos{\zeta}\right) \in H[[\hbar]]^{|\bos{\eta}|-1}\oplus \sC[[\hbar]]^{|\bos{\eta}|-2}$
such that
$$
\eqalign{
\bos{x}^\pr - \bos{x} &=-\bos{\k}\bos{y}
,\cr
\bos{\l}^\pr - \bos{\l}&=\bos{f}(\bos{y}) + \bos{K}\bos{\zeta}
.
}
$$
\end{proposition}

Consider any $\bos{\eta} \in \sC[[\hbar]]$ satisfying $\bos{K}\bos{\eta}=0$.
We shall show that (i) there is a pair $(\bos{x},\bos{\l})$ such that
$\bos{\eta} =\bos{f}(\bos{x}) +\bos{K}\bos{\l}$,
(ii) $\bos{\k}\bos{x}=0$ and (iii) for any other $(\bos{x}^\pr, \bos{\l}^\pr)$
satisfying $\bos{\eta} =\bos{f}(\bos{x}^\pr) +\bos{K}\bos{\l}^\pr$ 
there is a pair $(\bos{y},\bos{\eta})$ such that 
$\bos{x}^\pr - \bos{x} =-\bos{\k} \bos{y}$  and $\bos{\l}^\pr - \bos{\l}=\bos{f}(\bos{y}) + \bos{K}\bos{\zeta}$.
Denote $\bos{\eta}=\eta^{(0)} +\hbar \eta^{(1)} +\cdots$. 

We begin with proving the proposition modulo $\hbar$.
Note that
the condition $\bos{K}\bos{\eta}=0$ modulo $\hbar$ is $Q \eta^{(0)} =0$.
Then,

(i) It follows that
$
\eta^{(0)}=f\left(\left[\eta^{(0)}\right]\right) + Q \l^{(0)}
$
for some $\l^{(0)} \in \sC$. Set $x^{(0)} =\left[\eta^{(0)}\right]$, i.e.,
\eqn\tangoa{
\eta^{(0)}=f\left(x^{(0)}\right) + Q \l^{(0)}.
}
Set
$$
\eqalign{
\bos{x}=x^{(0)} \mod \hbar
,\cr
\bos{\l}=\l^{(0)}\mod\hbar
.
}
$$
Then the relation \tangoa\ is equivalent to $\bos{\eta} =\bos{f}(\bos{x}) +\bos{K}\bos{\l}\mod \hbar$.

(ii) It is obvious that $\bos{\k} \bos{x} =0 \mod \hbar$ since $\bos{\k}=0 \mod \hbar$.

(iii) Let $(x^{\pr(0)},\l^{\pr (0)})$ be any other pair satisfying 
$\eta^{(0)}=f\left(x^{\pr(0)}\right) + Q \l^{\pr(0)}$. Then, by a comparison with  \tangoa,
we have 
$$
f\left( x^{\pr(0)}-x^{(0)}\right)= -Q\left(\l^{\pr(0)}-\l^{(0)}\right).
$$
By taking the $Q$-cohomology class to the above, we conclude that
$x^{\pr(0)}=x^{(0)}$ as well as $Q\left(\l^{\pr(0)}-\l^{(0)}\right)=0$,
which implies that there is a pair $(y^{(0)}, \zeta^{(0)})$ such that
$\l^{\pr(0)}-\l^{(0)}= f\left(y^{(0)}\right) + Q\zeta^{(0)}$:
$$
\eqalign{
x^{\pr(0)}-x^{(0)}&=0
,\cr
\l^{\pr(0)}-\l^{(0)}&= f\left(y^{(0)}\right) + Q\zeta^{(0)}.
}
$$
The above set of relations is equivalent to 
$$
\eqalign{
\bos{x}^\pr - \bos{x} &=-\bos{\k}\bos{y}\mod \hbar
,\cr
\bos{\l}^\pr - \bos{\l}&=\bos{f}(\bos{y}) + \bos{K}\bos{\zeta}
,\cr
}
$$
where
$$
\eqalign{
\bos{x}^\pr=x^{\pr(0)} \mod \hbar
,\cr
\bos{\l}^\pr=\l^{\pr(0)}\mod\hbar
,
}\qquad
\eqalign{
\bos{y}=y^{(0)} \mod \hbar
,\cr
\bos{\zeta}=\zeta^{(0)}\mod\hbar
.
}
$$

The above demonstration is too trivial to be an useful example, though it  is  a necessary step
for mathematical induction. Now the next order is demonstrated for
pedagogical purposes---since it has all the essential features of the
proposition, some readers may read it and skip the actual proof.

Consider the proposition modulo $\hbar^2$.
Then
the condition $\bos{K}\bos{\eta}=0 \mod\hbar^2$ is $Q \eta^{(0)} =0$
and $Q\eta^{(1)} + K^{(1)}\eta^{(0)}=0$, which becomes
$$
Q\eta^{(1)} +K^{(1)}f\left(x^{(0)}\right) + K^{(1)}Q \l^{(0)}=0,
$$
after using \tangoa. From $\bos{K}\bos{f}=\bos{f}\bos{\k}$ and $\bos{K}^2=0$, we have
$K^{(1)}f=-Q f^{(1)} +f\k^{(1)}$ and $ K^{(1)}Q =-Q K^{(1)}$. Thus the above condition is equivalent to
\eqn\tangob{
Q\left(\eta^{(1)} -f^{(1)}\left(x^{(0)}\right) - K^{(1)} \l^{(0)}\right)=-f\left(\k^{(1)}x^{(0)}\right).
}
By taking the $Q$-cohomology class to the above, we have
\eqn\tangobx{
0= \k^{(1)}x^{(0)}.
}
Then \tangob\ also implies that
\eqn\tangoby{
Q\left(\eta^{(1)} -f^{(1)}\left(x^{(0)}\right) - K^{(1)} \l^{(0)}\right)=0.
}

(i) From \tangoby, It follows that
$\eta^{(1)} -f^{(1)}\left(x^{(0)}\right) - K^{(1)} \l^{(0)}=f\left(x^{(1)}\right) + Q \l^{(1)}$ 
for unique $x^{(1)}\in H$ and some $\l^{(1)} \in \sC$. Then, together with \tangoa, we have
\eqn\tangoc{
\eqalign{
\eta^{(0)} &=f\left(x^{(0)}\right) + Q \l^{(0)}
,\cr
\eta^{(1)} &=f^{(1)}\left(x^{(0)}\right) +f\left(x^{(1)}\right) + K^{(1)} \l^{(0)}+ Q \l^{(1)}.
}
}
Let
$$
\eqalign{
\bos{x}:=x^{(0)} +\hbar x^{(1)} \mod \hbar^2
,\cr
\bos{\l}:=\l^{(0)}+\hbar\l^{(1)}\mod\hbar^2.
}
$$
Then the relation \tangoc\ is equivalent to
$\bos{\eta} =\bos{f}(\bos{x}) +\bos{K}\bos{\l}\mod \hbar^2$.

(ii) From \tangobx. it follows  $\bos{\k} \bos{x} =\k^{(1)}x^{(0)}=0 \mod \hbar^2$ since $\bos{\k}=0 \mod \hbar$.

(iii) Recall that $\l^{(0)}$ is defined modulo $\Ker Q$ in \tangoa. Let $\l^{\pr(0)} =\l^{(0)} +\xi^{(0)}$
for any $\xi^{(0)}\in \Ker Q$. Then we also have
$$
\eta^{(0)}=f\left(x^{(0)}\right) + Q \l^{\pr(0)}.
$$
Repeating the same steps of deriving \tangob\ to \tangoby, we conclude that
$$
Q\left(\eta^{(1)} -f^{(1)}\left(x^{(0)}\right) - K^{(1)} \l^{\pr(0)}\right)=0.
$$
It follows that
$\eta^{(1)} -f^{(1)}\left(x^{(0)}\right) - K^{(1)} \l^{\pr(0)}=f\left(x^{\pr(1)}\right) + Q \l^{\pr(1)}$ 
for unique $x^{\pr(1)}\in H$ and
some $\l^{\pr(1)} \in \sC$, i.e.,
\eqn\tangoaz{
\eta^{(1)} =f^{(1)}\left(x^{(0)}\right) +f\left(x^{\pr(1)}\right) + K^{(1)} \l^{\pr(0)}+ Q \l^{\pr(1)}.
}
By comparing the above with \tangoc\ we have
\eqn\tangoaw{
f\left(x^{\pr(1)}-x^{(1)}\right) =- K^{(1)} \xi^{(0)}- Q \left(\l^{\pr(1)}-\l^{(1)}\right).
}
Recall that $ \xi^{(0)}= \l^{\pr(0)}- \l^{(0)}$ is an element in $\Ker Q$. Hence
\eqn\taox{
\xi^{(0)} = f\left(y^{(0)} \right)+ Q \zeta^{(0)}
}
for  $y^{(0)}=\left[\xi^{(0)}\right]\in H$ and some $\zeta^{(0)}\in \sC$.  Recall also that $\xi^{(0)}$
can be an arbitrary element in $\Ker Q$ so that $y^{(0)}$ is an arbitrary element in $H$
accordingly. From 
$$
\eqalign{
K^{(1)}\xi^{(0)} 
&=K^{(1)} f\left(y^{(0)} \right)+ K^{(1)}Q \zeta^{(0)}
\cr
&= f\left(\k^{(1)}y^{(0)} \right) -Q f^{(1)}\left(y^{(0)} \right)-Q K^{(1)} \zeta^{(0)},
}
$$
the relation
\tangoaw\ becomes
\eqn\tangoav{
f\left(x^{\pr(1)}-x^{(1)}+\k^{(1)}y^{(0)}\right) =- Q \left(\l^{\pr(1)}-\l^{(1)}
-f^{(1)}\left(y^{(0)} \right)- K^{(1)} \zeta^{(0)}
\right).
}
By taking the $Q$-cohomology class of the above, we have
\eqn\taoz{
x^{\pr(1)}=x^{(1)}-\k^{(1)}y^{(0)}.
}
Hence \tangoav\ also implies that
$$
Q \left(\l^{\pr(1)}-\l^{(1)}
-f^{(1)}\left(y^{(0)} \right)- K^{(1)} \zeta^{(0)}
\right)=0,
$$
so that there is unique $y^{(1)} \in H$ and some $\zeta^{(1)} \in \sC$ such that
\eqn\taoy{
\l^{\pr(1)}-\l^{(1)}=
f^{(1)}\left(y^{(0)} \right)
+f^{(1)}\left(y^{(1)} \right)
+ K^{(1)} \zeta^{(0)}
+Q \zeta^{(1)}.
}
Let $\bos{y} =y^{(0)} +\hbar y^{(1)}\mod \hbar^2$. From \taoz,
we conclude that the solution $\bos{x}=x^{(0)}+\hbar x^{(1)} \mod \hbar^2$ is 
unique up to $\bos{\k}\bos{y}=\k^{(1)}y^{(0)}  \mod \hbar^2$
for some $\bos{y}=y^{(0)} +\hbar y^{(1)}\mod \hbar^2$ in $H[[\hbar]]$. 
Combining together \taox\ and \taoy, we also conclude that
$$
\bos{\l}^\pr -\bos{\l} =\bos{f}\left(\bos{y}\right)+ \bos{K}\bos{\zeta} \mod \hbar^2
$$
where $\bos{\zeta} =\zeta^{(0)} +\hbar\zeta^{(1)}\mod \hbar^2$.

\begin{proof}
Consider  $\bos{\eta}=\eta^{(0)} +\hbar \eta^{(1)} +\cdots \in \sC[[\hbar]]$ satisfying $\bos{K}\bos{\eta}=0$.
Fix $n \geq 1$ and assume that 

(1) there is a pair $\left\{\bos{x},\bos{\l}\right\}$ which satisfies
$\bos{\eta} =\bos{f}(\bos{x}) +\bos{K}\bos{\l} \mod \hbar^n$, where
$$
\eqalign{
\bos{x}&=x^{(0)}+\hbar x^{(1)}+\cdots +\hbar^{n-1}x^{(n-1)}, \cr
\bos{\l}&=\l^{(0)}+\hbar \l^{(1)}+\cdots +\hbar^{n-1}\l^{(n-1)},\cr
}
$$

(2) $\bos{\k x}=0\mod \hbar^n$,

(3) for any other pair $\left\{\bos{x}^\pr,\bos{\l}^\pr\right\}$
satisfying 
$\bos{\eta} =\bos{f}(\bos{x}^\pr) +\bos{K}\bos{\l}^\pr \mod \hbar^n$, there is
a pair $\left\{\bos{y},\bos{\zeta}\right\}$,
 where
$$
\eqalign{
\bos{y}&=y^{(0)}+\hbar y^{(1)}+\cdots +\hbar^{n-1}y^{(n-1)}, \cr
\bos{\zeta}&=\zeta^{(0)}+\hbar \zeta^{(1)}+\cdots +\hbar^{n-1}\zeta^{(n-1)},\cr
}
$$
such that
$$
\eqalign{
\bos{x}^\pr -\bos{x} &= -\bos{\k}\bos{y} \mod \hbar^n
,\cr
\bos{\l}^\pr -\bos{\l} &= \bos{f}\left(\bos{y}\right) +\bos{K}\bos{\zeta} \mod \hbar^n
.
}
$$

 From the given condition
$\bos{K}\bos{\eta}=0$, we have
\eqn\tangoz{
Q\eta^{(n)} + \sum_{\ell=0}^{n-1}K^{(n-\ell)} \eta^{(\ell)}=0.
}
From  assumption $(1)$ we have, for $0\leq \ell \leq n-1$,
$$
\eta^{(\ell)}= \sum_{j=0}^\ell f^{(j)}\left(x^{(\ell-j)}\right) 
+ Q\l^{(\ell)} +\sum_{j=1}^\ell K^{(j)}\l^{(\ell-j)}.
$$
Hence \tangoz\ becomes
$$
Q\eta^{(n)} +\sum_{\ell=0}^{n-1}\sum_{j=0}^{\ell}
K^{(n-\ell)}f^{(\ell-j)}\left(x^{(j)}\right)
+\sum_{\ell=0}^{n-1}\sum_{j=1}^{\ell}
K^{(n-\ell)}K^{(j)}\l^{(\ell-j)}
+\sum_{\ell=0}^{n-1}
K^{(n-\ell)}Q \l^{(\ell)}
=0,
$$
which is equivalent to
\eqn\tangoy{
\eqalign{
Q\left(\eta^{(n)} - \sum_{\ell=0}^{n-1}f^{(\ell)\g}\left(x^{(n-\ell)}\right)
- \sum_{\ell=0}^{n-1} K^{(n-\ell)}\l^{(\ell)}\right)
=&
- \sum_{\ell=1}^{n-1}  f^{(n-\ell)}\left( \sum_{j=1}^{\ell} \k^{(j)}x^{(\ell-j)}\right)
\cr
&
-f\left(\sum_{i=1}^n\k^{(i)}x^{(n-i)}\right),
}
}
where we have used $\bos{K}^2=0$ and $\bos{K}\bos{f}=\bos{\k}\bos{f}$ with some ressummations.
From assumption $(2)$ we have $ \sum_{j=1}^{\ell} \k^{(j)}x^{(\ell-j)}=0$ for all $0\leq\ell\leq n-1$.
Hence \tangoy\ reduces to 
\eqn\tangox{
\eqalign{
Q\left(\eta^{(n)} - \sum_{\ell=0}^{n-1}f^{(\ell)\g}\left(x^{(n-\ell)}\right)
- \sum_{\ell=0}^{n-1} K^{(n-\ell)}\l^{(\ell)}\right)
=&
-f\left(\sum_{i=1}^n\k^{(i)}x^{(n-i)}\right).
}
}
Then, by taking the $Q$-cohomology class of the above relation, we have
\eqn\tangow{
\sum_{i=1}^n\k^{(i)}x^{(n-i)}=0,
}
as well as
\eqn\tangov{
Q\left(\eta^{(n)} - \sum_{\ell=0}^{n-1}f^{(\ell)\g}\left(x^{(n-\ell)}\right)
- \sum_{\ell=0}^{n-1} K^{(n-\ell)}\l^{(\ell)}\right)=0.
}
It follows, from \tangov\ that  
there is a $x^{(n)}\in H$ and some $\l^{(n)}\in \sC$ such that
$$
\eta^{(n)} - \sum_{\ell=0}^{n-1}f^{(\ell)\g}\left(x^{(n-\ell)}\right)
- \sum_{\ell=0}^{n-1} K^{(n-\ell)}\l^{(\ell)}
=f\left(x^{(n)}\right) + Q\l^{(n)}.
$$
Equivalently, we have
\eqn\tangou{
\eta^{(n)} = \sum_{\ell=0}^{n}f^{(\ell)}\left(x^{(n-\ell)}\right)
+ \sum_{\ell=0}^{n-1} K^{(n-\ell)}\l^{(\ell)} + Q\l^{(n)}.
}
Set 
$$
\eqalign{
\tilde\bos{x}&=\bos{x}+\hbar^n x^{(n)},\cr
\tilde\bos{\l}&=\bos{\l}+\hbar^n \l^{(n)}.\cr
}
$$
Then \tangou\ together with assumption $(1)$ is equivalent to the following:

$\tilde{ (1)}$. The pair $\left\{\tilde\bos{x},\tilde\bos{\l}\right\}$  satisfies
$\bos{\eta} =\bos{f}(\tilde\bos{x}) +\bos{K}\tilde\bos{\l} \mod \hbar^{n+1}$.

The relation \tangow\ together with assumption $(2)$ is equivalent to the following:

$\tilde{ (2)}$. $\bos{K}\tilde\bos{x} =0 \mod \hbar^{n+1}$.

Let $\left\{\bos{x}^\pr,\bos{\l}^\pr\right\}$ be any other pair satisfying
$\bos{\eta} = \bos{f}\left(\bos{x}^\pr\right)+ \bos{K}\bos{\l}^\pr \mod \hbar^n$
such that, by assumption (3),
$$
\eqalign{
\bos{x}^\pr &=\bos{x} - \bos{\k}\bos{y},\cr
\bos{\l}^{\pr} &=\bos{\l} +\bos{f}(\bos{y}) + \bos{K}\bos{\zeta},\cr
}
$$
where $\bos{x}^\pr=x^{\pr(0)}+\hbar x^{\pr(1)}+\cdots +\hbar^{n-1}x^{\pr(n-1)}$
and
$\bos{\l}=\l^{\pr(0)}+\hbar \l^{\pr(1)}+\cdots +\hbar^{n-1}\l^{\pr(n-1)}$.
Then we also have we have, for $0\leq \ell \leq n-1$,
$$
\eta^{(\ell)}= \sum_{j=0}^\ell f^{(j)}\left(x^{\pr(\ell-j)}\right) 
+ Q\l^{\pr(\ell)} +\sum_{j=1}^\ell K^{(j)}\l^{\pr(\ell-j)},
$$
Repeating the same steps as the derivations of \tangoy\ to \tangov, we can conclude that
\eqn\tangot{
\eqalign{
Q\left(\eta^{(n)} - \sum_{\ell=0}^{n-1}f^{(\ell)\g}\left(x^{\pr(n-\ell)}\right)
- \sum_{\ell=0}^{n-1} K^{(n-\ell)}\l^{\pr(\ell)}\right)
=&0.
}
}
Hence
there is a $x^{\pr(n)}\in H$ and some $\l^{\pr(n)}\in \sC$ such that
\eqn\tangos{
\eta^{(n)} = \sum_{\ell=0}^{n}f^{(\ell)}\left(x^{\pr(n-\ell)}\right)
+ \sum_{\ell=0}^{n-1} K^{(n-\ell)}\l^{\pr(\ell)} + Q\l^{\pr(n)}.
}
By comparing the above with \tangou\ we have
\eqn\tangor{
f\left(x^{\pr(n)}-x^{(n)}\right)+ Q\left(\l^{\pr(n)}-\l^{\pr(n)}\right)
=
- \sum_{\ell=0}^{n-1}f^{(n-\ell)}\left(x^{\pr(\ell)}-x^{(\ell)}\right)
- \sum_{\ell=0}^{n-1} K^{(n-\ell)}\left(\l^{\pr(\ell)} -\l^{(\ell)} \right).
}
The RHS of the above can be rewritten as follows:
$$
RHS = f\left(\sum_{\ell=1}^n \k^{(\ell)}y^{(n-\ell)} \right)
+ Q\sum_{\ell=1}^n\Big(f^{(\ell)}\left(y^{(n-\ell)}\right) + K^{(\ell)}\zeta^{(n-\ell)}\Big),
$$
where we have used assumption (3) in components, that is, for $0\leq \ell \leq n-1$,
$$
\eqalign{
x^{\pr(\ell)}-x^{(\ell)} &=\sum_{j=1}^{\ell}\k^{(j)}y^{(\ell-j)}
,\cr
\l^{\pr(\ell)} -\l^{(\ell)} &=\sum_{j=0}^{\ell}f^{(j)}\left(y^{(\ell-j)}\right)+ \sum_{j=0}^{\ell}K^{(j)}y^{(\ell-j)},
}
$$
and $\bos{K}^2=\bos{K}\bos{f}-\bos{f}\bos{\k}=0$.
Hence \tangor\ becomes
\eqn\tangoq{
f\left(x^{\pr(n)}-x^{(n)}+\sum_{\ell=1}^n \k^{(\ell)}y^{(n-\ell)} 
\right)
=-Q\left(\l^{\pr(n)}-\l^{\pr(n)}
-\sum_{\ell=1}^n\Big(f^{(\ell)}\left(y^{(n-\ell)}\right) + K^{(\ell)}\zeta^{(n-\ell)}\Big)
\right).
}
By taking the $Q$-cohomology class of the above, we have
\eqn\tangop{
x^{\pr(n)}-x^{(n)}=-\sum_{\ell=1}^n \k^{(\ell)}y^{(n-\ell)}.
}
Then, \tangop\ is reduced to 
$$
Q\left(\l^{\pr(n)}-\l^{\pr(n)}
-\sum_{\ell=1}^n\Big(f^{(\ell)}\left(y^{(n-\ell)}\right) + K^{(\ell)}\zeta^{(n-\ell)}\Big)
\right)=0.
$$
It follows that there is a pair $\left\{y^{(n)},\zeta^{(n)}\right\}$ such that
$$
\l^{\pr(n)}-\l^{\pr(n)}
-\sum_{\ell=1}^n\Big(f^{(\ell)}\left(y^{(n-\ell)}\right) + K^{(\ell)}\zeta^{(n-\ell)}\Big)
=f\left(y^{(n)}\right)+Q\zeta^{(n)}.
$$
Equivalently
\eqn\tangoo{
\l^{\pr(n)}-\l^{\pr(n)}
=f\left(y^{(n)}\right) +\sum_{\ell=1}^n f^{(\ell)}\left(y^{(n-\ell)}\right) 
+Q\zeta^{(n)}+\sum_{\ell=1}^n K^{(\ell)}\zeta^{(n-\ell)}.
}
Set
$$
\eqalign{
\tilde\bos{y} &= \bos{y}+\hbar^n y^{(n)}
,\cr
\tilde\bos{\zeta} &= \bos{\zeta}+\hbar^n \zeta^{(n)}
.\cr
}
$$
Then the relations \tangop\ and \tangoo\ together with assumption $(3)$ imply the following:

$\tilde{(3)}$.
For the another pair $\left\{\tilde\bos{x}^\pr,\tilde\bos{\l}^\pr\right\}$  satisfying
$\bos{\eta} = \bos{f}\left(\tilde\bos{x}^\pr\right)+ \bos{K}\tilde\bos{\l}^\pr \mod \hbar^{n+1}$,
we have
$$
\eqalign{
\tilde\bos{x}^\pr -\tilde\bos{x} &= -\bos{\k}\tilde\bos{y} \mod \hbar^{n+1}
,\cr
\tilde\bos{\l}^\pr -\tilde\bos{\l} &= \bos{f}\left(\tilde\bos{y}\right) +\bos{K}\tilde\bos{\zeta} \mod \hbar^{n+1}
.
}
$$
Hence by mathematical induction, we have the proposition.
\qed

\end{proof}

Now we turn to proposition \ref{flemmb}:

\begin{proposition}
A pair $\left\{\bos{x}, \bos{\l}\right\} \in H[[\hbar]]^{|\bos{x}|}\oplus \sC[[\hbar]]^{|\bos{x}|-1}$ satisfies
$$
\bos{f}(\bos{x}) =\bos{K}\bos{\l}
$$
if and only if there is a pair $\left\{\bos{y},\bos{\zeta}\right\} 
\in H[[\hbar]]^{|\bos{x}|-1}\oplus \sC[[\hbar]]^{|\bos{x}|-2}$
such that
$$
\eqalign{
\bos{x}&=\bos{\k}\bos{y}
,\cr
\bos{\l}&=\bos{f}\left(\bos{y}\right)+\bos{K}\bos{\zeta}
}
$$
\end{proposition}

\begin{proof}
Assume that $\bos{x}=\bos{\k}\bos{y}$ 
and $\bos{\l}=\bos{f}\left(\bos{y}\right)+\bos{K}\bos{\zeta}$.
Then $\bos{K}\bos{\l}=\bos{K}\bos{f}\left(\bos{y}\right)=\bos{f}\left(\bos{\k}\bos{y}\right)$. 
Hence $\bos{K}\bos{\l}=\bos{f}\left(\bos{x}\right)$. 
It remains to show that there is 
there is a pair $\left\{\bos{y},\bos{\zeta}\right\}$  
such that $\bos{x}=\bos{\k}\bos{y}$ and $\bos{\l}=\bos{f}\left(\bos{y}\right)+\bos{K}\bos{\zeta}$
if $\bos{f}(\bos{x}) =\bos{K}\bos{\l}$.

\begin{itemize}
\item
The condition $\bos{f}\left(\bos{x}\right)=\bos{K}\bos{\l}$ modulo $\hbar$ is
\eqn\mambo{
f\left(x^{(0)}\right)=Q\l^{(0)}.
}
It follows that $x^{(0)}=0$ and $Q\l^{(0)}=0$,
which implies that there is a pair $\left(y^{(0)},\zeta^{(0)}\right\}$ such that 
$\l^{(0)}=f\left(y^{(0)}\right) +Q \zeta^{(0)}$:
\eqn\mamba{
\eqalign{
x^{(0)}&=0,\cr
\l^{(0)}&=f\left(y^{(0)}\right) +Q \zeta^{(0)},\cr
}
}
Set $\bos{y}=y^{(0)}\mod \hbar$ and $\bos{\zeta}= \zeta^{(0)}\mod\hbar$.
Then \mamba\ is equivalent to
\eqn\mambb{
\eqalign{
\bos{x}&=\bos{\k}\bos{y}\mod \hbar
,\cr
\bos{\l}&=\bos{f}\left(\bos{y}\right)+\bos{K}\bos{\zeta}\mod \hbar
}
}

\item
The condition $\bos{f}\left(\bos{x}\right)=\bos{K}\bos{\l}$ modulo $\hbar^2$ is equivalent
to \mambo\ and 
\eqn\mambc{
f\left(x^{(1)}\right)+f^{(1)}\left(x^{(0)}\right)=K^{(1)}\l^{(0)}+ Q\l^{(1)}.
}
From \mamba, we have $f^{(1)}\left(x^{(0)}\right)=0$
and
$$
\eqalign{
K^{(1)}\l^{(0)}
&=K^{(1)}f\left(y^{(0)}\right) + K^{(1)}Q \zeta^{(0)}
\cr
&=f\left(\k^{(1)}y^{(0}\right) -Q\left(f^{(1)}\left(y^{(0)}\right) + K^{(1)} \zeta^{(0)}\right)
..
}
$$
Thus, \mambc\ becomes
$$
f\left(x^{(1)}-\k^{(1)}y^{(0)}\right)= Q\left(\l^{(1)}-f^{(1)}\left(y^{(0)}\right) - K^{(1)} \zeta^{(0)}\right).
$$
It follows that
\eqn\mambd{
x^{(1)}=\k^{(1)}y^{(0)}
}
and
$Q\left(\l^{(1)}-f^{(1)}\left(y^{(0)}\right) - K^{(1)} \zeta^{(0)}\right)=0$,
which implies that there is a pair $\left\{y^{(1)},\zeta^{(1)}\right\}$ such that 
$\l^{(1)}-f^{(1)}\left(y^{(0)}\right)- K^{(1)} \zeta^{(0)}=f\left(y^{(1)}\right) +Q \zeta^{(1)}$,
i.e.,
\eqn\mambe{
\l^{(1)}=f^{(1)}\left(y^{(0)}\right)+f\left(y^{(1)}\right) + K^{(1)} \zeta^{(0)}+Q \zeta^{(1)}.
}
Set $\bos{y}=y^{(0)}+\hbar y^{(1)}\mod \hbar^2$ and 
$\bos{\zeta}= \zeta^{(0)}+\hbar \zeta^{(1)}\mod\hbar^2$.
Then \mambd\ and \mambe\ together with  \mambb\ are equivalent to
\eqn\mambf{
\eqalign{
\bos{x}&=\bos{\k}\bos{y}\mod \hbar^2
,\cr
\bos{\l}&=\bos{f}\left(\bos{y}\right)+\bos{K}\bos{\zeta}\mod \hbar^2.
}
}

\item (Assumption): Fix $n > 2$ and assume that there is 
a pair $\left\{\bos{y},\bos{\zeta}\right\}$,
 where
$$
\eqalign{
\bos{y}&=y^{(0)}+\hbar y^{(1)}+\cdots +\hbar^{n-1}y^{(n-1)}, \cr
\bos{\zeta}&=\zeta^{(0)}+\hbar \zeta^{(1)}+\cdots +\hbar^{n-1}\zeta^{(n-1)},\cr
}
$$
such that
$$
\eqalign{
\bos{x}&=\bos{\k}\bos{y}\mod \hbar^n
,\cr
\bos{\l}&=\bos{f}\left(\bos{y}\right)+\bos{K}\bos{\zeta}\mod \hbar^n.
}
$$

\item
From the condition $\bos{f}\left(\bos{x}\right)=\bos{K}\bos{\l}$, we have
\eqn\mambg{
f\left(x^{(n)}\right)-Q\l^{(n)}
=-\sum_{\ell=0}^{n-1} f^{(n-\ell)}\left(x^{(\ell)}\right)
+\sum_{\ell=0}^{n-1} K^{(n-\ell)} \l^{(\ell)}.
}
From the assumption we have, for $0\leq \ell \leq n-1$,
$$
\eqalign{
x^{(\ell)} &=  \sum_{j=1}^{\ell} \k^{(j)}y^{(\ell-j)} 
,\cr
\l^{(\ell)} &= 
f\left(y^{(\ell)}\right) 
+ \sum_{j=1}^{\ell} f^{(j)}\left(y^{(\ell-j)}\right) 
+ Q \zeta^{(\ell)}
+ \sum_{j=1}^{\ell} K^{(j)}\zeta^{(\ell-j)}
.
}
$$
Make the substitution as the above to \mambg, we obtain that
\eqn\mambp{
\eqalign{
f\left(x^{(n)}\right)-Q\l^{(n)}= A^{(n)} + B^{(n)},
}
}
where
$$
\eqalign{
A^{(n)}
=&
-\sum_{\ell=0}^{n-1}\sum_{j=1}^{\ell} f^{(n-\ell)}\left(\k^{(j)}y^{(\ell-j)}\right)
+\sum_{\ell=0}^{n-1} K^{(n-\ell)} f\left(y^{(\ell)}\right) 
+\sum_{\ell=0}^{n-1} \sum_{j=1}^{\ell}  K^{(n-\ell)} f^{(j)}\left(y^{(\ell-j)}\right) 
,\cr
B^{(n)}
=&
+\sum_{\ell=0}^{n-1} K^{(n-\ell)} Q\zeta^{(\ell)}
+\sum_{\ell=0}^{n-1}\sum_{j=1}^{\ell} K^{(n-\ell)} K^{(j)}\zeta^{(\ell-j)}.
}
$$
From $\bos{K}\bos{f}=\bos{f}\bos{\k}$ and $\bos{K}^2=0$, we obtain that
$$
\eqalign{
A^{(n)}&=f\left(\sum_{\ell=1}^n\k^{(\ell)}y^{(n-\ell)}\right)
-Q\sum_{\ell=1}^n f^{(\ell)}\left(y^{(n-\ell)}\right)
,\cr
B^{(n)}&=-Q\sum_{\ell=1}^n K^{(\ell)} \zeta^{(n-\ell)}
.
}
$$
Hence \mambp\ reduces to 
\eqn\mambh{
f\left(x^{(n)}-\sum_{\ell=1}^n\k^{(\ell)}y^{(n-\ell)}\right)
= Q\left(\l^{(n)}-\sum_{\ell=1}^n f^{(\ell)}\left(y^{(n-\ell)}\right) -\sum_{\ell=1}^n K^{(\ell)} \zeta^{(n-\ell)}\right).
}
By taking the $Q$-cohomology class of the above relation we have
\eqn\mambi{
x^{(n)}=\sum_{\ell=1}^n\k^{(\ell)}y^{(n-\ell)},
}
as well as
$$
 Q\left(\l^{(n)}-\sum_{\ell=1}^n f^{(\ell)}\left(y^{(n-\ell)}\right) -\sum_{\ell=1}^n K^{(\ell)} \zeta^{(n-\ell)}\right)=0,
 $$
 which implies that there is a pair $\left\{y^{(n)}, \zeta^{(n)}\right\}$ such that
 $$
 \l^{(n)}-\sum_{\ell=1}^n f^{(\ell)}\left(y^{(n-\ell)}\right) -\sum_{\ell=1}^n K^{(\ell)} \zeta^{(n-\ell)}
 =f\left(y^{(n)}\right) + Q \zeta^{(n)}.
$$
Equivalently, we have
\eqn\mambf{
 \l^{(n)}= f\left(y^{(n)}\right)
 +\sum_{\ell=1}^n f^{(\ell)}\left(y^{(n-\ell)}\right) 
 + Q \zeta^{(n)}
 +\sum_{\ell=1}^n K^{(\ell)} \zeta^{(n-\ell)}.
}
Set $\tilde\bos{y} =\bos{y}+\hbar^n y^{(n)}$ and $\tilde\bos{\zeta} =\bos{\zeta}+\hbar^n \tilde \zeta^{(n)}$.
Then the relation \mambi\ and \mambh\ together with our assumption is equivalent to the following:
$$
\eqalign{
\tilde\bos{x}&=\bos{\k}\tilde\bos{y}\mod \hbar^{n+1}
,\cr
\tilde\bos{\l}&=\bos{f}\left(\tilde\bos{y}\right)+\bos{K}\bos{\tilde\zeta}\mod \hbar^{n+1}.
}
$$
Hence we have proved the proposition by the mathematical induction.
\qed

\end{itemize}

\end{proof}

\newsec{Appendix}

Let $\big(\sC, Q, (\bullet, \bullet)\big)$ be a DG0LA over $\Bbbk$ and let the cohomology
$H$ of the correspond cochain complex $(\sC, Q)$ is finite dimensional for each ghost number 
as a $\Z$-graded $\Bbbk$-vector space. Let  
$\{e_\a\}$ be a basis of $H$ and let $t_H=\{t^\a\}$ be the dual basis. We denote 
$\left\{\rd_\a =\Fr{\rd}{\rd t^\a}\right\}$
be the corresponding formal partial derivative acting on $\Bbbk[[t_H]]$ as derivations.
Assume that we have a {\it fixed} versal solution $\Theta \in \sC[[t_H]]^0$ to 
the Maurer-Cartan equation of the DG0LA 
$$
Q \Theta + \Fr{1}{2}\left(\Theta, \Theta\right)=0,
$$ 
where
$$
\Theta = t^\a O_\a + \sum_{n=2}^\infty \Fr{1}{n!} t^{\a_n}\cdots t^{\a_1} O_{\a_1\cdots \a_n}
$$
such that the set  $\{O_\a\}$ from the leading term is a set of representative of the basis $\{e_\a\}$ of 
$H$.  The Maurer-Cartan equation implies that 
$$
Q_\Theta:= Q +\left(\Theta,\bullet\right): 
\sC[[t_H]]^i \longrightarrow \sC[[t_H]]^{i+1}
$$
satisfies $Q_\Theta^2 =0$. By applying $\rd_\g$ to the Maurer-Cartan equation we obtain
that
$$
Q\Theta_\g + (\Theta,\Theta_\g)=0,
$$
where $\Theta_\g:= \rd_\g \Theta \in \sC[[t_H]]^{|\g|}$.

We recall that any homogeneous element $X \in \sC^{|X|}$ satisfying $Q X =0$ 
can be expressed as $X = c^{\g}O_{\g} + Q Y$ with unique set of constants $\{c^{\g}\}$ in $\Bbbk$ and
some $Y \in \sC^{|X|-1}$ defined modulo $\Ker Q$. Also for any equality in the form
$Q Y = c^{\g}O_{\g}$ implies that $c^{\g}=0$ for all $\g$, since
by taking the $Q$-cohomology class we have $c^{\g}[O_{\g}]\equiv c^{\g}e_{\g}=0$
and $\{e_{\g}\}$ are linearly independent, as well as that $Q Y=0$. 
The purpose of this appendix is to
confirm that the similar
properties involving $Q_\Theta=Q +\left(\Theta,\hbox{ }\right)$ and the set $\{\Theta_\g\}$.

\begin{proposition}
Any homogeneous element $\CX \in \sC[[t_{H}]]^{|\CX|}$
satisfying
$$
Q_{\!\Theta} \CX =0$$
can be expressed as
$$
\CX = B^{\g}\Theta_{\g} + Q_{\!\Theta} \CY,
$$
where $\{B^{\g}\}\in \Bbbk[[t_{H}]]$ is  defined uniquely  and 
$\CY \in \sC[[t_{H}]]^{|\CX|-1}$ is defined modulo 
$\Ker Q_\Theta$.
\end{proposition}

\begin{proposition}
Assume that we have the following equality
$$
Q_{\Theta} \CY = B^\g \Theta_\g,
$$
where $\{B^{\g}\}\in \Bbbk[[t_{H}]]$. Then $B^{\g}=0$ for all $\g$ and
$Q_{\Theta} \CY =0$.
\end{proposition}

For our purpose we shall establish the above propositions
modulo $t_H^{n+1}$ and take the limit $n\rightarrow \infty$.

Let $\Theta \in \sC[[t_{H}]]^0 \mod t_H^{n+1}$ be 
a versal solution of the Maurer-Cartan equation
modulo $t_H^{n+1}$:
$$
Q\Theta +\Fr{1}{2}(\Theta,\Theta)=0\mod t_H^{n+1}.
$$
Let $Q_\Theta = Q +(\Theta,\bullet) \mod t_H^{n+1}$. Then,
for any $\sX \in \sC[[t_H]] \mod t_H^{n+1}$,
we have $Q_\Theta^2 \sX =0 \mod t_H^{n+1}$. 
Let $\Theta_\g = \rd_\g \Theta  \in  \sC[[t_{H}]]^{|\g|} \mod t_H^{n}$. 
Then $Q_\Theta\Theta_\g =0 \mod t_H^n$.
We  use decompositions of $\Theta \mod t_H^{n+1}$ and $\Theta_{\g}\mod t_H^{n}$ in terms of
the word length in $t_H$:
$$
\eqalign{
\Theta&= \Theta^{[1]}+\cdots + \Theta^{[n]} \mod t^{n+1}
,\cr
\Theta_{\g}&= \Theta^{[0]}_{\g}+\Theta^{[1]}_{\g}+\cdots + \Theta^{[n-1]}_{\g} \mod t_H^{n},
}
$$ 
where
$$
\eqalign{
\Theta^{[k]}&= \Fr{1}{k!}t^{\a_{1}}\cdots t^{\a_{k}}O_{\a_{k}\cdots \a_{1}}\quad
\hbox{for } k=1,2,\cdots, n.
,\cr
\Theta^{[j]}_{\g}&= \Fr{1}{j!}t^{\a_{1}}\cdots t^{\a_{j}}O_{\a_{j}\cdots \a_{1}\g}
\quad\hbox{for }  j=0,1,\cdots,n-1.
}
$$
 In particular $\Theta^{[1]}= t^{\a}O_{\a}$ and
$\Theta^{[0]}_{\g}= O_{\g}$, where $Q O_{\g}=0$ for all $\g$ and the $Q$-cohomology
classes $\{[O_{\g}]\}$ of $\{O_{\g}\}$ form a basis $\{e_{\g}\}$ of the $Q$-cohomology
group $H$. 
\begin{property} 
\label{appprop1}
The condition
$Q\Theta +\Fr{1}{2}\left(\Theta,\Theta\right) =0 \mod t_H^{n+1}$ is
equivalent to the following sequence of equations
$$
\eqalign{
Q\Theta^{[1]}&=0,\cr
Q\Theta^{[2]}+\Fr{1}{2}\left(\Theta^{[1]},\Theta^{[1]}\right)&=0,\cr
&\vdots\cr
Q\Theta^{[n]}+\Fr{1}{2}\sum_{j=1}^{n-1}\left(\Theta^{[j]},\Theta^{[n-j]}\right)&=0.\cr
}
$$
\end{property}

\begin{property} 
\label{appprop2}
The condition that
$Q\Theta_{\g} +\left(\Theta,\Theta_{\g}\right) =0 \mod t_H^{n}$ is
equivalent to the following sequence of equations
$$
\eqalign{
Q\Theta^{[0]}_{\g}&=0,\cr
Q\Theta^{[1]}_{\g}+\left(\Theta^{[1]},\Theta^{[0]}_{\g}\right)&=0,\cr
\vdots
\cr
Q\Theta^{[n]}_{\g}+\sum_{\ell=1}^{n}\left(\Theta^{[\ell]},\Theta^{[n-\ell]}_{\g}\right)&=0.\cr
}
$$
\end{property}

\begin{proposition}\label{atech1}
Let $\Theta \in  \sC[[t_{H}]]^0 \mod t_H^{n+1}$ be 
a versal solution to the MC equation modulo $t_H^{n+1}$
and let $\Theta_\g =\rd_\g \Theta \in \sC[[t_{H}]]^{|\g|} \mod t_H^{n}$
so that
$Q\Theta_\g +\left(\Theta,\Theta_\g\right)=0 \mod t_H^{n}$.
Then
any homogeneous element $\CX \in  \sC[[t_{H}]]^{|\CX|} \mod t_H^{n}$
 satisfying
$$
Q \CX +\left(\Theta, \CX\right)=0\mod t_H^{n}
$$
can be expressed as
$$
\CX = B^{\g}\Theta_{\g} + Q \CY +\left(\Theta,\CY\right) \mod t_H^{n}
$$
where $B^{\g}\in \Bbbk[[t_{H}]]$ defined uniquely modulo $t_H^{n}$ and 
$\CY  \in  \sC[[t_{H}]]^{|\CX|-1} \mod t_H^n$ defined modulo 
$\Ker Q_\Theta$.
\end{proposition}

\begin{proof}
Decompose $\CX\mod t_H^{n}$ in terms of the word length in $t_H$:
$$
\eqalign{
\CX&= \CX^{[0]}+\CX^{[1]}+\cdots + \CX^{[n-1]} \mod t_H^{n}.
}
$$
Then the condition
$Q \CX +\left(\Theta, \CX\right)=0\mod t_H^{n}$ is equivalent to the following
sequence of equations:
$$
\eqalign{
Q\CX^{[0]}&=0,\cr
Q\CX^{[1]}+\left(\Theta^{[1]},\CX^{[0]}\right)&=0,\cr
\vdots
\cr
Q\CX^{[n-1]}+\sum_{\ell=0}^{n-2}\left(\Theta^{[n-\ell-1]},\CX^{[\ell]}\right)&=0.\cr
}
$$
Then we need to find
following sequence of solutions with the claimed properties: 
$$
\eqalign{
\CX^{[0]}&=B^{[0]\g}\Theta^{[0]}_{\g} + Q \CY^{[0]},
\cr
\CX^{[1]}&=B^{[0]\g}\Theta^{[1]}_{\g} +B^{[1]\g}\Theta^{[0]}_{\g} + Q \CY^{[1]}
+\left(\Theta^{[0]},\CY^{[1]}\right)
,\cr
\vdots
\cr
\CX^{[n-1]}&=\sum_{\ell=0}^{n-1}B^{[\ell]\g}\Theta^{k-\ell}_{\g} +Q\CY^{[n-1]} 
+\sum_{i=1}^{n-1}\left(\Theta^{[i]},\CY^{n-i-1}\right),
}
$$
for unique $B^{\g}= B^{[0]\g}+B^{[1]\g}+\cdots +B^{[n-1]\g}\mod t_H^{n}$,
where $B^{[j]\g}= \Fr{1}{j!}t^{\a_{1}}\cdots t^{\a_{j}}b_{\a_{j}\cdots \a_{1}}{}^{\g}$
and $b_{\a_{j}\cdots \a_{1}}{}^{\g}\in \Bbbk$, and for some
$\CY= \CY^{[0]}+\CY^{[1]}+\cdots +\CY^{[n-1]}\mod t_H^{n}$,
where $\CY^{[j]}= \Fr{1}{j!}t^{\a_{1}}\cdots t^{\a_{j}}Y_{\a_{j}\cdots \a_{1}}$
and $Y_{\a_{j}\cdots \a_{1}} \in \sC^{|\a_{j}|+\cdots +|\a_{1}|-1}$.

We are going to use mathematical induction: 
consider
the condition $Q_{\!\Theta}\CX =0\mod t_H^{1}$, which  is equivalent to $Q\CX^{[0]}=0$.
Then we have
$$
\CX^{[0]}=B^{[0]\g}\Theta^{[0]}_{\g} + Q \CY^{[0]}
$$
for unique $B^{[0]\g}\in \Bbbk$ and for some $\CY^{[0]}\in \sC^{|\CX|}$
defined modulo $\Ker Q$, since $\CX^{[0]}\in \sC$ and $\Theta^{[0]}_{\g}=O_{\g}$.
Hence our claim is true modulo $t_H^1$.

Fix $k$ such that $1< k < n-1$ and
assume that our claim is true for modulo $t_H^{k+1}$:
\eqn\asumn{
\CX = B^{\g}\Theta_{\g} + Q \CY +\left(\Theta,\CY\right) \mod t_H^{k+1}
}
where $B^{\g}=B^{[0]\g} +B^{[1]\g}\cdots +B^{[k]\g}\mod t_H^{k+1}$ defined uniquely and 
$\CY=\CY^{[0]}+\CY^{[1]}+\cdots +\CY^{[k]} \mod t_H^{k+1}$ defined modulo 
$\Ker Q_\Theta$.
We have, in components, for 
all $j=0,1,\cdots, k$,
$$
\CX^{[j]}=\sum_{\ell=0}^{j}B^{[\ell]\g}\Theta^{j-\ell}_{\g} +Q\CY^{[j]} 
+\sum_{i=1}^{j}\left(\Theta^{[i]},\CY^{[j-i]}\right).
$$
Then, we need to show that
there exist unique $B^{[k+1]\g}$ and $\CY^{[k+1]}$  satisfying
$$
\CX^{[k+1]}=\sum_{\ell=0}^{k+1}B^{[\ell]\g}\Theta^{k+1-\ell}_{\g} +Q\CY^{[k+1]} 
+\sum_{i=1}^{k+1}\left(\Theta^{[i]},\CY^{[k+1-i]}\right),
$$
such that 
$$
\CX = B^{\g}\Theta_{\g} + Q \CY +\left(\Theta,\CY\right) \mod t_H^{k+2}
$$
for unique $B^{\g}=B^{[0]\g} +B^{[1]\g} +\cdots+ B^{[k]\g} +B^{[k+1]\g}\mod t_H^{k+2}$
and for some $\CY= \CY^{[0]}+\CY^{[1]}+ \cdots +\CY^{[k]}+\CY^{[k+1]}\mod t_H^{k+2}$
is defined modulo $\Ker Q_\Theta$.

We shall proceed as follows:
\begin{itemize}

\item (Claim 1):
we claim that the following expression:
$$
\CJ^{[k+1]}:= \CX^{[k+1]}-\sum_{\ell=1}^{k+1}B^{[k+1-\ell]\g}\Theta^{[\ell]}_{\g}
-\sum_{\ell=1}^{k+1}\left(\Theta^{[\ell]},\CY^{[k+1-\ell]}\right)
$$
satisfies $Q\CJ^{[k+1]}=0$.

\item (Claim 2):
Note that the last term of the expression
$\CJ^{[k+1]}$
 is ambiguous due
to  the ambiguities of $\CY =\CY^{[0]}+\cdots +\CY^{[k]}\mod t_H^{k+1}$, which
is assumed to be  defined modulo $\Ker Q +(\Theta, \hbox{ })$.  
We claim that the resulting ambiguity of $\CJ^{[k+1]}$ is $Q$-exact.
That is,
if $\CY^\pr = \CY^{\pr[0]}+\cdots \CY^{\pr[k]}\mod t_H^{k+1}$
denote any other possible choice for \asumn, and let
$$
\CJ^{\pr[k+1]}:= \CX^{[k+1]}-\sum_{\ell=1}^{k+1}B^{[k+1-\ell]\g}\Theta^{[\ell]}_{\g}
-\sum_{\ell=1}^{k+1}\left(\Theta^{[\ell]},\CY^{\pr[k+1-\ell]}\right),
$$
then
$$
\CJ^{[k+1]}-\CJ^{\pr[k+1]}
=\sum_{\ell=0}^{k}\left(\Theta^{[k+1-\ell]},\CY^{\pr[\ell]}-\CY^{[\ell]}\right)
\in \hbox{ Im } Q.
$$
Hence both $\CJ^{[k+1]}$ and $\CJ^{\pr[k+1]}$ belong to the same $Q$-cohomology class.

\end{itemize}

From the above claims, it follows that
there is unique set $\{B^{[k+1]\g}\}$, where 
$B^{[k+1]\g}=\Fr{1}{(k+1)!}t^{\r_{1}}\cdots t^{\r_{k+1}}b_{\r_{k+1}\cdots\r_{1}}{}^{\g}$,
such that
\eqn\mgioa{
\eqalign{
\CJ^{[k+1]}&= B^{[k+1]\g}\Theta^{[0]}_{\g} +Q\CY^{[k+1]}
,\cr
\CJ^{\pr[k+1]}&= B^{[k+1]\g}\Theta^{[0]}_{\g} +Q\CY^{\pr [k+1]},
}
}
for some $\CY^{[k+1]},\CY^{\pr[k+1]}$ defined modulo $\Ker Q$.
{}From the above and the definitions of $\CJ^{[k+1]}$ and $\CJ^{\pr[k+1]}$ we obtain that 
\eqn\mgiob{
\eqalign{
\CX^{[k+1]}
=&\sum_{\ell=0}^{k+1}B^{[k+1-\ell]\g}\Theta^{[\ell]}_{\g}
+Q\CY^{[k+1]}
+\sum_{\ell=1}^{k+1}\left(\Theta^{[\ell]},\CY^{[k+1-\ell]}\right)
\cr
=&\sum_{\ell=0}^{k+1}B^{[k+1-\ell]\g}\Theta^{[\ell]}_{\g}
+Q\CY^{\pr[k+1]}
+\sum_{\ell=1}^{k+1}\left(\Theta^{[\ell]},\CY^{\pr[k+1-\ell]}\right).
}
}
{}From eq.\ \mgioa\ we also obtain that
$\CJ^{[k+1]}-Q \CY^{[k+1]}= \CJ^{\pr[k+1]}-Q \CY^{\pr[k+1]}$,
which implies that
\eqn\mgioc{
Q\left(\CY^{\pr [k+1]}-\CY^{[k+1]}\right)
+\sum_{\ell=0}^{k}\left(\Theta^{[k+1-\ell]},\CY^{\pr[\ell]}-\CY^{[\ell]}\right)
=0.
}
Let $\CY := \CY^{[0]}+\cdots +\CY^{[k]}+\CY^{[k+1]}\mod t_H^{[k+2]}$
and $\CY^\pr := \CY^{\pr[0]}+\cdots +\CY^{\pr[k]}+\CY^{\pr[k+1]}\mod t_H^{[k+2]}$.
Combined with our assumption, the relations in eq.\ \mgiob\ imply that
$$
\eqalign{
\CX &= B^\g\Theta_\g + Q \CY +\left(\Theta, \CY\right) \mod t_H^{k+2}
\cr
&= B^\g\Theta_\g + Q \CY^\pr +\left(\Theta, \CY^\pr\right) \mod t_H^{k+2}.
}
$$
Combined with our assumption, the relation in eq.\ \mgioc\ implies that
$$
Q(\CY^\pr-\CY)+\left(\Theta, \CY^\pr-\CY\right)=0\mod t_H^{k+2}.
$$
Thus, we shall have a proof of our proposition by mathematical
induction the two claims above are checked.

\noindent
{\it Proof of Claim 1.}
We shall show  that
\eqn\masia{
\sum_{j=0}^{k}\left(\Theta^{[k+1-j]},\CX^{[j]}\right)=
-Q\left(\sum_{\ell=1}^{k+1}B^{[k+1-\ell]\g}\Theta^{[\ell]}_{\g}
+\sum_{\ell=1}^{k+1}\left(\Theta^{[\ell]},\CY^{[k+1-\ell]}\right)\right)..
}
From the condition $Q\CX +\big(\Theta,\CX\big) =0\mod t_H^{n+1}$, which implies that
$$
Q\CX^{[k+1]}+\sum_{\ell=1}^{k+1}\left(\Theta^{[\ell]},\CX^{[k+1-\ell]}\right)=0,
$$
we see that \masia\ is equivalent to  the claim that $Q \CJ^{[k+1]}=0$. 

Consider the left hand side of  eq.\ \masia:
$$
\sum_{j=0}^{k}\left(\Theta^{[k+1-j]},\CX^{[j]}\right).
$$
From the assumption  \asumn\ we have
$$
\eqalign{
\sum_{j=0}^{k}\left(\Theta^{[k+1-j]},\CX^{[j]}\right)
=&
\sum_{j=0}^{k}\sum_{\ell=0}^{j}B^{[\ell]\g}\left(\Theta^{[k+1-j]},
\Theta^{[j-\ell]}_{\g}\right) 
\cr
&
+\sum_{j=0}^{k}\left(\Theta^{[k+1-j]},Q\CY^{[j]}\right) 
+\sum_{j=0}^{k}\sum_{i=1}^{j}\left(\Theta^{[k+1-j]},\left(\Theta^{[i]},\CY^{[j-i]}\right)\right).
}
$$
We consider the first and second lines in the right hand side of the above separately:
\begin{itemize}
\item For the first line we have
$$
\eqalign{
\sum_{j=0}^{k}\sum_{\ell=0}^{j}B^{[\ell]\g}\left(\Theta^{[k+1-j]},
\Theta^{[j-\ell]}_{\g}\right) 
&=\sum_{\ell=1}^{k+1}B^{[k+1-\ell]\g}\left(
\sum_{i=1}^{\ell}
\left(\Theta^{[i]},
\Theta^{[\ell-i]}_{\g}\right) \right)
\cr
&= -Q\left(\sum_{\ell=1}^{k+1}B^{[k+1-\ell]\g}\Theta^{[\ell]}_{\g}\right)
}
$$
where we did re-summations for the first equality and used Property \ref{appprop2}
for the second equality.

\item For the second line we have
$$
\eqalign{
\sum_{j=0}^{k}&\left(\Theta^{[k+1-j]},Q\CY^{[j]}\right) 
+\sum_{j=0}^{k}\sum_{i=1}^{j}\left(\Theta^{[k+1-j]},\left(\Theta^{[i]},\CY^{[j-i]}\right)\right)
\cr
=&
-Q\sum_{j=0}^{k}\left(\Theta^{[k+1-j]},\CY^{[j]}\right) 
\cr
&
+\sum_{j=0}^{k}\left(Q\Theta^{[k+1-j]},\CY^{[j]}\right) 
+\sum_{j=0}^{k}\sum_{i=1}^{j}\left(\Theta^{[k+1-j]},\left(\Theta^{[i]},\CY^{[j-i]}\right)\right)
\cr
=&
-Q\sum_{\ell=1}^{k+1}\left(\Theta^{[\ell]},\CY^{[k+1-\ell]}\right)
\cr
 &
+\sum_{j=1}^{k+1}\left(Q\Theta^{[j]}+\sum_{i=1}^{j-1}\left(\Theta^{[i]},\Theta^{[j-i]}\right)
,\CY^{[k+1-j]}\right) 
\cr
=&
-Q\sum_{\ell=1}^{k+1}\left(\Theta^{[\ell]},\CY^{[k+1-\ell]}\right),
}
$$
where we have used the property that $Q$ is a derivation of the BV bracket in the first equality,
re-summations in the second equality and the Jacobi identity of the BV bracket in the third equality,
and the MC equation modulo $t_H^{n+1}$ (Property \ref{appprop1}) for the last equality. 
Thus we have established the identity in eq.\ \masia.
\end{itemize}

\noindent
{\it Proof of Claim 2.}
Consider the last term of $\CJ^{[k+1]}$ defined in claim 1, which can be
re-summed as 
$$
-\sum_{\ell=0}^{k}\left(\Theta^{[k+1-\ell]},\CY^{[\ell]}\right)
$$
Let $\CY = \CY^{[0]}+\cdots \CY^{[k]}\mod t_H^{k+1}$. By the assumption
that $\CY\mod t_H^{k+1}$ is defined modulo $\Ker Q +(\Theta, \hbox{ })$.
Let $\CY^\pr = \CY^{\pr[0]}+\cdots \CY^{\pr[k]}\mod t_H^{k+1}$
denote any other possible choice such that
$(\CY^{\pr}-\CY) \mod t_H^{k+1}$ belongs to $\Ker Q +(\Theta,\hbox{ })$.
 Then, by the assumption, we have
$$
\CY^{\pr}-\CY = C^{\g}\Theta_{\g} + Q \CW +(\Theta,\CW)\mod t_H^{k+1}
$$
for some $C^{\g} \in \Bbbk[[t_{H}]]/t_H^{k+1}$ and
some 
$\CW\in \left(\sC\otimes  \Bbbk[[t_{H}]]/t_H^{k+1}\right)^{|\CY|-1}    $.
Thus for each $0\leq\ell \leq k$, we have
$$
\CY^{[\ell]\pr}-\CY^{[\ell]} = 
\left(Y^{\g}\Theta_{\g}\right)^{[\ell]} + Q \CW^{[\ell]} +\left(\Theta,\CW\right)^{[\ell]}.
$$
Let 
$$
\CJ^{\pr[k+1]}:= \CX^{[k+1]}-\sum_{\ell=1}^{k+1}B^{[k+1-\ell]\g}\Theta^{[\ell]}_{\g}
-\sum_{\ell=1}^{k+1}\left(\Theta^{[\ell]},\CY^{\pr[k+1-\ell]}\right),
$$
then
$$
\eqalign{
\CJ^{[k+1]}-\CJ^{\pr[k+1]}
=&\sum_{\ell=0}^{k}\left(\Theta^{[k+1-\ell]},\CY^{\pr[\ell]}-\CY^{[\ell]}\right)
\cr
=&\left(\Theta,Y^{\g}\Theta_{\g}\right)^{[k+1]}
+\left(\Theta,  Q \CW +(\Theta,\CW)\right)^{[k+1]}.
}
$$

It remains to show that the right hand side of the above is $Q$-exact.
The first  term can be rearranged as follows
$$
\eqalign{
\left(\Theta,Y^{\g}\Theta_{\g}\right)^{[k+1]}
=&\sum_{j=0}^{k}Y^{[j]\g}
\left(\Theta,\Theta_{\g}\right)^{[k+1-j]}.
}
$$
We then use the identity
$Q\Theta + \left(\Theta, \Theta_{\g}\right)=0 \mod t_H^{n}$  (Property \ref{appprop2})
to obtain that
$$
 \left(\Theta,Y^{\g}\Theta_{\g}\right)^{[k+1]}
=-Q\sum_{j=0}^{k}Y^{[j]\g}\Theta^{[k+1-j]}_{\g}.
$$
For the second term, 
we have
$$
\eqalign{
\left(\Theta,  Q \CW +(\Theta,\CW)\right)^{[k+1]}
&=-Q(\left(\Theta,\CW\right)^{[k+1]}
+\left(Q\Theta+\Fr{1}{2}\left(\Theta,\Theta\right), \CW\right)^{[k+1]}
\cr
&=-Q\left(\Theta,\CW\right)^{[k+1]}
}
$$
where we used the property that $Q$ is a derivation of the BV bracket and the Jacobi
identity of the BV bracket for the first equality and the 
MC equation 
$Q\Theta +\Fr{1}{2}\left(\Theta,\Theta\right)=0\mod t_H^{n+1}$
for the second equality.
Combined together we have
$$
\eqalign{
\CJ^{[k+1]}-\CJ^{\pr[k+1]}
=&\sum_{\ell=0}^{k}\left(\Theta^{[k+1-\ell]},\CY^{\pr[\ell]}-\CY^{[\ell]}\right)
\cr
=&-Q\left(\sum_{j=0}^{k}Y^{[j]\g}\Theta^{[k+1-j]}_{\g}+\left(\Theta,\CW\right)^{[k+1]}\right),
}
$$
to establish the claim.\qed
\end{proof}

\begin{proposition}\label{atech2}
Let $\Theta \in \sC[[t_H]]^0 \mod t_H^{n+1}$ be a versal solution 
to the MC equation modulo $t_H^{n+1}$
and let $\Theta_\g =\rd_\g \Theta \in \sC[[t_H]]^{|\g|} \mod t_H^{n}$.
Let $m\leq n$.
Assume that we have the following equality
$$
Q \CX +\left(\Theta, \CX\right)=C^{\g}\Theta_{\g}\mod t_{H}^{m}
$$
where $\CX \in \sC[[t_H]]^{|\CX|}mod t_H^{m}$ and $C^{\g}\in \Bbbk[[t_H]]\mod t_H^{m}$.
Then
$$
\eqalign{
C^{\g}&=0\mod t_{H}^{m} \quad\forall \g
,\cr
Q\CX +\left(\Theta, \CX\right)&=0\mod t_{H}^{m}
.
}
$$
and

\end{proposition}

\begin{proof}

It is obvious, for the case $m=1$,  since the condition
$Q \CX +\left(\Theta, \CX\right)= C^{\g}\Theta_{\g}\mod t_{H}^{1}$
is equivalent to
$$
Q\CX^{[0]} =C^{[0]\g}\Theta^{[0]}_{\g}=C^{[0]\g}O_{\g}
$$
so that $C^{[0]\g}=0$ for all $\g$ and
$Q\CX^{[0]}=0$.  It follows that 
$$
\CX^{[0]}= B^{[0]\g}\Theta^{[0]}_{\g} +Q \CY^{[0]}.
$$

Fix $j$ such that $2\leq j\leq m-1$ 
and assume 
that 
\eqn\assumna{
\eqalign{
Q\CX +\left(\Theta, \CX\right)&=0\mod t_{H}^{j}
,\cr
C^{\g}&=0\mod t_{H}^{j}\quad \forall \g.
}
}
Then the condition
$$
Q \CX +\left(\Theta, \CX\right)=C^{\g}\Theta_{\g}\mod t_{H}^{j+1}
$$
is equivalent to
\eqn\auytx{
Q \CX^{[j+1]} +\left(\Theta, \CX\right)^{[j+1]}
=C^{[j+1]\g}\Theta^{[0]}_{\g},
}
in addition to \assumna, and  we need to show that $C^{[j+1]\g}=0$ for all $\g$.

From proposition \ref{atech1}, the assumption that 
$Q\CX +\left(\Theta, \CX\right)=0\mod t_{H}^{j}$
implies that
$$
\CX = B^{\g}\Theta_{\g} + Q \CY + \left(\Theta,\CY\right)\mod t_H^{j}
$$
with some $B^{\g}\in \Bbbk[[t_{H}]]\mod t_H^j$  and some 
$\CY\in\sC[[t_{H}]]^{|\CX|-1}\mod t_H^j$. 
Then the term $\left(\Theta, \CX\right)^{[j+1]}$ in \auytx\ becomes
$$
\eqalign{
\left(\Theta, \CX\right)^{[j+1]}
&=\left(\Theta, B^{\g}\Theta_{\g} + Q \CY + \left(\Theta,\CY\right)\right)^{[j+1]}
\cr
&=
\left(\Theta, B^{\g}\Theta_{\g}\right)^{[j+1]} - Q\left(\Theta,\CY\right)^{[j+1]}
+\left(Q\Theta,\CY\right)^{[j+1]} +\left(\Theta, \left(\Theta,\CY\right)\right)^{[j+1]}
\cr
&=
-Q\left(  B^{\g}\Theta_{\g}  +\left(\Theta,\CY\right) \right)^{[j+1]},
}
$$
where we have used the property that $Q$ is a derivation of the bracket, i.e.,
$$
Q \left(\Theta,\CY\right)^{[j+1]}= \left(Q\Theta,\CY\right)^{[j+1]}- \left(\Theta,Q\CY\right)^{[j+1]},
$$
for the $2$nd equality and the assumptions that $Q\Theta_\g + (\Theta, \Theta_\g)=0\mod t_H^n$
and $Q\Theta + \Fr{1}{2}(\Theta, \Theta)=0\mod t_H^{n+1}$ after the Jacobi law of the bracket for the
last equality. Hence \auytx\ implies that
$$
Q\left(\CX- B^{\g}\Theta_{\g}  -\left(\Theta,\CY\right) \right)^{[j+1]}=C^{[j+1]\g}\Theta^{[0]}_{\g}.
$$
It follows that $C^{[j+1]\g}=0$ for $\forall \g$ and $Q \CX^{[j+1]} +\left(\Theta, \CX\right)^{[j+1]}
=0$. Thus we have established that
$C^{\g}=0\mod t_{H}^{j+1}$ for all $\g$ and
$Q \CX +\left(\Theta, \CX\right)=0\mod t_{H}^{j+1}$. Our proposition follows 
by mathematical induction.
\qed
\end{proof}

\end{document}